\documentclass[11pt]{article}
\usepackage[margin=2.5cm]{geometry}
\usepackage{graphicx}
\usepackage{float}
\usepackage{times}
\usepackage{url}
\usepackage[shortlabels]{enumitem}
\usepackage[table]{xcolor}
\usepackage{csquotes}
\usepackage{multicol}
\usepackage{titlesec}
\usepackage[
  backend=biber,
  minnames=999,
  maxnames=999,
  minbibnames=999,
  maxbibnames=999,
  style=ieee,
  giveninits=false,
  citestyle=numeric,
  sortcites=true,
]{biblatex}

\DeclareBibliographyDriver{misc}{%
  \usebibmacro{bibindex}%
  \usebibmacro{begentry}%
  \usebibmacro{author}%
  \setunit{\labelnamepunct}\newblock
  \usebibmacro{title}%
  \newunit\newblock
  \iffieldundef{number}
    {%
      \usebibmacro{publisher+location+date}%
      \newunit\newblock
      \usebibmacro{doi+eprint+url}%
    }
    {%
      \printfield{year}%
      \setunit{\addcomma\space}%
      \printfield{number}%
      \setunit{\addcomma\space}%
      \printtext{Available:}%
      \setunit{\space}%
      \iffieldundef{doi}
        {\printfield{url}}%
        {\printfield{doi}}%
    }%
  \usebibmacro{finentry}%
}

\DeclareFieldFormat[misc]{title}{\mkbibquote{#1}}

\usepackage[bitstream-charter]{mathdesign}
\usepackage[T1]{fontenc}
\usepackage{setspace}

\usepackage{multirow}
\usepackage{array}
\usepackage[font={normal,it},labelfont=bf]{caption}
\usepackage{subcaption}
\usepackage{booktabs}
\usepackage{longtable}
\usepackage{changepage}
\usepackage{textcomp}
\usepackage{ragged2e}
\usepackage{amsmath}
\usepackage[table]{xcolor}
\usepackage{xltabular}
\usepackage{array}
\usepackage{xurl}
\usepackage[
  colorlinks=true,
  linkcolor=blue!50!black,
  citecolor=blue!50!black,
  urlcolor=blue!50!black
]{hyperref}
\usepackage[nameinlink]{cleveref}
\usepackage[titletoc]{appendix}
\usepackage{afterpage}
\usepackage{microtype}
\usepackage{newfloat}
\usepackage{etoolbox}
\usepackage{tocloft}
\usepackage{svg}

\usepackage{tabularx}
\usepackage{colortbl}
\usepackage{xcolor}

\usepackage[most]{tcolorbox}

\newtcolorbox{infobox}{
  colback=gray!10,
  colframe=gray!50,
  boxrule=0.5pt,
  arc=3pt,
  left=6pt,
  right=6pt,
  top=6pt,
  bottom=6pt
}

\definecolor{tablegray}{RGB}{217,217,217}

\newlist{tabitemize}{itemize}{1}
\setlist[tabitemize]{label=\textbullet, leftmargin=*, nosep, before=\vspace*{-0.5\baselineskip}, after=\vspace*{-0.5\baselineskip}}

\definecolor{headergray}{RGB}{238, 238, 235}

\definecolor{bg_green}{RGB}{217, 234, 211}
\definecolor{text_green}{RGB}{56, 118, 29}

\definecolor{bg_yellow}{RGB}{255, 242, 204}
\definecolor{text_yellow}{RGB}{180, 130, 0}

\definecolor{bg_red}{RGB}{244, 204, 204}
\definecolor{text_red}{RGB}{166, 28, 0}

\definecolor{bg_negative}{RGB}{255, 230, 153}
\definecolor{bg_positive}{RGB}{217, 234, 211}

\newcommand{\fullyincluded}{\cellcolor{bg_green}\textcolor{text_green}{Fully included}}
\newcommand{\partiallyincluded}{\cellcolor{bg_yellow}\textcolor{text_yellow}{Partially included}}
\newcommand{\notincluded}{\cellcolor{bg_red}\textcolor{text_red}{Not included}}

\usepackage{tocloft}

\renewcommand\cftsecfont{\vskip-5pt}
\renewcommand\thesection{\arabic{section}}

\usepackage{titlesec}
\titleformat{\section}
  {\huge}{\thesection}{2em}{}
\usepackage{titlesec}
\titleformat{\subsection}
  {\Large}{\thesubsection}{1em}{}

\renewcommand\thesection{\arabic{section}}
\title{\huge Frontier AI Auditing:\\Toward Rigorous Third-Party Assessment of Safety and Security Practices at Leading AI Companies}

\author{Miles Brundage$^{1}$\footnote{Listed authors contributed significant writing, research, and/or review for one or more sections. The sections cover a wide range of empirical and normative topics, so with the exception of the corresponding author (Miles Brundage, miles.brundage@averi.org), inclusion as an author does not entail endorsement of all claims in the paper, nor does authorship imply an endorsement on the part of any individual's organization.}
\quad Noemi Dreksler$^{2}$ \quad Aidan Homewood$^{2}$ \quad Sean McGregor$^{1}$\\
Patricia Paskov$^{3}$ \quad Conrad Stosz$^{4}$ \quad Girish Sastry$^{5}$ \quad  A. Feder Cooper$^{1}$\\
George Balston$^{1}$ \quad Steven Adler$^{6}$ \quad Stephen Casper$^{7}$ \quad  Markus Anderljung$^{2}$\\
Grace Werner$^{1}$ \quad Sören Mindermann$^{5}$ \quad Vasilios Mavroudis$^{8}$ \quad Ben Bucknall$^{9}$\\
Charlotte Stix$^{10}$ \quad Jonas Freund$^{2}$ \quad Lorenzo Pacchiardi$^{11}$ \quad José Hernández-Orallo$^{11}$\\
Matteo Pistillo$^{10}$ \quad Michael Chen$^{12}$ \quad Chris Painter$^{12}$ \quad Dean W. Ball$^{13}$\\
Cullen O'Keefe$^{14}$ \quad Gabriel Weil$^{15}$ \quad  Ben Harack$^{3}$ \quad Graeme Finley$^{5}$ \quad Ryan Hassan$^{16}$\\
Scott Emmons$^{5}$ \quad Charles Foster$^{12}$ \quad Anka Reuel$^{17}$ \quad Bri Treece$^{18}$ \quad Yoshua Bengio$^{19}$ \\
Daniel Reti$^{20}$ \quad Rishi Bommasani$^{17}$ \quad Cristian Trout$^{21}$ \quad Ali Shahin Shamsabadi$^{22}$ Rajiv Dattani$^{21}$ \\
Adrian Weller$^{11}$ \quad Robert Trager$^{3}$ \quad Jaime Sevilla$^{23}$ \quad Lauren Wagner$^{24}$ \quad Lisa Soder$^{25}$\\ 
Ketan Ramakrishnan$^{26}$ \quad Henry Papadatos$^{27}$ \quad Malcolm Murray$^{27}$ \quad Ryan Tovcimak$^{28}$
\vspace{0.3in}\\
\small{$^1$AVERI\quad $^2$GovAI\quad $^3$Oxford Martin AI Governance Initiative \quad $^4$Transluce}\\
\small{$^5$Independent \quad $^6$Clear-Eyed AI \quad $^7$MIT CSAIL \quad $^8$Alan Turing Institute}\\
\small{$^9$University of Oxford \quad $^{10}$Apollo Research \quad $^{11}$University of Cambridge \quad $^{12}$METR}\\
\small{$^{13}$Foundation for American Innovation \quad $^{14}$Institute for Law and AI \quad $^{15}$Touro University Law Center}\\
\small{$^{16}$New Science \quad $^{17}$Stanford University \quad $^{18}$Fathom \quad $^{19}$Mila, Université de Montréal}\\
\small{$^{20}$Exona Lab \quad $^{21}$AI Underwriting Company \quad $^{22}$Brave Software \quad $^{23}$Epoch AI}\\
\small{$^{24}$Abundance Institute \quad $^{25}$interface \quad $^{26}$Yale University \quad  $^{27}$SaferAI \quad $^{28}$UL Solutions}
}
\date{January 2026}

\crefformat{section}{\textbf{#2Section~#1#3}}
\Crefformat{section}{\textbf{#2Section~#1#3}}

\crefmultiformat{section}{\textbf{#2Sections~#1#3}}{ and \textbf{#2#1#3}}{, \textbf{#2#1#3}}{, and \textbf{#2#1#3}}
\Crefmultiformat{section}{\textbf{#2Sections~#1#3}}{ and \textbf{#2#1#3}}{, \textbf{#2#1#3}}{, and \textbf{#2#1#3}}

\crefmultiformat{appendix}{\textbf{#2Appendices~#1#3}}{ and \textbf{#2#1#3}}{, \textbf{#2#1#3}}{, and \textbf{#2#1#3}}
\Crefmultiformat{appendix}{\textbf{#2Appendices~#1#3}}{ and \textbf{#2#1#3}}{, \textbf{#2#1#3}}{, and \textbf{#2#1#3}}
 
\titlespacing*{\subsubsection}{0pt}{0.25\baselineskip}{0.25\baselineskip}
\titlespacing*{\subsection}{0pt}{0.4\baselineskip}{0.4\baselineskip}

\makeatletter
\newcommand{\appsection}[1]{
  \refstepcounter{section}
  \addcontentsline{toc}{appsec}{\protect\numberline{\thesection}#1}
  \section*{\thesection \quad #1}
}

\let\l@appsec\l@section
\patchcmd{\l@appsec}{\cftsecfont}{\cftappsecfont}{}{}
\patchcmd{\l@appsec}{#2}{\cftappsecpagefont #2}{}{}

\apptocmd{\l@appsec}{\setlength{\cftbeforesecskip}{.1ex}}{}{}

\patchcmd{\l@appsec}{\numberline{}{#1}}{\numberline{#1}}{}{}

\makeatother
\newcommand{\cftappsecfont}{\cftsecfont}
\newcommand{\cftappsecpagefont}{\cftsecpagefont}

\crefformat{appsec}{\textbf{#2Appendix~#1#3}}
\Crefformat{appsec}{\textbf{#2Appendix~#1#3}}
\crefmultiformat{appsec}{\textbf{#2Appendices~#1#3}}{ and \textbf{#2#1#3}}{, \textbf{#2#1#3}}{, and \textbf{#2#1#3}}
\Crefmultiformat{appsec}{\textbf{#2Appendices~#1#3}}{ and \textbf{#2#1#3}}{, \textbf{#2#1#3}}{, and \textbf{#2#1#3}}
\renewcommand\thesubsection{\thesection.\arabic{subsection}}

\makeatletter
\renewcommand\@makefntext[1]{%
    \parindent 1em%
    \noindent
    \hb@xt@1.8em{\hss\@makefnmark}\hspace{0.3em}#1}
\makeatother

\begin{document}

\maketitle

\newpage 
\section*{Executive Summary} \label{sec:executive}

\subsection*{Key paper takeaways} \label{ssec:keytakeaways}

\begin{itemize}[leftmargin=0.25in]

    \item Despite their rapidly growing importance, AI systems are subject to less rigorous third-party scrutiny than many of the other social and technological systems that we rely on daily such as consumer products, corporate financial statements, and food supply chains. This gap is becoming increasingly untenable as AI becomes more capable and widely deployed, and it inhibits confident deployment of AI in high-stakes contexts.

    \item Transparency alone cannot enable well-calibrated trust in the most capable (``frontier'') AI systems and the companies that build them: many safety- and security-relevant details are legitimately confidential and require expert interpretation, and third parties are right to be skeptical of companies ``checking their own homework'' given the track record of that approach in other industries.

    \item We outline a vision for \textit{frontier AI auditing}, which we define as rigorous third-party verification of frontier AI developers' safety and security claims, and evaluation of their systems and practices against relevant standards, based on deep, secure access to non-public information.

    \item Frontier AI audits should not be limited to a company's publicly deployed products, but should instead consider the full range of organization-level safety and security risks, including internal deployment of AI systems, information security practices, and safety decision-making processes. 
    
    \item We describe four AI Assurance Levels (AALs), the higher levels of which provide greater confidence in audit findings. We recommend AAL-1 as a baseline for frontier AI generally, and AAL-2 as a near-term goal for the most advanced subset of frontier AI developers.

    \item Achieving the vision we outline will require (1) ensuring high quality standards for frontier AI auditing, so it does not devolve into a checkbox exercise or lag behind changes in the industry; (2) growing the ecosystem of audit providers at a rapid pace without compromising quality; (3) accelerating adoption of frontier AI auditing by clarifying and strengthening incentives; and (4) achieving technical readiness for high AI Assurance Levels so they can be applied when needed.
\end{itemize}

\subsection*{Frontier AI auditing motivations} \label{ssec:motivation}

Artificial intelligence (AI) is rapidly becoming critical societal infrastructure. Every day, AI systems inform decisions that affect billions of people. Increasingly, they also make consequential decisions autonomously. 
Although these technologies hold incredible promise, the pace of development and deployment has outpaced the creation of institutions that ensure AI works safely and as advertised.

This institutional gap is especially important for the most capable (``frontier'') systems --- general-purpose AI models and systems whose performance is no more than a year behind the state-of-the-art --- which many experts expect to exceed human performance across most tasks within the coming years. Already, developers of frontier AI systems need to prevent harmful system failures (e.g., outputting false medical information or buggy code), weaponization by malicious parties (e.g., to carry out cyberattacks), and theft of or tampering with sensitive data. The magnitude of risks that need to be managed is growing rapidly.\looseness=-1

AI users, policymakers, investors, and insurers need reliable ways to verify that promised technical safeguards exist and to detect when they do not. This is challenging because the technology is complex, fast-moving, and often proprietary. Public transparency alone cannot solve this problem since many key details are --- and often should remain --- confidential, and require expert judgment to interpret. Many industries outside of AI already address similar challenges through independent auditors who review sensitive, non-public information and publish trustworthy conclusions that outsiders can rely on. We argue that similar practices are needed in the AI industry: broad, sustainable adoption of AI over time requires a solid foundation of trust built on credible scrutiny by independent experts.

Toward this end, we propose institutions designed to give stakeholders --- including those who are uncertain about or even strongly skeptical of frontier AI companies --- justified confidence that this critical technology is being developed safely and securely. Specifically, we describe and advocate for \textbf{frontier AI auditing}: rigorous third-party verification of frontier AI developers' safety and security claims, and evaluation of their systems and practices against relevant standards, based on deep, secure access to non-public information.\looseness=-1

An ecosystem of private sector frontier AI auditors (both for-profit and non-profit) would enable widespread confidence that frontier AI systems can be adopted broadly and would avoid reliance on companies ``grading their own homework,'' an approach with a checkered track record in many industries. It would also avoid relying entirely on governments to have the technical expertise, capacity, and agility to ensure high standards for frontier AI safety and security. If well-executed and scaled, frontier AI auditing would improve safety and security outcomes for users of AI systems and other affected parties, create a system to learn and update standards based on real-world outcomes, and enable more confident investment in and deployment of frontier AI, especially in high-stakes sectors of the economy.

\subsection*{Summary of the proposal} \label{ssec:proposal}

Drawing on our analysis of current practices in AI and lessons from other industries with more mature assurance regimes, we recommend eight interlinked design principles for a long-term vision for frontier AI auditing. This vision is deliberately ambitious to match the rising stakes as frontier AI capabilities advance:\looseness=-1

\begin{itemize}[leftmargin=.25in]
        \item \textbf{Scope of risks: Comprehensive coverage of four key risk categories.} Frontier AI auditing should focus on four risk categories: risks from (1) intentional misuse of frontier AI systems (e.g., for cyberattacks); (2) unintended frontier AI system behavior (e.g., errors harming the user, their property, or third parties due to pursuing the wrong goal or having an unreliable performance profile); (3) information security (e.g., theft of an AI model or user data); and (4) emergent social phenomena (e.g., addiction to AI or facilitation of self-harm). For each category of risks, auditors should (a) verify company claims and (b) evaluate the company's systems and practices against its stated safety and security policies, applicable regulations, and industry best practices.

    \item \textbf{Organizational perspective: Auditing companies' safety and security practices as a whole, not just individual models and systems.} Auditors should use an organization‑level perspective to avoid abstraction errors (i.e., forming the wrong conclusion by treating a partial or simplified unit of analysis, such as evaluating a specific component in isolation, as if it were sufficient to assess overall system and organizational risk). 
    Risk does not come from AI models alone; it emerges from the interaction of three overarching components: 
    digital systems, computing hardware, and governance practices, and harm can arise even when a model is never deployed in external-facing systems. 
    Rigorous, but isolated, model and system evaluations are therefore insufficient to evaluate all safety and security claims on their own. 
    And while individual audits may focus on particular domains depending on their goals, the ecosystem as a whole should ensure comprehensive coverage across all three components in assessing safety and security claims.

    \begin{figure}[t!]
    \centering
    \includegraphics[width=0.8\textwidth]{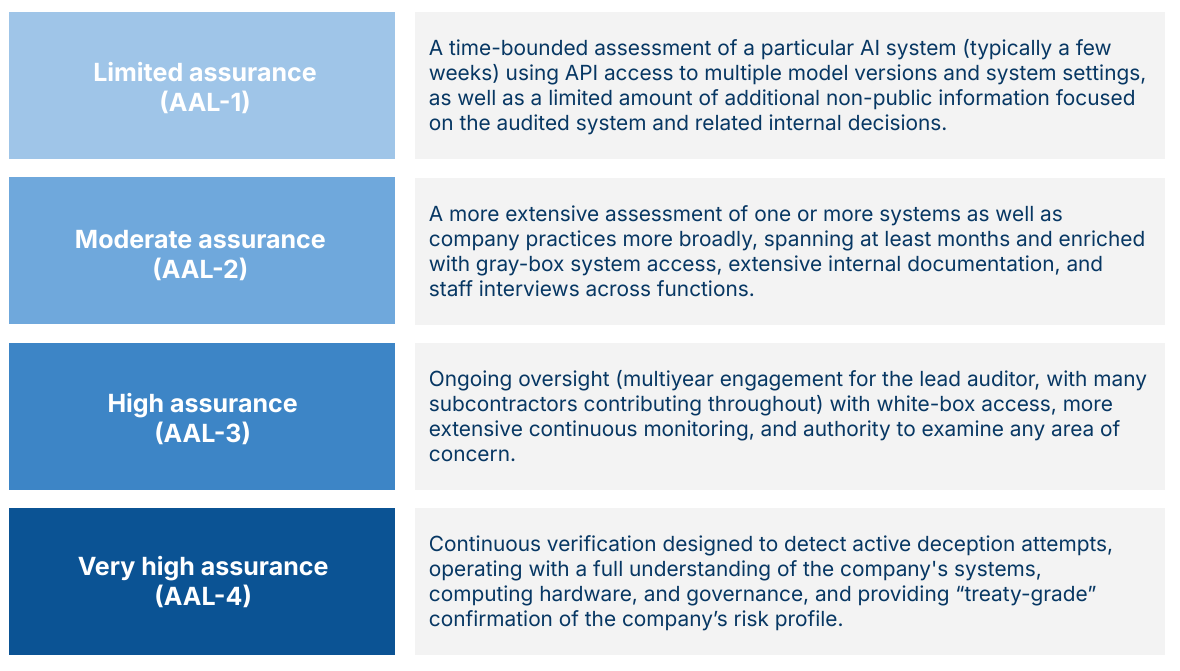}
    \caption{Four AI Assurance Levels (AALs) for different frontier AI audits.}
    \label{fig:execsum-aals}
\end{figure}

    \item \textbf{Levels of assurance: A framework for calibrating and communicating confidence in audit conclusions.} Not all audits provide the same level of certainty, and stakeholders need to understand these differences. We propose \textbf{AI Assurance Levels (AALs)} as a means of clarifying what kind of assurance particular frontier AI audits provide (Figure~\ref{fig:execsum-aals}). At lower levels, auditors and other stakeholders rely more heavily on information provided by the company and can primarily speak to a particular system's properties. At higher levels, auditors take fewer assumptions for granted, and assess the full range of relevant company systems, organizational processes, and risks. At the highest level, auditors can rule out the possibility of materially significant deception by the auditee. Determining the appropriate AAL for different contexts and purposes is complex, but we recommend AAL-1 (the peak of current practices in AI) as a starting point for frontier AI generally, and AAL-2 as a near-term goal for the companies closest to the state-of-the-art. AAL-2 involves greater access to non-public information, less reliance on companies' statements, and a more holistic assessment of company-level risks. 
    The two highest assurance levels (AAL-3 and AAL-4) are not yet technically and organizationally feasible, but we outline research directions to change this.\looseness=-1

    \item \textbf{Access: Deep enough to assure auditors and other stakeholders, secure enough to reassure auditees.} Frontier AI auditors should receive deep, secure access to non-public information of various kinds --- including model internals, training processes, compute allocation, governance records, and staff interviews --- proportional to the audit's scope and the level of assurance being sought for the audit. Access arrangements should protect intellectual property and security‑sensitive information using mechanisms imported from other domains (e.g., sharing certain information with a subset of the auditing team on-site under a restrictive nondisclosure agreement) and newly-developed techniques (e.g., AI-powered summarization or analyses of information that is too sensitive to be directly shared).

    \item \textbf{Continuous monitoring: Living assessments, not stale PDFs.} AI systems change constantly, including through adjustments to the underlying model(s), surrounding software, and shifts in user behavior. An audit conclusion that was accurate at the time of the assessment may become misleading in some respects within days or weeks. Audit findings should therefore carry explicit assumptions and validity conditions, and should be automatically deprecated when key underlying assumptions no longer hold. A mature auditing ecosystem will combine periodic deep assessments of slower-moving elements (e.g., governance, safety culture) with event-triggered reviews of major changes (e.g., new releases, serious incidents) and continuous automated monitoring of fast-changing surfaces (e.g., API behavior, configuration drift), enabling timely detection of changes that could invalidate prior conclusions.\looseness=-1 

    \item \textbf{Independent experts: Trustworthy results through rigorous independence safeguards and deep expertise.} Auditors must be genuinely independent third parties, free from commercial or political influence, and have deep expertise across AI evaluation, safety, security, and governance. Safeguarding independence requires mandatory disclosure of financial relationships, standardized terms of engagement that prevent companies from shopping for favorable auditors, and cooling-off periods when moving, in both directions, between industry and audit roles. Alternative payment models that reduce auditor dependence on auditees should also be urgently explored. Where single auditing organizations lack sufficient expertise, subcontracting and consortia models can enable the necessary breadth across AI evaluation, safety, security, and governance.

    \item \textbf{Rigor: Processes that are methodologically rigorous, traceable, and adaptive.} Audits should follow a standardized process while giving auditors the autonomy to flexibly determine specific methods and adjust scope as issues emerge. Auditors should be able to define evaluation metrics and criteria rather than simply validating companies' preselected approaches. Wherever feasible, audit procedures should be automated, transparent, and reproducible to support consistent application across engagements and enable continuous monitoring as systems evolve. Auditors need to safeguard evaluation construct and ecological validity, and audit criteria should be protected against gaming. Finally, audits should incorporate procedural fairness, giving companies structured opportunities to correct factual errors while preventing undue influence on conclusions.

    \item \textbf{Clarity: Clear communication of audit results.} Stakeholders must be able to understand the audit results. These should be communicated in audit reports with a standardized structure, covering the audit's scope, level of assurance, conclusions, reasoning, and recommendations. Results should be communicated appropriately to different stakeholders: to protect sensitive information, auditors and companies can publish summarized or redacted versions for external stakeholders while sharing full, unredacted audit reports with boards, company executives, and, in some cases, regulatory bodies.\looseness=-1

\end{itemize}

\subsection*{Challenges and next steps} \label{ssec:challenges}

Our long-term vision will require concrete efforts by several categories of stakeholders to both achieve and maintain. The most urgent challenges are:

\begin{itemize}[leftmargin=0.25in]
    \item \textbf{Ensuring high quality standards} for frontier AI auditing, so it does not devolve into a checkbox exercise or lag behind changes in the AI industry.

    \item \textbf{Growing the ecosystem} of audit providers at a rapid pace without compromising quality.

    \item \textbf{Accelerating adoption} of frontier AI auditing by clarifying and strengthening incentives.

    \item \textbf{Achieving technical readiness} for high AI Assurance Levels so they can be applied when needed.
\end{itemize}

These challenges are substantial but not unprecedented. Companies routinely share sensitive information with financial auditors, potential acquirers, penetration testers, and consumer product testing laboratories under carefully controlled terms. We believe similar practices for AI safety and security are both achievable and urgently needed. For each of the challenges we describe, we recommend specific next steps:

\begin{figure} [H]
    \centering
    \includegraphics[trim={0cm 0cm 0cm 0cm},clip,width=0.9\linewidth]{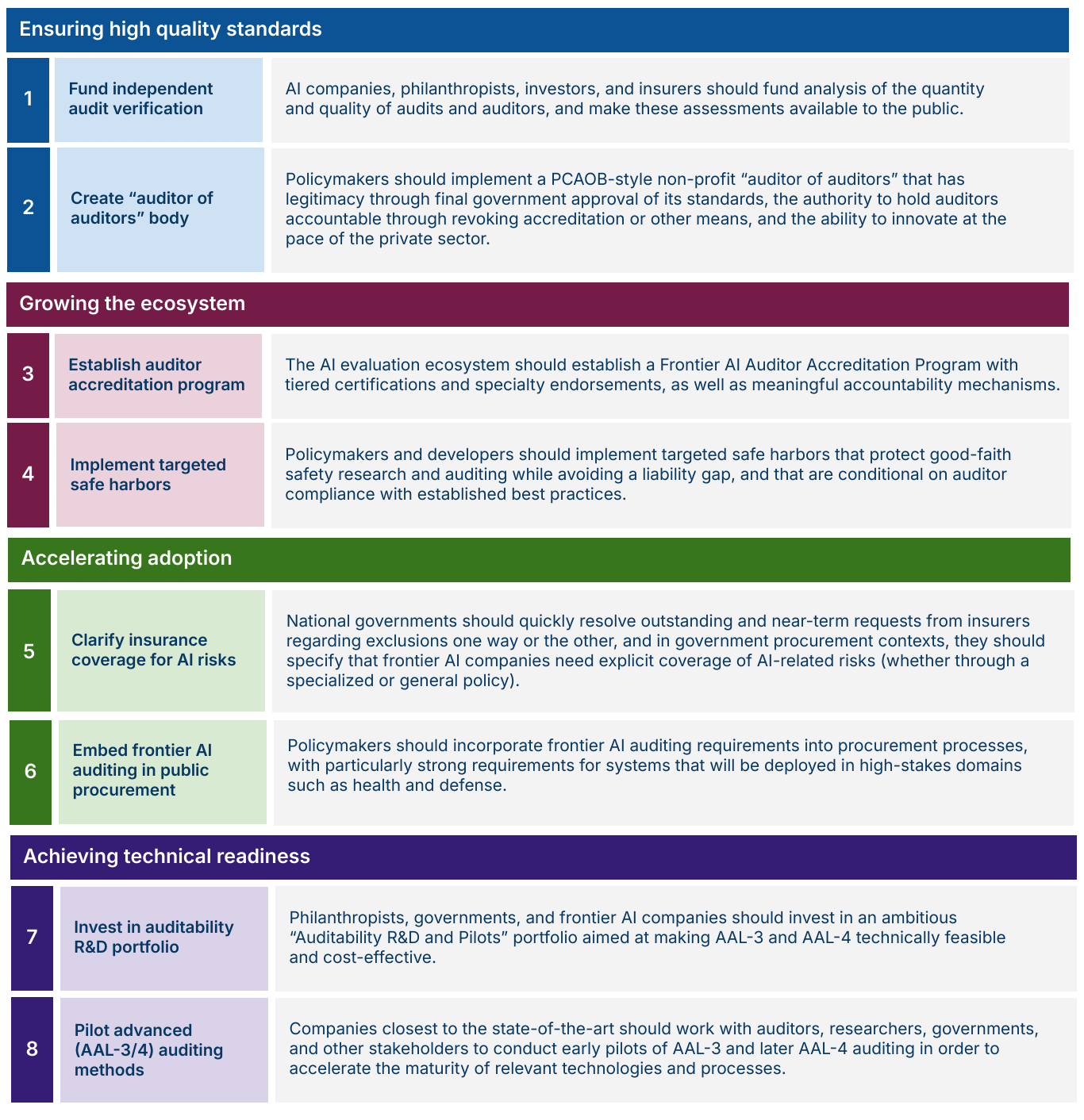}
    \caption{Recommendations for next steps across four challenges in frontier AI auditing.}
    \label{fig:recommendations}
\end{figure}

Keeping up with the rapid pace of AI progress and deployment requires quickly importing best practices from more mature industries and immediate investment in auditing pilots, technical research, and policy research. Moving with urgency is essential if frontier AI auditing is to reach maturation and scale alongside AI development.

\newpage 
{\begin{spacing}{0.95}
\tableofcontents
\end{spacing}}

\newpage 
\section{Introduction}\label{sec:introduction}

\hyperlink{gls:frontierai}{Frontier AI systems} are rapidly transforming society \autocite{bengio_international_2025, maslej_artificial_2025}. While many \hyperlink{gls:frontieraidevelopers}{frontier AI developers} invest substantially in \hyperlink{gls:safety}{safety} and \hyperlink{gls:security}{security}, the details of how these systems are built, evaluated, and safeguarded remain largely opaque to external stakeholders \autocite{wang_2025_foundation_2025} --- users, insurers, investors, and policymakers alike. And yet, external stakeholders are the ones most affected by the systems' downstream impacts, bearing both the benefits and risks of outcomes that they have little ability to scrutinize or contest. 

Today, most third-party \hyperlink{gls:assessment}{assessments} rely on public information and publicly accessible products \autocite{reuel_who_2025}. These are valuable but insufficient. Such public information rarely provides the detail necessary to offer meaningful assurances about system behavior, including the precise conditions under which an \hyperlink{gls:evaluation}{evaluation} was conducted, whether conclusions drawn from evaluations meaningfully generalize, or whether negative findings were softened or withheld. As frontier AI systems become more capable and more widely deployed, assurance provided solely by analyzing information and systems that frontier AI companies choose to disclose publicly will be less and less tolerable.\looseness=-1

Meaningful assurance requires \hyperlink{gls:independence}{independent} access to non-public technical and organizational information \autocite{casper_black-box_2024}. When developers do grant such access, they typically do so through bespoke contracts with terms rarely visible to the public (see \autocite{openai_openai_2023}). Standards for how companies should work with third-party AI assessment organizations are nascent, and developers can substantially influence the timing, scope, and publication of any assessments conducted that involve non-public information.\looseness=-1

This opacity contrasts sharply with oversight of other critical technologies and institutions in at least two important respects. First, frontier AI is advancing at unprecedented speed, outpacing the development of corresponding governance frameworks and structures \autocite{reuel_open_2025, maslej_artificial_2025}. Second, unlike earlier transformative technologies that were deployed within more specific, well-defined domains, frontier AI is being integrated horizontally across many critical sectors at once. By contrast, many sector-specific technologies --- cars, airplanes, food supply chains --- are often subject to more rigorous independent scrutiny \autocite{hodgkinson_iosa_2005, koppel_how_2008, warriner_understanding_2013}. In many cases, such oversight began only after major disasters \autocite{johnson_process_2005, baker_institutional_2006, sutton_sems_2014}, and avoiding the greatest risks of AI systems requires greater foresight to forestall catastrophe. By forming “soft law” \autocite{wallach_soft_2023}, early norms around access, evaluation, and disclosure can play a crucial role in shaping what more formal, future oversight can realistically demand.\looseness=-1

To meet the demands of this critical moment, this paper presents our vision: \hyperlink{gls:frontieraiauditing}{frontier AI auditing}. We define frontier AI auditing as rigorous third-party \hyperlink{gls:verification}{verification} of frontier AI developers' safety and security claims about systems throughout development and deployment, combined with the ongoing evaluation of their systems and practices against relevant standards, based on deep, secure access to non-public information.\looseness=-1

Rather than suggesting a one-size-fits-all approach to all audits, our goal is to provide an initial vocabulary for this emerging field --- one that can be adapted based on context, including the \hyperlink{gls:level}{level of assurance} required in a particular case. Toward this end, we propose standardized \hyperlink{gls:levels}{AI Assurance Levels (AALs)} that clarify the assumptions required in order to trust a particular audit's conclusions.

Auditing complements internal safety practices and external regulation --- it doesn't replace them. Critically, there needs to be a meaningful set of safety and security standards against which companies can be audited and for which there are strong incentives to comply. Auditing also entails real costs that rise with the assurance level being sought. While some costs may fall through automation and other efficiency improvements, some costs will always remain, making it important to efficiently allocate auditing activity while maintaining high-quality standards.

Our paper proceeds as follows:
\begin{itemize}[leftmargin=0.25in]
    \item \cref{sec:key} defines the key terminology used in the paper and clarifies the scope of the discussion.

    \item \cref{sec:motivations} explains why frontier AI auditing is an appropriate response to the growing gap between AI's impact and the external scrutiny applied to it. 

    \item \cref{sec:lessons} draws lessons from other industries with mature third-party assurance ecosystems and examines current assessment practices in the AI sector.

    \item \cref{sec:vision} details our vision for frontier AI auditing, covering scope, organization-level focus, assurance levels, access needs, \hyperlink{gls:continuousmonitoring}{continuous monitoring}, independence, rigor, and communication of outputs.

    \item \cref{sec:challenges} describes the challenges in achieving this vision and proposes directions for addressing them.

    \item \cref{sec:conclusion} takes stock of the paper's contributions and summarizes proposed next steps.

    \item \hyperref[apx:glossary]{\textbf{Appendices}} cover a range of topics: a glossary of key terms; additional motivations for frontier AI auditing not covered in Section 3; details on access requirements; placing AI auditing in context of other AI policy topics; more details on our lessons learned from other domains and from contemporary AI assessment practices; and a more detailed discussion of different ways that frontier AI auditing could be funded.
\end{itemize}

\subsection*{How to read this paper} \label{ssec:how_to_read}

Those less familiar with the terminology used above should read \cref{sec:key} and consult \cref{apx:glossary} as needed. Most readers should read \cref{sec:motivations} in order to understand the specific problems we think frontier AI auditing would help address. Those familiar with current AI assessment limitations can skip to \cref{ssec:lessons_from_established_domains} for lessons from other industries, then proceed directly to \cref{sec:vision} for our proposal. Researchers, engineers, industry executives, policymakers, and philanthropists interested in ways they can accelerate frontier AI auditing may be especially interested in \cref{sec:challenges} (Challenges and Next Steps). \hyperref[apx:glossary]{\textbf{Appendices}} may be of interest to researchers or policymakers interested in various specific details.\looseness=-1

\newpage 
\section{Key Terminology and Scope}\label{sec:key}

\begin{figure} [H]
    \centering
    \includegraphics[trim={0cm 0cm 0cm 0cm},clip,width=0.5\linewidth]{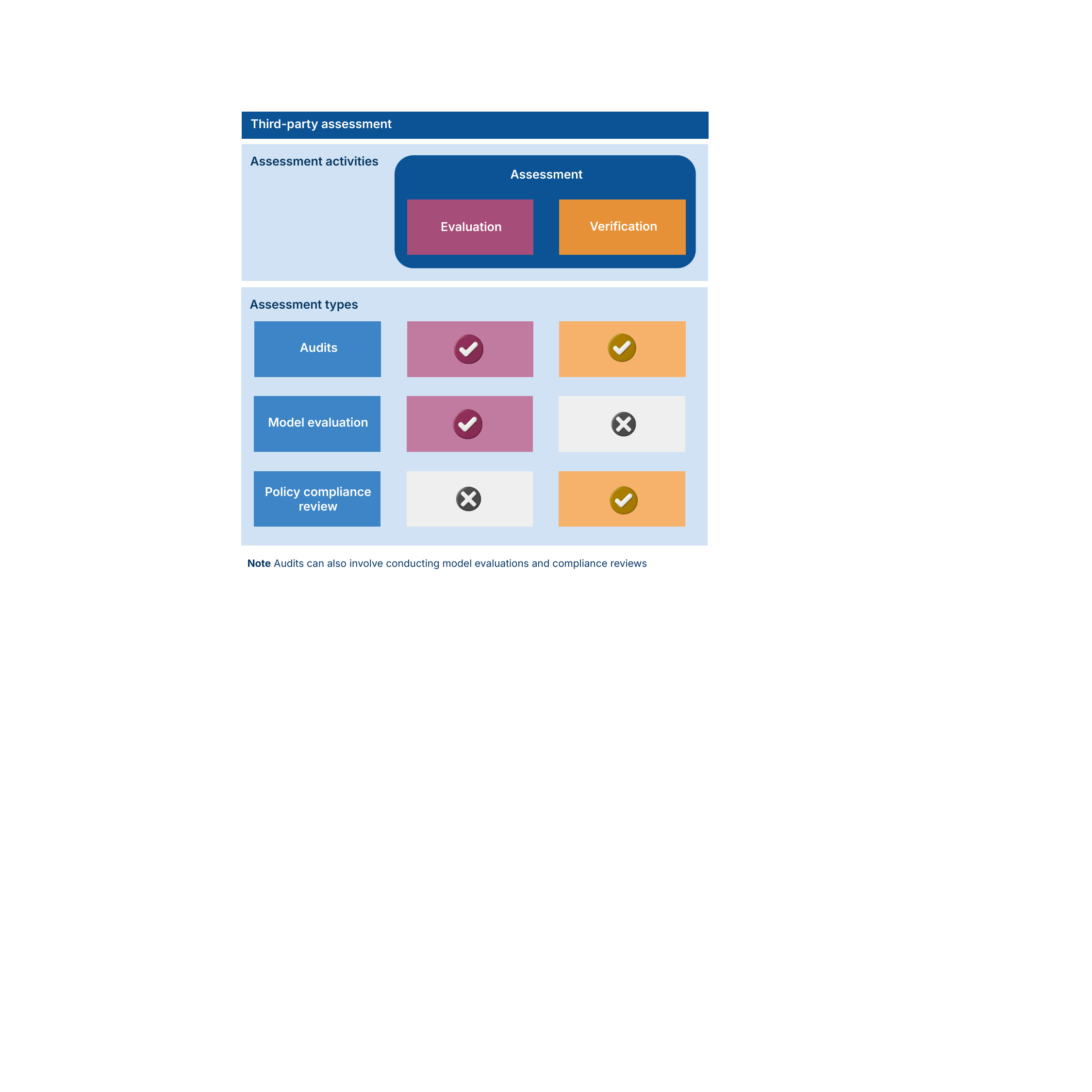}
    \caption{Understanding key concepts: how assessments relate to evaluation, verification, and audits.}
    \label{fig:assessment}
\end{figure}

In this report, we describe and advocate for \hyperlink{gls:frontieraiauditing}{frontier AI auditing}, which we define as (1) rigorous third-party evaluation of frontier AI developers' systems and practices against relevant standards and (2) rigorous third-party verification of frontier AI developers' safety and security claims, both based on deep, secure access to non-public information. In general, when we refer to \hyperlink{gls:verification}{verification}, we mean the activity of confirming whether a specific claim, commitment, or property (e.g., a training compute figure) is true. \hyperlink{gls:evaluation}{Evaluation} refers to any activity that measures, characterizes, or analyzes properties of AI models or systems and the organizations operating them. More generally, assessments are activities that involve evaluation, verification, or both (see Figure \ref{fig:assessment}). And so the frontier AI audits we describe are a particular type of third-party \hyperlink{gls:assessment}{assessment} concerning AI systems and the companies that build them.

In this section, we clarify the specific scope of our focus on frontier AI auditing.

\begin{figure}
    \centering
    \includegraphics[trim={0cm 0cm 0cm 0cm},clip,width=0.5\linewidth]{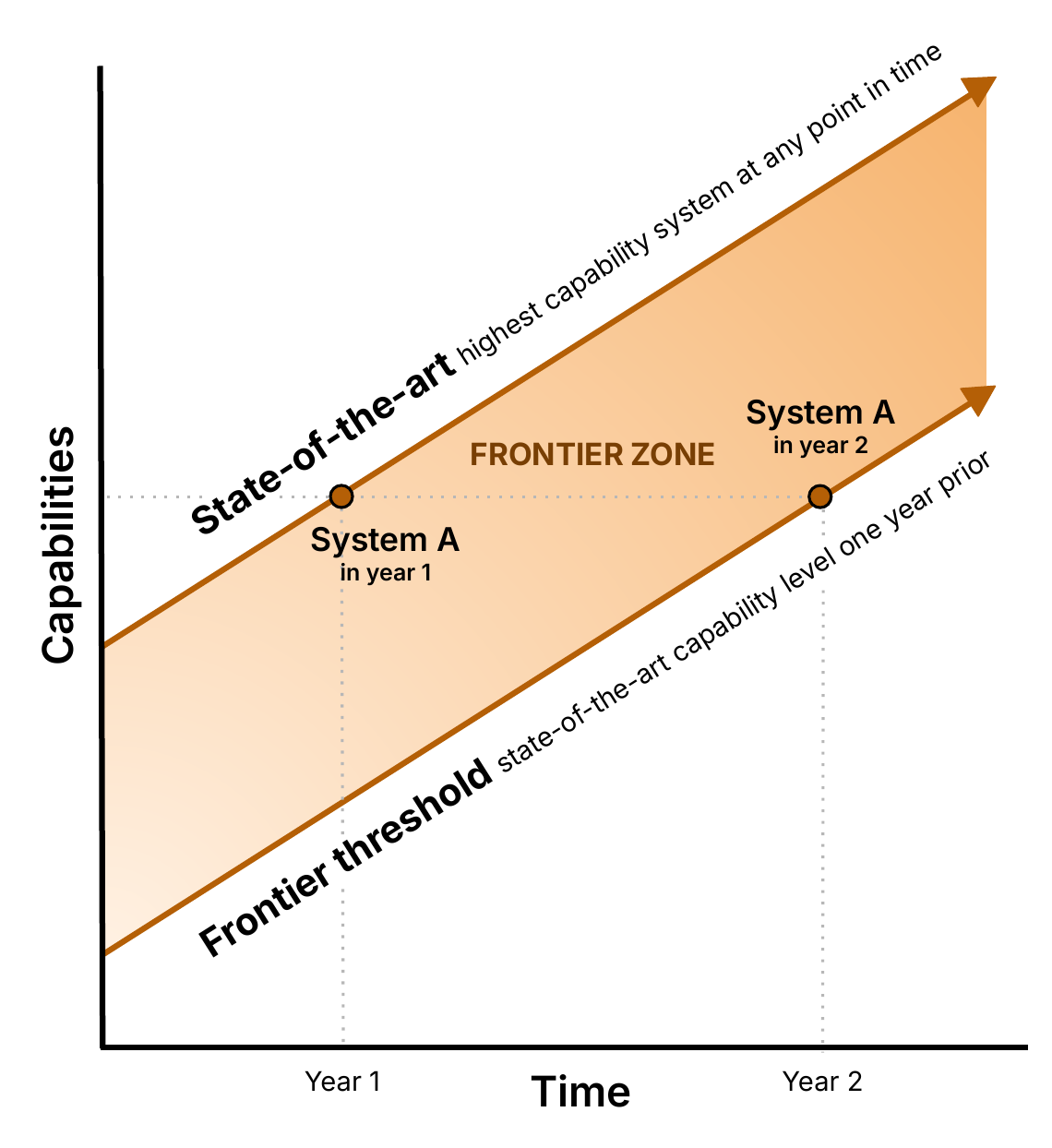}
    \caption{Understanding frontier AI. The frontier advances over time as the state-of-the-art capability level increases, so a system is considered frontier if it remains within a fixed lag (e.g., 1 year) of the state-of-the-art at a given point in time.}
    \label{fig:frontier-nature}
\end{figure}

\textbf{We focus specifically on frontier AI.} Frontier AI includes general-purpose AI models and systems whose performance is no more than a year behind the state-of-the-art on a broad suite of general capability benchmarks (see Figure~\ref{fig:frontier-nature}). The risks and capabilities of the most capable systems, all else equal, are the least well understood by virtue of being new \autocite{ganguli_predictability_2022, cooper_report_2023}, making rigorous auditing particularly important. With respect to the AI components of frontier AI systems, we focus on \hyperlink{gls:closedweight}{closed-weight models}, or those whose weights are not publicly released and thus cannot be freely copied or redistributed. \hyperlink{gls:open}{Open-weight models} have distinct challenges and affordances, and only a subset of what we discuss here is relevant to them.\footnote{In short, our auditing framework is intended to work around access challenges that are not always in the open-weight context, and to assess mitigations and governance processes that are also quite different in an open-weight context. Related questions about open-weight model safety and evaluation are discussed in more detail in \autocite{bommasani_considerations_2024, seger_open_2024, francois_different_2025, srikumar_risk_2024, casper_open_2025, kapoor2024societal}.} We define frontier AI in terms of capabilities --- rather than through characteristics like computational power or money spent on research and development --- because capabilities are more directly related to the risks resulting from the production and deployment of frontier AI systems. 

There are many other ways to define frontier AI, as well as different ways to determine whether an AI system or company should be audited, besides frontier status. While important, we don't think these differences significantly change the basic ideas in the rest of the paper, so we reserve more detailed discussion to \cref{apx:defs}.

\textbf{We focus on companies that develop frontier AI.} Frontier AI developers are the companies that produce frontier AI systems: the entities that train models from scratch themselves, or significantly extend model capabilities (e.g., via further training of a preexisting frontier AI model, or the addition of significant new system-level features to a model, such as tool use). Developers often (though not always) deploy these models themselves, embedding models in software systems for both internal and external use. Many companies only deploy or integrate systems built by others, but do not significantly extend the underlying models’ capabilities, and we do not discuss auditing of those companies here.\footnote{A company ``deploys'' AI systems if they make these systems available for direct, economically relevant use by internal or external parties, even if the company does not train their own models. Companies may develop frontier AI models for internal research, only deploying them internally (see \autocite{openai_better_2019}). A developer could also make AI systems available for deployment exclusively by others without deploying these models themselves. If customers conduct fine-tuning themselves on pre-trained models, the company hosting the training service wouldn’t be considered a developer under our use of the term.} We focus on developers because they control the basic nature of AI systems’ capabilities and behaviors. Unless otherwise specified, we use ``developer'' and ``company'' interchangeably to refer to the frontier AI developer.\footnote{This is admittedly imprecise, since a developer could be a non-profit organization or a government, though for-profit companies are the most frequent type of frontier AI developer as well as the most frequent type of frontier AI deployer.} 

\textbf{We focus on frontier AI systems throughout their development and deployment.} Organizations and software systems evolve over time. This is particularly the case for frontier AI. Even with a seemingly stable set of companies at the frontier, new developers can enter the market at any time. Beyond rapid advances in underlying model capabilities, frontier AI systems are continually reshaped by new product offerings, deployment contexts, and design paradigms, such as the integration of models into increasingly autonomous, agentic scaffolds. As a result, assessment of organizations and system behaviors at a single point in time (e.g., pre-deployment) can quickly become outdated or incomplete. Ongoing assessment enables tracking how organizations and systems change in practice, how new uses and incentives change the risk landscape, and how emergent behaviors arise as technologies are combined and scaled in new ways.\looseness=-1

\textbf{We focus on third-party assessments of frontier AI that involve non-public information.} When people make decisions on which AI system to license or use, they typically rely on a combination of assessments produced by the company itself and third-party organizations. While the company statements may benefit from substantial insider access, they lack the credibility afforded by statements free of conflicts of interest (i.e., statements from third-parties). However, while third-party assessors may be free of conflicts of interest, they often lack access adequate for assessing risk. Unprivileged assessments have value, and there are various respects in which frontier AI companies could and should share more information publicly than they do today \autocite{ball_four_2024}. But external information is often misleading \autocite{mcgregor_risk_2025} and may not reveal major risks or vulnerabilities that lie below the surface. We expect that non-public information will become increasingly important over time.\footnote{There are at least four reasons for this: (1) Developers might only deploy certain dangerous systems internally \autocite{stix_ai_2025}, or only give access to them to certain customers, as in the case of systems that have fewer guardrails \autocite{anthropic_claude_2025, openai_strengthening_2025-1}, limiting what can be learned from testing public products. (2) Since AI systems are already being misused by criminals and state actors \autocite{anthropic_disrupting_2025, openai_disrupting_2025, google_threat_intelligence_group_gtig_2025}, developers may respond by publishing more limited information about some aspects of these systems and their risks in order to prevent this information from itself being misused, e.g., to more easily ``break'' system safeguards. Indeed, some details are already explicitly withheld on that basis, such as the details of chemical, biological, radiological, and nuclear (CBRN) threat models \autocite{noauthor_ai_2025, gemini_gemini_2025, google_threat_intelligence_group_gtig_2025}. (3) AI systems will likely become more complex (e.g., with multiple interacting models solving complex problems whose solutions are difficult to understand even with the benefit of access to the models' chains-of-thought, let alone without it), thus reducing the information value of the final output as a means of understanding how it was produced and what potential failure modes might exist in that system. (4) AI systems will become more capable, likely increasing the worst-case risks associated with outputs that may be intentionally deceptive, as well as increasing the likelihood of strategically low-performing behavior by systems \autocite{noauthor_more_2025}. }

\textbf{We develop a ``menu'' of different assurance levels, based on the effort applied to auditing.} Audits are a particular kind of assessment: a systematic, evidence-based process where a qualified party examines an organization’s activities, records, technologies, and claims in order to provide assurance that stated information is accurate and/or that applicable standards are being met. This definition leaves open the question of how much effort should be applied in order to build the auditor’s confidence and skeptical third parties’ confidence in the audit’s findings. We propose four different AI Assurance Levels (AALs), corresponding to different degrees of assurance gained through an audit process. Higher levels will tend to be more costly and time-consuming because the auditors take fewer assumptions for granted and verify more claims directly, with the goal of ultimately attaining enough confidence to rule out deception, not just error, and to identify subtle errors that take careful investigation to detect.

We think that there will ultimately be some frontier AI systems that merit the highest AALs: as capabilities improve, the standard of analysis should increase proportionally, and stakeholders will rightly want more confidence that critical risks will not materialize. We therefore recommend a range of directions aimed at making it feasible to reach these high AALs. 

\clearpage
\textbf{Frontier AI auditing takes place ``upstream'' in the supply chain by design.} Safety and security improvements ``upstream'' in the supply chain improve outcomes across many ``downstream'' applications built on top of frontier AI systems, and are therefore highly leveraged \autocite{anderljung_frontier_2023}. An audit of a downstream company would be fully credible only if the frontier model or system on which they are building has itself been audited. But achieving effective AI safety and security risk management will likely require a layered approach combining frontier AI auditing with context-specific and sector-specific risk management and deployment-level assessments, to evaluate how these systems act in particular high-risk settings. Frontier AI auditing is an essential piece of the puzzle, but needs to be viewed in a larger context (see \cref{apx:frontier}).\looseness=-1

\newpage 
\section{Motivations: Why Frontier AI Auditing is Needed}\label{sec:motivations}

In this section, we articulate why frontier AI auditing is necessary for meeting the challenges of this current moment --- to provide meaningful assurance to external stakeholders that bear the benefits and risks of frontier AI systems. We focus on two motivations for frontier AI auditing: how it can lead to significantly improved safety and security outcomes (\cref{ssec:improving_outcomes}), and enable more confident investment and deployment (\cref{ssec:enabling}). These motivations often point toward similar auditing practices, but sometimes they diverge in emphasis or design (\cref{ssec:different_audit}). This is one reason why we propose a “menu” of assurance levels, which specify how much effort should be applied in different contexts to instill confidence in an audit’s findings. These motivations also point to the need for a private sector-based auditing regime (\cref{ssec:government_vs_private}).

\cref{apx:additional_motivations} discusses three additional motivations (enabling risk price discovery through insurance, maintaining international stability, and ensuring accountability for risk creation).

\subsection{Improving safety and security outcomes} \label{ssec:improving_outcomes}

Internal evaluations of frontier AI systems tend to be insufficient along two distinct but reinforcing dimensions: (1) limits in frontier AI developers’ abilities to fully understand, anticipate, and characterize the external risks of systems they develop, and (2) misaligned incentives. Auditing can directly address these issues, and the resulting safety and security risks they bring about, by introducing perspectives that challenge internal company narratives, encouraging better internal practices for system assessment, and sharing critical advancements in AI safety and security knowledge across organizations. 

First, relative to self-assessment by AI developers, external auditing provides fresh perspectives, which can offer healthy skepticism (i.e., guard against groupthink\footnote{Organizational psychology research documents “groupthink” as a pervasive risk in cohesive teams --- e.g., self-censorship of doubts and collective rationalization of warnings. These dynamics can render the possibility of failure unthinkable or at least unspeakable \autocite{reason_life_2013}. Independent third-party auditors provide a structural countervailing force against these tendencies.}), while also expanding the range of expertise brought to bear on development and deployment decisions. There is already evidence that third-party assessment can surface safety and security issues that developers subsequently remedy \autocite{raji_actionable_2019}. For example, the UK AI Security Institute (AISI) and US Center for AI Standards and Innovation (CAISI)’s pre-deployment testing efforts \autocite{aisi_pre-deployment_2024} identified safety issues that developers then addressed before release \autocite{anthropic_strengthening_nodate, openai_working_2025}. Developers have also noted the value of external review in strengthening internal evaluation processes. \hyperlink{gls:systemcards}{System cards} documenting behaviors and risks frequently reference third-party benchmarks and findings \autocite{rottger_safetyprompts_2025}, sometimes produced with non-public information \autocite{ghosh_ailuminate_2025, chollet_arc_2025}. Anticipating independent review may encourage investing in more robust mitigations earlier in development, before potential risks translate to concrete safety and security failures. 

Auditing also mitigates a distinct institutional failure of self-assessment: potential misalignment between deployment incentives and judgments about sufficient safety precautions. Frontier AI developers are simultaneously optimizing for capabilities, speed, and market position, while contending with how to determine the conditions under which their own systems are too risky to release or scale. This creates a structural conflict of interest, including internal pressure on safety teams to narrow scope or provide premature sign-off to meet deployment timelines (e.g., \autocite{GoldmanKahn2024OpenAIResigns}). Independent auditing can separate evaluation and verification of safety properties from commercial incentives. 

Beyond individual firms, external auditing enables learning at the level of the ecosystem, rather than just the level of individual developers. Without shared assessment, safety practices remain difficult to compare across organizations, making systemic risk hard to detect until failures occur. Auditors working with multiple companies can identify patterns, disseminate best practices, and share effective mitigations between developers with different levels of maturity. This affords the ability to make direct comparisons across developers, for example, allowing insights from state-of-the-art frontier systems to inform evaluations and \hyperlink{gls:safeguard}{safeguards} for less capable models that may later encounter similar risks.\footnote{This is one of several reasons to focus particular governance attention on frontier AI, as discussed further in \autocite{anderljung_frontier_2023}.}

Notably, while some of the benefits above can be achieved even if only some companies participate, wide participation is important in order to capture the full benefits. Having more participating companies helps broaden the amount of experience that others can learn from, and wide participation can discourage companies from cutting corners in order to gain a short-term advantage at the expense of the larger industry and society \autocite{askell_role_2019}. Even if auditing is made as efficient as possible through technical and process innovations, it will always have some costs, so there is a risk that selective participation will disadvantage responsible developers who incur those costs, while exposing the public to systemic risks from the industry’s weakest links. 

\subsection{Enabling confident investment and deployment} \label{ssec:enabling}

Frontier AI systems are unusually difficult to responsibly invest in and deploy because uncertainty, liability, and information asymmetries compound. Credible third-party auditing unlocks broader AI adoption by giving potential investors and deployers of AI systems better-founded confidence in safety and security claims.\looseness=-1 

When credible third-party audits play a central role in the deployment ecosystem, enterprises and government agencies can rely on shared, independent assessments rather than attempting to evaluate frontier AI systems on their own. Audits provide a common reference point that enables adoption decisions to scale beyond a small number of technically sophisticated firms. This lowers the cost and complexity of due diligence, particularly for organizations that lag behind the frontier in terms of deep internal AI safety expertise.

Auditing also plays a stabilizing role in legal and regulatory environments that are still in flux. Because standards of care for frontier AI deployment are unsettled, adopters face the risk that decisions made under uncertainty may later be judged negligent after harm occurs. Credible third-party audits help mitigate this risk (or at least bound the scope of it) by documenting that deployment decisions were made in accordance with recognized, independent assessment practices. This makes relevant aspects of reasonable care demonstrable ex ante rather than contestable only after the fact, reducing uncertainty for adopters and investors.

As a consequence of more reliable information, developers that pass rigorous audits gain competitive advantages, as do downstream companies building on audited systems. Audit credentials can differentiate providers in procurement, particularly with governments and regulated industries. Without such mechanisms, frontier AI markets are prone to adverse selection: responsible developers bear higher internal safety costs, while less cautious actors can make similar claims at lower expense. Auditing allows safety, security, and governance quality to become more observable, enabling competition to reward genuinely higher standards rather than marketing alone. Over time, this creates incentives for wider participation, reinforcing auditing as a normal part of market entry rather than an exceptional burden, as it has done in other sectors \autocite{pcaob_investor_2025}.

From frontier AI developers’ perspectives, rigorous third-party auditing provides concrete benefits: it can identify safety and security issues before they become costly incidents; build trust with enterprise customers and government agencies hesitant to adopt AI; provide legal clarity, potentially in the form of evidentiary support in court; and differentiate products in competitive procurement processes.

These effects extend to insurance and capital markets. Frontier AI risks are difficult to insure because they are novel, potentially catastrophic, and poorly characterized, leading some insurers to exclude AI-related harms altogether \autocite{yang_insurance_2025, noauthor_insurers_2025}. Audits can help unlock two distinct insurance markets: (1) For frontier AI developers, audits provide the standardized, quantifiable risk data that insurers need to underwrite coverage, lowering the cost of capital and clarifying accountability in the event of harm. (2) For businesses building on frontier AI models, audits of the underlying models give insurers visibility into risks that would otherwise be opaque. This allows insurers to differentiate based on audit status (e.g., as shown by a recently proposed underwriting standard from AIUC \autocite{aiuc_aiuc-1_2026}), offering better terms to businesses that choose audited models over unaudited alternatives. See \cref{apx:additional_motivations} for more discussion of insurance.

As with the safety and security benefits described above, confidence in frontier AI investment and deployment will be greater to the extent that there is wide adoption of auditing, rather than just a few firms participating. High-profile safety incidents, such as the Three Mile Island Accident \autocite{wikipedia_threemileisland_accident}, can set back an entire industry \autocite{baron_public_2020} even if there are safer companies or products in the market. There is growing interest in AI-related risks among investors \autocite{tangen_responsible_2023}, and frontier AI auditing can help manage such risks.

\subsection{Different audit requirements based on motivation}
\label{ssec:different_audit}

Although the motivations for frontier AI auditing share the common foundations of independence, varying levels of non-public access, and standardized frameworks for comparison, they place fundamentally different demands on what an audit must accomplish. In some settings, audits are primarily tools for reducing uncertainty and supporting private decision-making; in others, they are mechanisms for enabling credible commitments where trust, enforcement, or risk-sharing are limited \autocite{harack_verification_2025, baker_verifying_2025, mitre_artificial_2025} (see also \cref{ssec:reaching_full}). These roles cannot be served equally well by a single, undifferentiated notion of “an audit,” and treating these cases as requiring the same level of assurance would either hollow out audits where they must be strongest or make them overly burdensome where lighter-touch approaches would suffice.\looseness=-1

This combination of convergence and variation is why we present a single overall vision that includes multiple AI Assurance Levels (AALs); \cref{sec:vision} describes how appropriate AALs can be selected in practice.\looseness=-1

\subsection{Government vs. private auditing} \label{ssec:government_vs_private}

In principle, at least some of the positive outcomes described above are achievable with governments assuming an auditing role. Public agencies can and do play valuable roles in system evaluation, as illustrated by UK AISI and US CAISI's pre-deployment testing efforts \autocite{aisi_pre-deployment_2024}. However, relying primarily on governments to conduct frontier AI audits faces structural limitations that are especially binding in this domain. Frontier AI systems evolve rapidly, require deep technical specialization, and often demand sustained access to non-public model details, internal processes, and proprietary data. Most governments face persistent challenges in building and retaining the requisite expertise at scale, adapting quickly to new model architectures and risk profiles, and matching the pace of innovation in the private sector. These constraints are not unique to AI, but they are particularly acute given the speed and complexity of frontier model development.\looseness=-1

At the same time, a purely private auditing ecosystem without public involvement would be inadequate. Governments have a critical role to play in providing oversight, setting baseline standards, and ensuring democratic accountability. In practice, this includes defining minimum requirements for auditor independence and competence, accrediting or supervising auditing organizations, and enforcing consequences when audits are negligent or misleading. This division of labor mirrors established practice in other domains, such as financial auditing, where private auditors perform evaluations while public authorities set the rules and provide backstop enforcement. In the context of frontier AI, such oversight is essential to ensure that audits retain substantive value rather than devolving into compliance theater.

A largely private-sector auditing regime also offers an additional governance advantage: it limits the concentration of power over AI oversight in any single institutional actor. Governments will inevitably play a central role in AI governance through regulation, enforcement, and national security policy. Assigning primary responsibility for auditing to the private sector helps distribute governance functions across institutions with different incentives, expertise, and failure modes. Taken together, these considerations point toward an auditing ecosystem that is predominantly private in execution but publicly overseen.

\newpage 
\section{Lessons from Related Domains and Current AI Assessment}\label{sec:lessons}

Before we detail our vision for frontier AI auditing, we briefly survey two bodies of practice that informed it: more established auditing and assurance practices in other industries and current third‑party assessment in the AI industry. \cref{apx:lessons} and \cref{apx:contemporary} provide more extensive discussions of these topics.

Historically, many industries introduced rigorous third-party oversight only after serious incidents compelled action. Pharmaceutical pre-market approval, for example, became mandatory only after the 1937 sulfanilamide disaster killed over 100 people, and the 1957–1961 thalidomide tragedy prompted additional efficacy requirements \autocite{ballantine_sulfanilamide_1981}. A degree of aviation certification became mandatory through the Air Commerce Act of 1926, championed by industry leaders who believed ``the airplane could not reach its full commercial potential without federal action to improve and maintain safety standards'' \autocite{federal_aviation_administration_brief_nodate}. Decades later, standards were ratcheted up significantly after high‑profile accidents such as the Grand Canyon crash in 1956 \autocite{FAA_N6902C}. 

The ultimate success of frontier AI auditing would be enabling dramatic safety and security progress without catastrophe as a necessary catalyst.

\subsection{Key lessons from more established assessment domains} \label{ssec:lessons_from_established_domains}

Third-party audits are common across many industries \autocite{anderson-samways_ai-relevant_2024}, where carefully designed frameworks facilitate external assessment of sensitive technologies and institutions while protecting intellectual property. We draw lessons from four domains, discussed below and summarized in \Cref{tab:key_lessons}. See \cref{apx:lessons} for detailed discussion and examples.

\textbf{Food safety and consumer product testing.} These domains demonstrate that effective safety culture requires ``\hyperlink{gls:defense}{defense in depth}'': testing at multiple stages of a product’s lifecycle and for multiple failure modes. Independent testing organizations like Underwriters Laboratories \autocite{underwriters_laboratories_inc_engineering_2016} show that companies can opt in and pay for certification if people avoid products lacking trusted third-party assurance. Critically, safety system failures --- such as the 2008 Chinese milk scandal \autocite{gossner_melamine_2009} --- produce widespread distrust that propagates across companies and can persist for years \autocite{li_consumer_2021}. For frontier AI, these precedents suggest (1) continuous testing throughout the lifecycle, (2) joint industry investment in testing infrastructure, and (3) recognition that a single high‑profile failure can damage the entire industry’s standing.

\textbf{Safety-critical systems engineering and aviation safety.} Industries like aviation and nuclear power treat safety as an emergent property of complex sociotechnical systems, employing structured methodologies --- including hazard analysis, safety cases, and continuous lifecycle risk management \autocite{leveson_introduction_2023} --- to proactively identify and manage risks. Aviation’s strong safety record involves interlocking elements providing defense in depth: pre-approval of designs, mandatory incident reporting, and criminal liability in some cases. However, at the same time, the Boeing 737 MAX disasters highlighted the catastrophic risks of excessive self-certification and deference, where commercial pressures overrode safety concerns. Key lessons include: (1) systems-level analysis provides greater evidence for safety decisions than component-level analysis alone; (2) near-misses are often early warning signs of eventual failures; (3) effective safety reporting requires structural independence and protection from retaliation; (4) self-certification and delegation of audits create dangerous conflicts of interest; and (5) auditing must be technically rigorous, rather than relying on company attestations. (see \cref{ssec:safety-critical,ssec:aviation_safety} for in-depth discussions)

\textbf{Penetration testing.} Penetration testing demonstrates that security attributes are often best assessed through active adversarial testing rather than static checklists. Instead of checking only whether documented requirements are met, testers creatively search for unexpected failure modes and chain together subtle weaknesses. The field shows that an adversarial analytical posture can coexist with a collaborative relationship --- auditors and companies iteratively fix issues rather than treating audits as one-off pass/fail exercises. Bug bounty programs \autocite{hackerone_bug_nodate} extend this into ongoing, market-based mechanisms with clear incentives. For frontier AI, adversarial testing should be a core component of \hyperlink{gls:misuse}{misuse} and security audits.

\begingroup
\hyphenpenalty=10000
\exhyphenpenalty=10000
\sloppy
\begin{table}[t]
\caption{Key lessons drawn from the domains discussed in this section}
\label{tab:key_lessons}
\centering
\renewcommand{\arraystretch}{2}
\setlength{\tabcolsep}{8pt}

\begin{tabularx}{\textwidth}{|>{\RaggedRight}p{2.5cm}|>{\RaggedRight}p{4.5cm}|X|}
\hline
\rowcolor{headergray} 
\textbf{Principle} & \textbf{Source Domains} & \textbf{Implication for Frontier AI} \\ 
\hline

%\rowcolor{tablegray} 
Independence & 
Financial auditing, aviation & 
\begin{tabitemize}
    \item Auditors need to be incentivized to meet very high standards in their analysis through mechanisms such as regulation, liability, and market pressures that reward rigor
    \item Conflicts of interest need to be managed carefully
    \vspace{0.3em}
\end{tabitemize} \\ 
\hline

%\rowcolor{tablegray} 
Defense in depth & 
Food safety, aviation, consumer products & 
\begin{tabitemize}
    \item Multiple layers of assessment are needed at different lifecycle stages
\end{tabitemize} \\ 
\hline

%\rowcolor{tablegray} 
Continuous monitoring & 
Safety-critical systems, consumer products & 
\begin{tabitemize}
    \item One-off, static certifications are insufficient
    \item Audits must account for systems changing over time
    \vspace{0.3em}
\end{tabitemize} \\ 
\hline

%\rowcolor{tablegray} 
Adversarial testing & 
Penetration testing & 
\begin{tabitemize}
    \item Adaptive red-teaming is needed, not just checking off a list
\end{tabitemize} \\ 
\hline

%\rowcolor{tablegray} 
Organizational assessment & 
Safety-critical systems, financial auditing & 
\begin{tabitemize}
    \item Culture, governance, and security matter, not just specific AI systems
\end{tabitemize} \\ 
\hline

\end{tabularx}
\end{table}
\endgroup

\textbf{Financial auditing.} Financial auditing offers perhaps the richest set of analogies --- both positive and cautionary --- for frontier AI auditing. On the positive side, it demonstrates the feasibility of professionalized processes allowing independent parties to review highly sensitive information, the value of standardized metrics for comparing risks across organizations, and the importance of combining verification of specific claims (e.g., financial statements) with broader assessment of internal controls. Financial auditing has developed crucial conceptual tools that are relevant to frontier AI: (1) clear norms for managing conflicts of interest \autocite{booker_cpas_2016}, (2) sharp distinctions between error and fraud, and (3) recognition that professional judgment by auditors is indispensable.

However, financial auditing also provides warnings. Catastrophic failures --- Enron, Wirecard \autocite{fbi_enron_nodate, heese_wirecard_2021} --- illustrate what happens when auditors derive most of their revenue from a small number of large clients \autocite{mary_locatelli_good_2002}. Even after reforms like Sarbanes–Oxley, the sector has struggled with conflicts of interest \autocite{pcaob_pcaob_2023}, procedural focus that risks missing systemic issues, and a persistent ``\hyperlink{gls:expectations}{expectations gap}'' between public belief that audits guarantee the absence of fraud and auditors’ more modest mandate. For frontier AI, these suggest the critical importance of auditor independence, clear communication of assurance levels, and avoiding criteria that devolve into box-ticking.

These domains illustrate both the achievements and the pitfalls of common assurance regimes. We do not present these examples as gold standards --- rather, we highlight constructive lessons for frontier AI auditing while encouraging thoughtful and deliberate effort to build self-correction mechanisms into the vision outlined in \cref{sec:vision}.

\subsection{The current state of third-party AI assessment} \label{ssec:current_state}

Although frontier AI auditing as we define it does not yet exist, a growing field of third-party assessment provides a foundation on which to build \autocite{staufer_audit_2025}. This section summarizes the current state as of December 2025. See \cref{apx:contemporary} for detailed discussion and examples.

\textbf{Overview.} Current third-party assessments of frontier AI vary substantially in scope, access, rigor, and transparency. Most evaluators receive only the same public access as ordinary users, with only a select few receiving early or privileged access. Public reporting is inconsistent: system cards sometimes mention third‑party evaluators only in the abstract; methodological details are often omitted; and evaluators are sometimes not named at all even when they are used \autocite{google_deepmind_gemini_2025}, making it difficult to follow up for more information or for other companies to seek to work with those evaluators. Assessments focus predominantly on capability evaluation and, increasingly, propensity evaluation (e.g., tendencies of AI models to deceive), with comparatively little attention to organizational risk governance, safety culture, or platform-level controls.

\textbf{Key Dimensions.} We assessed current practice across seven dimensions (a more detailed discussion can be found in \cref{apx:contemporary}):
\begin{itemize}[leftmargin=0.25in,topsep=.01in]
    \item \textbf{Reporting:} Public reporting is sparse and inconsistent. System cards vary substantially in how they describe third-party involvement, and methodological details are often absent \autocite{wang_2025_foundation_2025, gallifant_peer_2024}.

    \item \textbf{Access:} Most evaluators receive only black-box API access. A small but growing number of collaborations with government institutes have tested deeper access (e.g., \hyperlink{gls:cot}{chain-of-thought}, internal documentation \autocite{aisi_pre-deployment_2024}), but \hyperlink{gls:graybox}{gray-box} and \hyperlink{gls:whitebox}{white-box} access remain highly limited.

    \item \textbf{Rigor:} Methodology and effort vary substantially. Benchmark-based assessments face issues with quality \autocite{reuel_open_2025}, contamination \autocite{mcgregor_risk_2025}, and construct validity 
    \autocite{bean_measuring_2025, wallach_position_2025}. Red-teaming effectiveness is skill-dependent \autocite{mcgregor_err_2024}. Neither companies nor evaluators typically publish substantive threat models.

    \item \textbf{Standardization:} Standards remain nascent, though they are evolving rapidly \autocite{staufer_audit_2025, bommasani_foundationmodel_2024}. Evaluations are typically conducted under bespoke, confidential contracts with terms rarely visible to regulators or the public.

    \item \textbf{Continuous monitoring:} Assessments are one-off ``snapshots'' rather than continuous. Companies frequently update systems without providing third-party access for updated risk assessment.\looseness=-1

    \item \textbf{Scope:} Assessments focus heavily on technical systems (often just models) rather than organizational practices. Assessment of mitigations, platform-level controls, and safety culture is comparatively rare.\looseness=-1

    \item \textbf{Scale and independence:} Participation is voluntary and concentrated among a few developers. Evaluators depend on companies’ goodwill for access and sometimes funding, creating potential conflicts of interest.

\end{itemize}

\textbf{Emerging Developments.} Recent positive developments include proposed evaluation frameworks (e.g., \autocite{paskov_toward_2025, mccaslin_stream_2025, reuel_betterbench_2024}); initial best practices from the Frontier Model Forum \autocite{frontier_model_forum_issue_2024}; the establishment of the AI Evaluator Forum \autocite{noauthor_ai_nodate}; pilots with government AI safety institutes in the US and UK \autocite{openai_working_2025, anthropic_strengthening_nodate}; early examples of third‑party review of company risk assessments (e.g., METR’s review of Anthropic’s sabotage risk report \autocite{samuel_r_bowman_anthropics_2025}, which we consider to be among the first AAL-1 audits, and third-party review of the safety work conducted for OpenAI’s release of gpt-oss \autocite{openai_gpt-oss-120b_2025}); and OpenAI and Anthropic's reciprocal safety assessments of each other’s systems \autocite{openai_findings_2025, samuel_r_bowman_anthropics_2025}. These developments are promising but remain early‑stage compared to established assurance regimes.

\subsection{The gap between current practice and cross-industry best practices} \label{ssec:gap_current}

Current third-party AI assessment efforts provide a valuable starting point --- including a nascent ecosystem of organizations, both for-profit and non-profit, that have conducted increasingly rigorous assessments over time. Yet significant gaps remain between these practices and the best practices found in other industries. 

How much further improvement is needed depends in part on the risk profile that can be expected from AI at different points in time. Roughly speaking, those who expect faster progress in AI capabilities in the future --- and therefore greater safety and security risks, given AI's general-purpose nature --- should desire a faster rate of progress in third-party assessment along various dimensions discussed above, so that we are not caught unprepared. Furthermore, to the extent that one believes that risks are highly correlated with raw capabilities, then one might desire particular scrutiny to be applied to the very most capable AI systems and the companies building them. These insights inform the approach we take in the next section, where we suggest both general principles for how frontier AI auditing should work in general as well as a series of progressively stronger assurance levels that can be adapted to particular contexts.\looseness=-1

\newpage 
\section{A Vision for Frontier AI Auditing}\label{sec:vision}

In this section, we set out a long-term vision for what mature third-party auditing could look like --- auditing of both the most capable AI systems and the companies building them.\footnote{In addition to insights from other industries (\cref{ssec:lessons_from_established_domains}) and gaps in current AI assessment practices (\cref{ssec:gap_current}) (on which further details can be found in \cref{apx:lessons} and \cref{apx:contemporary}), our vision builds on prior work outlining frameworks for AI auditing, including field scans of the algorithmic auditing ecosystem \autocite{costanza-chock_who_2022}, proposals for third-party audit ecosystem design based on a survey of the challenges and existing practices in other industries \autocite{raji_outsider_2022, rismani_plane_2023}, internal algorithmic auditing frameworks \autocite{raji_closing_2020}, external scrutiny requirements for frontier LLMs \autocite{anderljung_towards_2023}, assurance audit frameworks modeled on financial auditing \autocite{lam_framework_2024}, and layered approaches combining governance, model, and application audits \autocite{mokander_auditing_2023, mokander_auditing_2023-1}.
} Some elements of this vision can be pursued now, while others will require years of investment and development before they become practical. 
We aim significantly beyond the status quo both because not all current assurance needs are being met by the current AI assurance ecosystem, and because we expect future AI systems to be far more capable and risky than those that exist today.

Our vision for frontier AI auditing is organized around eight interlinked design principles, which we discuss in turn:
\begin{itemize}[leftmargin=0.25in]
    \item \textbf{Scope of risks:} Comprehensive coverage of four key risk categories that can be linked to company actions.

    \item \textbf{Organizational perspective:} Auditing companies' safety and security practices as a whole, not just individual models and systems. 
    
    \item \textbf{Levels of assurance:} A framework for calibrating and communicating confidence in audit conclusions.

    \item \textbf{Access:} Deep enough to assure auditors and other stakeholders, secure enough to reassure auditees. 

    \item \textbf{Continuous monitoring:} Living assessments, not stale PDFs. 

    \item \textbf{Independent experts:} Trustworthy results through rigorous independence safeguards and deep expertise.

    \item \textbf{Rigor:} Processes that are methodologically rigorous, traceable, and adaptive. 

    \item \textbf{Clarity:} Clear communication of audit results. 
\end{itemize}

\subsection{Risk scope of audits} \label{ssec:risk_scope}

Frontier AI auditing should focus on risks for which an AI company's action or inaction can be directly linked to harmful outcomes, including at least the following risk categories (see Figure~\ref{fig:focus_areas}):

\begin{figure} [t]
    \centering
    \includegraphics[width=0.6\textwidth]{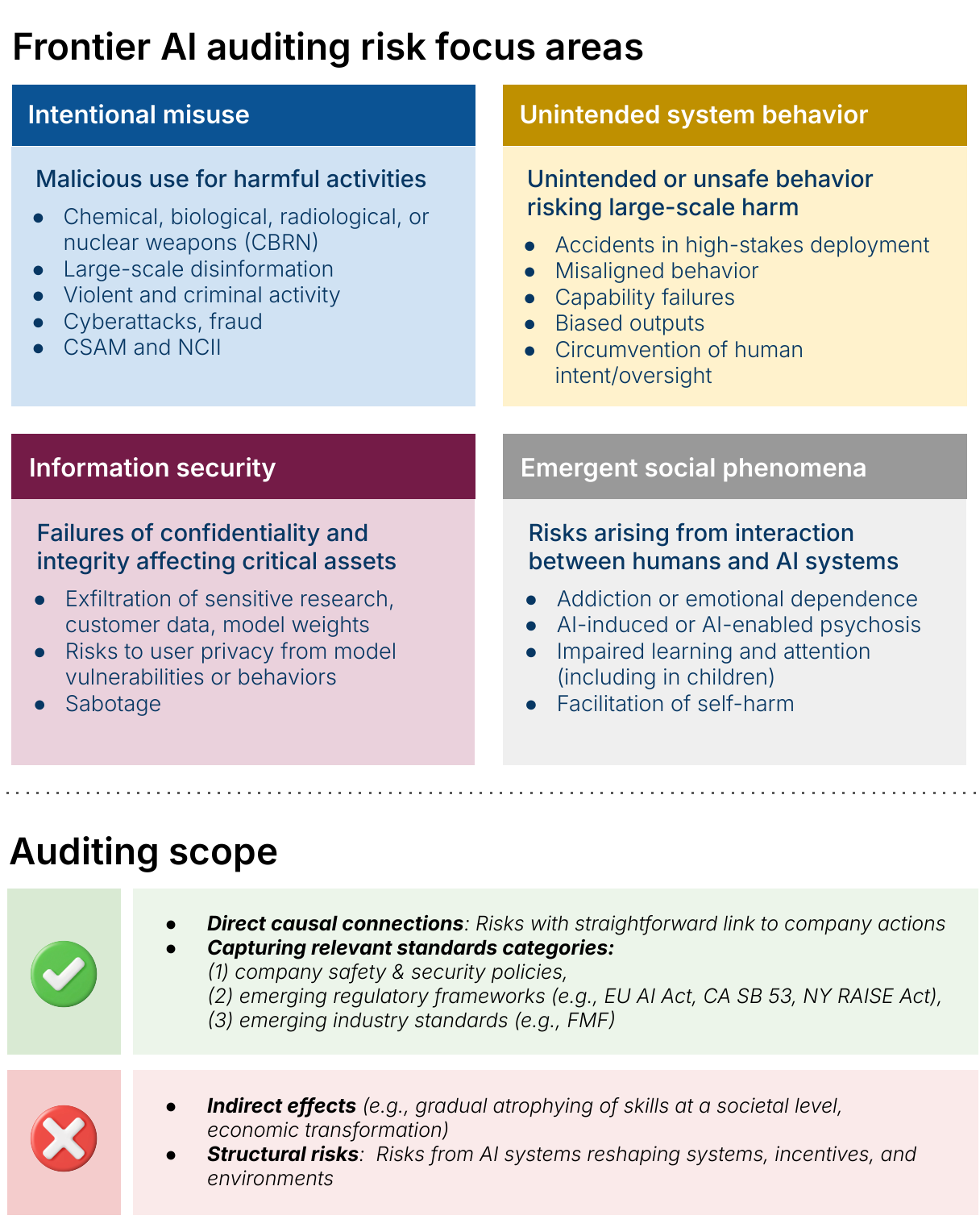}
    \caption{Proposed risk focuses and sources of relevant standards for frontier AI auditing.}
    \label{fig:focus_areas}
\end{figure}
\begin{itemize}[leftmargin=0.25in]
    \item \textbf{Intentional misuse.} The use of frontier AI systems by malicious actors to enable or scale harmful activities. This includes, but is not limited to, cyberattacks; the development and use of chemical, biological, radiological, or nuclear weapons (CBRN); large-scale disinformation; violent and criminal activity; fraud; and the generation of child sexual abuse material (CSAM) or nonconsensual intimate imagery (NCII) \autocite{nist_managing_2024}.\looseness=-1
    
    \item \textbf{Unintended system behavior.} AI systems behaving in ways that developers and users did not intend, or being unsafe in ways that could plausibly cause large-scale harm. This includes highly consequential \hyperlink{gls:accidents}{accidents} caused by inadequate capabilities, \hyperlink{gls:alignment}{alignment}, or \hyperlink{gls:safeguard}{safeguards} \autocite{tabassi_artificial_2023, center_for_ai_safety_ai_2023}. Examples include systems taking harmful, irreversible actions, e.g., permanently deleting critical files \autocite{atherton_incidentreplit_2025,atherton_incidentgoog_2025}.\footnote{We categorize misalignment and loss of control as ``unintended'' in the sense that humans did not intend for the system to behave in these ways, even where the system itself may be acting coherently in pursuit of goals that diverge from those intended. Loss of control can be passive (inability to monitor or correct system behavior) or active (systems resisting human oversight) \autocite{bengio_international_2025}. Some taxonomies treat misalignment and loss of control as a distinct risk category rather than a subset of accidents \autocite{shah_approach_2025, hendrycks_overview_2023}, and others consider misalignment a catalyst for loss of control \autocite{stix_loss_2025}.
}

    \item \textbf{Information security.} Failures of confidentiality or integrity affecting critical AI assets. This includes the exfiltration of model weights \autocite{carlini_stealing_2024}, exfiltration of sensitive research and customer data via internal or external threats \autocite{nasr_scalable_2023, cooper_extracting_2025, barbero_extracting_2025, ahmed_extracting_2026}, risks to user privacy arising from model vulnerabilities or behaviors \autocite{brown_what_2022, cooper_machine_2025}, as well as sabotage of highly capable AI systems \autocite{nevo_securing_2024, anderson_security_2020}.

    \item \textbf{Emergent social phenomena.} Risks that arise from interaction between humans and AI systems and do not fit neatly into ``misuse'' or ``unintended behavior,'' but can nevertheless cause significant harm if left unaddressed. Examples include addiction to or emotional dependence on AI systems, AI-induced or AI-enabled psychosis, and facilitation of self-harm \autocite{hu_how_2023, head_minds_2025, noauthor_emotional_2025, liu_chatbot_2025, saracini_techno-emotional_2025, apa_health_2025, haskins_people_2025, moore_expressing_2025, hate_fake_2025, choi_private_2025}.

\end{itemize}

In reviewing the most recent 300 AI incidents logged by the AI Incident Database \autocite{mcgregor_preventing_2021}, we found these risks to cover all incidents cataloged except (1) those that do not involve frontier AI systems under our definition, such as those involving Waymo self-driving cars,\footnote{See \autocite{aaid_waymo_2026}.} which are highly capable in their domain but not general-purpose; and (2) those that did not result in very significant harms, such as an instance of confabulation of citations in a machine learning book.\footnote{See \autocite{atherton_incidentuber_2026}.}

\hyperlink{gls:structural}{Structural risks} arising from how AI systems reshape systems, incentives, and environments in which they are deployed \autocite{zwetsloot_thinking_2023} are not a design target of our risk list. For example, gradual atrophying of skills at both individual and societal levels as more people rely on AI to perform analytical tasks~\autocite{kosmyna_your_2025}, economic transformation generally, and greater vulnerability of society to electricity disruptions as a result of heavy AI use throughout the economy are not within our design focus or listed risks for this framework. 
This does not mean that we are opposed to auditing with respect to such risks, or that there could not be fruitful transparency requirements at a company level that shed light on how best to address structural risks.\looseness=-1

For each category of risks, auditors should (1) independently verify company claims and (2) evaluate the company's systems and practices against its stated safety and security policies, applicable regulations, and industry best practices. Indeed, these risk categories largely map onto company safety and security policies, emerging industry standards (e.g., the Frontier Model Forum \autocite{FrontierModelForum2025_RiskTaxonomy}), and regulatory initiatives such as California SB 53, New York's RAISE Act, the EU AI Act, and the EU General-Purpose AI Code of Practice.\footnote{We expect the appropriate scope and emphasis of audits to evolve over time as threats, norms, and regulations change, but that there are common threads in how frontier AI auditing should work (e.g., careful management of sensitive information, ensuring auditor independence) that will not change significantly over time. We therefore think that one could endorse the vision discussed in this section, even if one would prefer a different scope.}\looseness=-1

\newcolumntype{C}{>{\centering\arraybackslash}X}

\begin{table}[t]
\caption{Risk categories in company policies (e.g., from OpenAI, Google DeepMind, Anthropic, xAI, Meta, Microsoft, and Amazon) and regulatory texts \autocite{metr_common_2025}.}
\centering
\renewcommand{\arraystretch}{2} 
\setlength{\tabcolsep}{8pt} 

\begin{tabularx}{\textwidth}{|p{4.5cm}|C|C|C|}
\hline

\rowcolor{headergray} 
\textbf{AI risk category} & \textbf{Company policies} & \textbf{CA SB 53 / NY RAISE} & \textbf{EU AI Act Code of Practice} \\ 
\hline

Intentional misuse & 
\partiallyincluded & 
\partiallyincluded & 
\fullyincluded \\ 
\hline

Unintended system behavior & 
\partiallyincluded & 
\partiallyincluded & 
\fullyincluded \\ 
\hline

Information security & 
\fullyincluded & 
\fullyincluded & 
\fullyincluded \\ 
\hline

Emergent social phenomena & 
\partiallyincluded & 
\notincluded & 
\partiallyincluded \\ 
\hline

\end{tabularx}
\end{table}

\subsection{Comprehensive, organizational-level perspective} \label{ssec:comprehensive_organizational}

To examine the risks we outline above, an audit could cover different parts of a company, or the company as a whole. In this subsection, we argue that frontier AI auditors should emphasize \textit{the company as a whole} as the most important level of analysis. Individual AI systems may be partially illustrative of or a big component of a company's risk management, but they are never the full story of the company's impact. Specific components of and artifacts produced by a company are important to audit and may even be the focus of specific audits, but should always be explicitly considered in --- and audit conclusions should be framed in relation to --- this larger context.

\textbf{Avoiding abstraction errors.} A central danger in auditing frontier AI developers is that an audit can be right about the specific artifact or process it examined while still being \textit{wrong in the way that matters} about the company's overall risk posture. 
This reflects an \hyperlink{gls:abserror}{abstraction error}: 
forming the wrong conclusion by treating a partial or simplified unit of analysis (e.g., evaluating a specific component in isolation) as if it were sufficient to assess overall system and organizational risk.
Such abstraction errors are especially likely in frontier AI because (1) risks such as those listed in \cref{ssec:risk_scope} are shaped by interactions across internal processes, AI systems, and other parts of the internal technology stack, (2) many relevant systems and decisions are non-public and fast-changing, and (3) it is easy to (often unintentionally) audit what is most legible rather than what is most risk-relevant. Put differently: auditing can miss the forest for the trees not because the trees are unimportant, but because the forest is not simply the sum of individually ``healthy-looking'' trees.\looseness=-1

There are at least four ways abstraction errors can arise in practice:
\begin{itemize}[leftmargin=0.25in]
    \item \textbf{Portfolio blindness: auditing the most visible or best-behaved system.} Frontier AI developers rarely operate a single model or a single system. They maintain \textit{portfolios}: multiple \hyperlink{gls:checkpoints}{checkpoints}; post-training variants; internal research models; preview builds for partners; fine-tunes for specific customers; custom model weights transferred to a datacenter controlled by a customer; and internal tools with broader permissions than the public product. It is therefore possible for an audit to establish that a flagship deployment is well-controlled, while missing a materially riskier surface elsewhere. In these cases, a favorable finding about one audited surface is not \textit{false} per se, but may become misleading if it is treated as representative of the organization's overall risk posture.

    \item \textbf{Configuration drift: outdated or incomplete audit results due to system-level changes.} Even when the same exact model checkpoint is being used in different cases, real-world behavior and risk depend on system-level configurations: system prompts, input and output filters, routing across multiple models, retrieval sources (e.g., search engine APIs or periodically updated databases from which knowledge is retrieved during operation), tool access, memory, rate limits, monitoring thresholds, user-specific personalization, UI features, public-facing API implementations, and downstream post-processing. Seemingly modest changes such as enabling a new tool, relaxing a filter threshold, swapping in a different safety classifier, or changing routing rules for a subset of users or at different times of day can materially alter misuse potential or the likelihood of harmful failures. An abstraction error occurs here when an audit treats a specific evaluation (or a staging configuration) as a proxy for the actual deployed system, without establishing that the audited configuration matches production deployments and will remain stable enough for conclusions to hold. The need to hedge against configuration drift is one reason why we emphasize \hyperlink{gls:continuousmonitoring}{continuous monitoring} for changes in \cref{ssec:continuous_assurance}.

    \item \textbf{Non-compositional safety and security: safe components, unsafe assembly.} Many safety and security properties do not necessarily compose together neatly. A model that refuses harmful requests in an isolated user chat setting may still enable harmful outcomes in another isolated user chat, or when embedded in an agentic scaffold that chains together multiple tool calls and operates over long horizons. A model with concerning raw capabilities and propensities (e.g., to deceive users) may be kept low-risk through strong system-level controls. For frontier AI, the risk-relevant question is often less ``what can the model do in isolation?'' and more ``what can the organization's integrated systems do, under realistic conditions, given the actual controls?'' Abstraction errors arise here when auditors over-weight component-level findings while under-weighting system-level or organization-level interactions that dominate the actual risk.

    \item \textbf{Boundary mismatch in security: strong product security, weak security of trade secrets.} A company may deploy a well-engineered public API (authentication, rate limits, abuse monitoring) while leaving training infrastructure, model weight storage, experiment tracking, or internal repositories comparatively exposed. Indeed, at least two frontier AI companies have had AI research-related intellectual property stolen from them \autocite{metz_hacker_2024, us_doj_chinese_2024}, and likely there are many similar cases that are not publicly known given what is known about these companies' security practices and the difficulty of defending against sophisticated attacks \autocite{nevo_securing_2024, mitch_governance_2025}. The resulting organizational risk can be dominated by the weaker boundary: if an adversary can exfiltrate model weights or tamper with training and deployment artifacts, the company's public-facing mitigations may become irrelevant (e.g., stolen weights can be used without those mitigations). Here, a ``system-level'' audit focused on the externally visible interface can substantially underestimate information security risks that sit behind the interface but govern the most consequential assets.
\end{itemize}

Abstraction errors are not rare edge cases to be aware of and carefully avoided. Rather, they demonstrate that the company level of analysis is best for forming confident conclusions, even if it is hard to achieve in practice \autocite{ball_entity-based_2025}. There are predictable ways audits focused on only a single component of a frontier AI company can mislead all stakeholders regarding that company's risk posture.

\begin{figure}[t]
    \centering
    \includegraphics[width=1\linewidth]{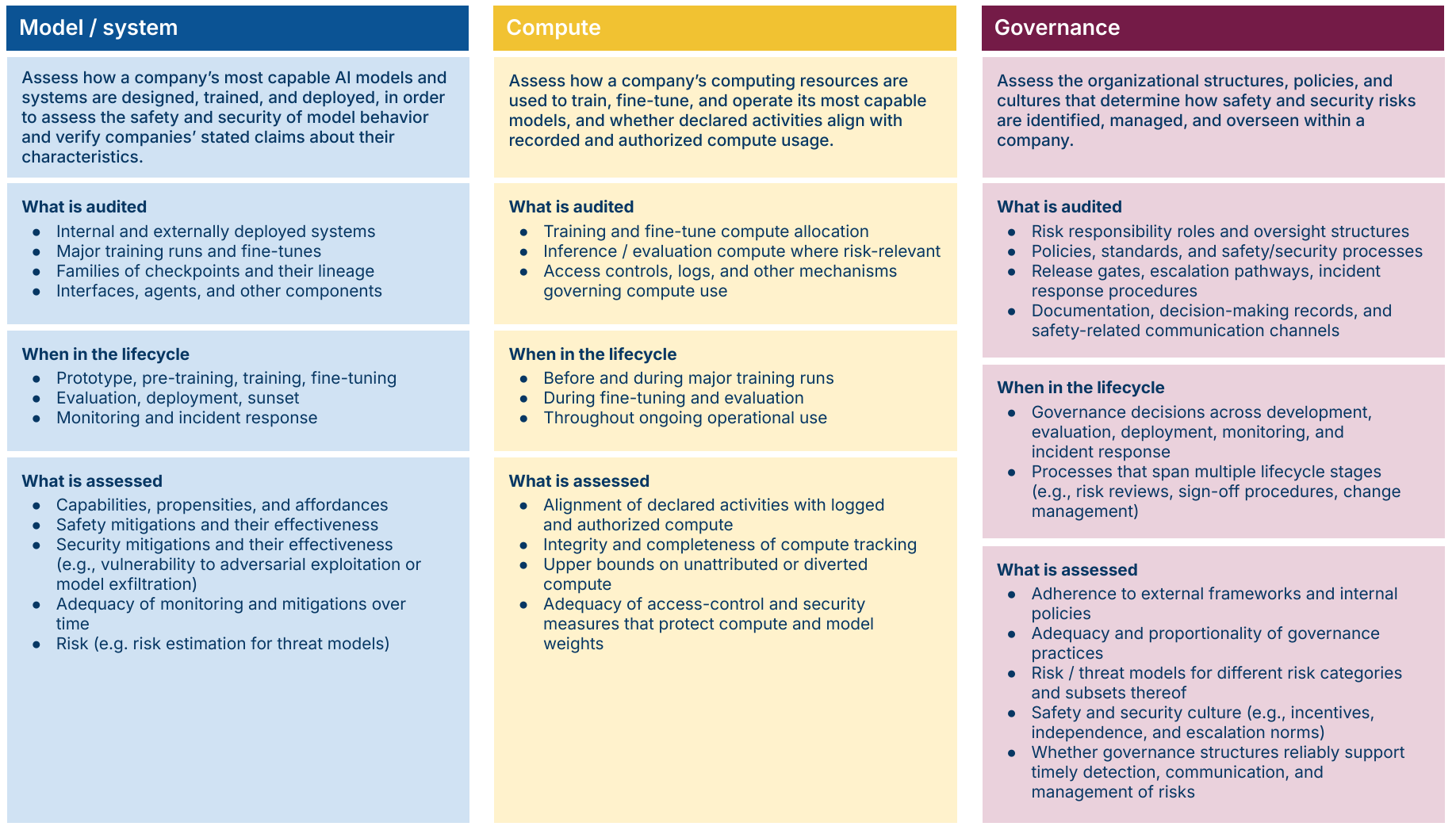}
    \caption{What should be audited?}
    \label{fig:what_should}
\end{figure}

\textbf{Three lenses.} In our vision for frontier AI auditing, lead auditors need to integrate three lenses: models and systems, which includes AI models, system features connected to those models (e.g., input and output classifiers, system prompts), and information security safeguards (e.g., user authentication); computing hardware, including its quantity and security, and how it is allocated across development and deployment efforts; and governance, including development and deployment decision-making processes, information security systems and protocols, incident response protocols, the safety and security culture of the organization, and the clarity with which responsibility is allocated within the company. Neglecting one of these lenses risks an incomplete picture of a company's risk profile (see Figure~\ref{fig:what_should}).

AI models and systems are the primary focal point of a frontier AI company's work. But critically, from an auditing perspective, it's important not to focus on a single model or system to the exclusion of others. All but the very most nascent companies have many different models and systems at a given time. This includes models and systems that are in development in addition to those that are deployed; models that are smaller, cheaper, and faster but less capable as well as those that are larger, more expensive, and slower but more capable; versions of systems that use more or less computing power while in use (''test-time compute''); versions that are produced and provided specifically for a given customer, such as a company or government agency, and may have different guardrails; information security systems, which are critical to ensuring that the other systems are not stolen or tampered with; and much more. At higher assurance levels, more of these systems are critically examined, and in more detail.

Auditors also need to understand the computing hardware that a company has access to and how it is using it. Physical or digital access to that computing hardware could be a weak link for the security of training infrastructure, weight storage, and internal repositories, as discussed above; major training runs or deployments that are not publicly announced could contribute disproportionately to a company's risk profile --- such internal deployment \autocite{stix_ai_2025} might be unknown to auditors by default, but shouldn't be if auditors are to be effective in characterizing risks. Gaps in a company's ability to comprehensively account for its own compute use could point to gaps in the company's understanding of its own activities, or could indicate efforts to mislead auditors (note that this is particularly important at higher assurance levels, where significant effort is made to rule out the possibility of deception).

Lastly, understanding the safety and security governance of all of these digital and physical systems is critical in order to put those systems in context. An auditor needs to know who is responsible for what, how documents are produced, what the incentives facing the document-writers were, etc. in order to meaningfully interact with non-public information and spot subtle errors and --- at the highest levels of assurance --- intentional deception. In short, information about governance helps indicate how much to trust other kinds of information. Furthermore, significant gaps in governance --- both formal (e.g., limited or non-existent policies governing internal AI deployments) and informal (e.g., a culture of corner-cutting in specific areas that comes up in staff interviews) --- may provide vital clues to gaps in risk mitigation at a system level.

\textbf{Practical implications of the organization-level perspective.} To achieve higher levels of assurance about a company's risk profile (\cref{ssec:levels}), deeper access to information (\cref{ssec:deep_access}) will tend to be required about each of these lenses in order to enable drawing confident conclusions about a company's risk profile in each of the four risk categories. This in turn will require standardized processes for ``mixing and matching'' subcontractors with different skill sets (\cref{ssec:independent_auditors}). In order to avoid committing abstraction errors due to configuration drift, continuous monitoring will be needed, and a range of different audit cadences will need to be conducted, corresponding to the different paces of change of different organizational components (\cref{ssec:continuous_assurance}). Rigorous, traceable processes are needed in order to allow those interpreting or replicating an audit to infer whether abstraction errors are likely (\cref{ssec:rigorous_processes}). Privately and publicly shared audit findings (see \cref{ssec:clear_communication}) need to enumerate the assumptions being made in order for analyses of artifacts to be representative of the company as a whole.\looseness=-1 

Over time, there needs to be research toward a standardized analytical framework (e.g., an ``organizational safety and security case'') that combines different inputs into a composite picture of a company's risk profile. 
Such research should draw on best practices from safety-critical systems engineering, such as \hyperlink{gls:safetycase}{safety cases}, which are structured arguments supported by evidence that justify the safety of a system \autocite{uk_ministry_of_defence_defence_2007}.\footnote{Existing work has proposed safety cases to verify that AI systems are safe enough to develop or deploy (see \autocite{buhl_safety_2024, clymer_safety_2024}).} 
We think our AI Assurance Level (AAL) framework, discussed next, is an early step.\looseness=-1

\subsection{Levels of assurance} \label{ssec:levels}

To address the different risk scopes for different depths of organization-level audits, we propose a framework for calibrating and communicating confidence in audit conclusions that we call AI Assurance Levels (AALs). This framework is intended to help those conducting and relying on audits to understand what conclusions they can reasonably draw, and what kinds of abstraction errors (\cref{ssec:comprehensive_organizational}) --- among other types of errors --- still cannot be ruled out.

\subsubsection{The meaning of levels of assurance in general}

Frontier AI audits should each be conducted at a specific ``level of assurance.'' 
A given level describes how confident the auditor is in their conclusions about a given company's safety and security practices \autocite{jelen_practical_1998, icaew_limited_nodate}. 
In principle, an auditor could reach a high assurance conclusion that safety and security safeguards are very poor; 
however, we will often use examples of audit findings that are positive with respect to safety and security risk management. 
We do so because, in practice, frontier AI audits involve active collaboration between the auditor and auditee and allow the possibility of remediation prior to publication of results. 

Explicit assurance levels help stakeholders understand how much they can rely on audit results and what assumptions remain untested. In other industries, reviewers and auditors provide either ``limited'' or ``reasonable'' assurance \autocite{iaasb_international_2013, noauthor_isoiec_2019, public_company_accounting_oversight_board_at_nodate}. Safety-critical industries (e.g., aviation, nuclear power) also use the concept of \hyperlink{gls:reasonable}{reasonable assurance} \autocite{bsee_oil_2015, dapas_key_nodate, jackson_internal_2021}, which implies a higher degree of confidence. The level of assurance required for different contexts may differ, depending in part on the costs of the audit and the costs of errors \autocite{harris_which_2023}.\looseness=-1

\subsubsection{Overview of AI Assurance Levels}

We use the term AI Assurance Levels (AALs) to refer to assurance levels in the sense above, as applied to the specific context of a frontier AI audit.\footnote{After completing most of this paper, the authors learned of a prior use of the term ``AI Assurance Level'' with a very different meaning \autocite{jain_ai-driven_2025}, as well as another use of the acronym AAL in a related context (Authenticator Assurance Levels \autocite{nist_authenticator_2026}). These collisions are unintentional.\looseness=-1} Higher levels more and more confidently assess the risk level associated with the frontier AI company as a whole, and progressively rule out abstraction errors such as those discussed in \cref{ssec:comprehensive_organizational} as well as other possible sources of error in the audit's findings. To achieve this, audits at higher levels will tend to require greater access to non-public information relative to lower levels, larger allocations of time and talent, and more sophisticated infrastructure and analysis.

Lower AI Assurance Levels (AAL-1 and 2) can detect some errors on the part of companies and verify the existence of significant compliance efforts, and they can achieve this using smaller expenditures of time and talent. While audits at these levels may be able to detect errors (i.e., unintentional misstatements or mistakes), they are less likely to be able to detect fraud (i.e., intentional deception) (see \autocite{hamilton_evaluating_2016}). 

Higher AI Assurance Levels (AAL-3 and 4) can provide stakeholders significantly more confidence that the conclusion of the audit is correct and that more subtle errors will be detected, and they aim to address the possibility of deliberate deception on the part of the company. Since we envision audits of companies rather than just systems, audits at higher AALs serve as better and better estimates of company-level risks (versus just system-level risks). At the same time, these audits are more costly because they will require more allocation of both company capacity and auditor capacity, and will involve greater access to (often sensitive) non-public information (see \autocite{harack_verification_2025}). 

Using lower AI Assurance Levels may be appropriate when risks of audit errors are less severe, making the cost of achieving higher AI Assurance Levels greater than the assurance that is obtained \autocite{uk_psctg_gross_2006}. 
In contrast, using higher assurance levels may be appropriate for auditing risks that stakeholders are more concerned about, or auditing in situations where there are strong incentives for the companies to cut corners \autocite{askell_role_2019}.\looseness=-1 

\subsubsection{AI Assurance Level details}

Drawing inspiration from the precedents above, as well as the specific context of frontier AI, we describe each of our proposed four \textbf{AI Assurance Levels (AALs)} below.
We begin with an overarching summary, then provide more details on each. 

\begin{itemize}[leftmargin=0.25in,topsep=0.01in]
    \item \textbf{Limited assurance (AAL-1)}. 
    A time-bounded audit of a particular frontier AI system (typically a few weeks), which makes use of API access to multiple model versions and system settings, as well as a limited amount of additional, non-public information focused on the audited system and related internal decisions.\looseness=-1 

    \item \textbf{Moderate assurance (AAL-2)}. 
    A more extensive assessment of one or more frontier AI systems, as well as company practices more broadly, which, at a minimum, spans months and makes use of gray-box system access, extensive internal documentation (e.g., unredacted safety cases), some continuous monitoring, and staff interviews across several functions.

    \item \textbf{High assurance (AAL-3)}. 
    Ongoing oversight (multiyear engagement for the lead auditor, with many subcontractors contributing throughout) with white-box access, more extensive continuous monitoring, and the authority to examine any area of concern. 

    \item \textbf{Very high assurance (AAL-4)}. 
    Continuous verification designed to detect active deception attempts, operating with a full understanding of the company's systems, computing hardware, and governance, and providing ``treaty-grade'' confirmation of the company's risk profile.
\end{itemize}

The table and paragraphs below summarize the level progression.
\renewcommand{\arraystretch}{1.5}

{
\footnotesize
\begin{xltabular}{\textwidth}{|>{\bfseries\RaggedRight}p{2cm}|>{\RaggedRight}p{2cm}|>{\RaggedRight}X|>{\RaggedRight}X|>{\RaggedRight\arraybackslash}p{2cm}|}

\caption{Summary of AI Assurance Levels. At higher levels, auditors are more confident in their conclusion.}
\label{tab:summary_assurance}\\

\hline
\rowcolor{headergray}
\textbf{Level} & \textbf{Duration} & \textbf{Typical access to information (cumulative)} & \textbf{Methods} & \textbf{Readiness} \\
\hline
\endfirsthead

\multicolumn{5}{c}{{\bfseries \tablename\ \thetable{} -- continued from previous page}} \\
\hline
\rowcolor{headergray}
\textbf{Level} & \textbf{Duration} & \textbf{Typical access to information (cumulative)} & \textbf{Methods} & \textbf{Readiness} \\
\hline
\endhead

\multicolumn{5}{|r|}{{Continued on next page}} \\ \hline
\endfoot

\hline
\endlastfoot

%\rowcolor{tablegray}
AAL-1: \newline Limited \newline assurance &
Time-bounded, typically a few weeks to a small number of months &
\textbf{System:} Black-box API access to multiple checkpoints/variants; access to chain-of-thought outputs and logits; ability to enable/disable safety classifiers that block certain inputs and outputs; overview of safety mitigations;
limited amount of non-public information scoped to the system under audit and related internal decisions. \vspace{0.5em} \newline
\textbf{Organization:} Written representations; organization chart; list of key staff members; attestations about training processes. &
\textbf{System:} Run private evaluation suites probing for dangerous capabilities (e.g., cyber, bio, manipulation); conduct limited red-teaming to probe system boundaries. \vspace{0.5em} \newline
\textbf{Organization:} Review provided documentation; interview key staff about governance structure and reporting lines. &
Achievable now; early pilots already conducted. \\
\hline

%\rowcolor{tablegray}
AAL-2: \newline Moderate \newline assurance &
Months (at least) &
\textbf{System:} Gray-box access to (multiple) key systems; samples of training and testing logs; compute allocation records; ability to remove and examine all mitigations. \vspace{0.5em} \newline
\textbf{Organization:} Safety/security documentation (e.g., safety cases); governance decision records; access to interview staff across safety, security, policy, and product; subsets of training data; information on previous incidents. &
\textbf{System:} Reproduce a sample of internal evaluations to verify they were conducted as described; conduct extensive, targeted red-teaming beyond standard probes; investigate sandbagging and other difficult-to-detect behaviors that could affect evaluation results, leveraging access to model internals; develop statistical model signatures in order to confirm whether audited models correspond to deployed models. \vspace{0.5em} \newline
\textbf{Organization:} Review governance artifacts (release checklists, escalation logs, decision records); interview staff across functions to understand how decisions are made in practice vs. policy. &
Early to mid-2026 (requires investment in IP protection protocols and method standardization). \\
\hline

%\rowcolor{tablegray}
AAL-3: \newline High \newline assurance &
Multiyear engagement for the lead auditor, with many subcontractors contributing throughout &
\textbf{System:} White-box access to all models; safety case with comprehensive supporting details; detailed logs and compute accounting with cryptographic provenance (e.g., ``proof of training''); continuous monitoring (e.g., automated drift and change-detection infrastructure); training data. \vspace{0.5em} \newline
\textbf{Organization:} Authority to examine any system, process, or decision; ability to interview all relevant personnel; full visibility into relevant governance processes and decision-making; continuous access to relevant internal communications related to safety and security. &
\textbf{System:} Continuous automated monitoring of model behavior, deployments, and compute usage; comprehensive interpretability research on models; automated detection of behavioral drift and undisclosed training. \vspace{0.5em} \newline
\textbf{Organization:} Review major decisions (training runs, deployments, safety exceptions) as they happen; ongoing governance review; cross-functional safety culture assessment. Certifications automatically degrade when material changes occur. &
Uncertain -- possibly early 2027 if significant investments begin soon. \\
\hline

%\rowcolor{tablegray}
AAL-4: \newline Very high \newline assurance &
Continuous &
\textbf{System:} Tamper-evident logging across infrastructure using formally verified open-source cryptographic provenance tooling; ability to inspect training data, model weights, and deployment configurations. \vspace{0.5em} \newline
\textbf{Organization:} Multiple independently selected audit teams with full organizational access; short notice inspections of physical facilities and interviewing of staff. &
\textbf{System:} Hardware attestation and verification; formally verified open-source cryptographic provenance tooling; destructive testing of hardware samples to detect possible tampering; continuous automated and manual monitoring of all risk-relevant systems. \vspace{0.5em} \newline
\textbf{Organization:} Independent verification by multiple teams with no shared dependencies; whistleblower bounties for disclosure of unaccounted compute or theft; regular unannounced inspections. Adversarial red-teaming targets verification mechanisms themselves. &
Uncertain -- possibly late 2027 if significant investments begin soon. \\
\end{xltabular}
}

\textbf{AAL-1 (limited assurance).} An audit at AAL-1 indicates to stakeholders that the auditor has some degree of confidence in their conclusion, as there were no glaring issues found with the claims, systems, and practices they assessed, though at this level, the auditor is still relying heavily on a company's representations (i.e., formal statements by company staff asserting certain facts) (see \autocite{harris_which_2023, pcaob_as_2000}) and knowingly runs the risk of multiple types of abstraction error (\cref{ssec:comprehensive_organizational}). 
With the time span (typically a few weeks) and limited breadth and depth of AAL-1, it is possible to only rule out a subset of possible abstraction errors --- e.g., conflating evaluation of a specific model with the overarching system being assessed. Conclusions have a short half-life and say relatively little about the company as a whole, with the exceptional case of the company having negligible additional activities beyond developing the one audited system. As a result, any company-level claims made by the auditor specifically relate to the company's processes as they are applied to the specific system being audited, rather than providing much confidence regarding how those processes are applied more broadly (e.g., to internal deployments or to other externally deployed systems that are out of scope).\looseness=-1

While these engagements are not enough to qualify as audits according to some standards in other domains (see \autocite{pcaob_as_2024}), they still provide meaningful evidence compared to self-assessment alone, so we treat them as the starting point for frontier AI auditing. Some very recent frontier AI assessments are at this level (see \autocite{metr_review_2025}).

\medskip 
\begin{infobox}
\textbf{What this can detect:} Dangerous capabilities that surface under industry standard evaluation methods (e.g., ability to generate working exploit code, synthesis instructions for controlled substances); glaring gaps between stated policies and documentation; basic failures in safety mitigations.

\medskip
\textit{Example conclusion:} ``Within our three-week evaluation using API access, we found no evidence that the system can reliably assist with novel cyberattacks.''

\medskip
\textbf{What this cannot detect:} Capabilities requiring sophisticated elicitation; gaps between documentation and actual practice; undisclosed systems or training runs; any intentional concealment. Auditors take company representations largely at face value.

\medskip
\textbf{Who audits:} Single evaluation organization or small team. Standard conflict-of-interest disclosures; lighter independence requirements than higher levels.
\end{infobox}
\medskip

\textbf{AAL-2 (moderate assurance).} Auditors use company-provided documentation to inform their analysis and make it more efficient, but ultimately draw their conclusions based primarily on direct evidence gathered over months rather than company representations (see \autocite{pcaob_as_2000}). A limited degree of continuous monitoring is achieved: auditors at this level will verify that the model and systems that they are examining are the same ones that are actually being deployed \autocite{south_verifiable_2024}, ruling out some kinds of configuration drift (\cref{ssec:comprehensive_organizational}). An audit at AAL-2 indicates to stakeholders that the auditor has ruled out several potential sources of abstraction errors compared to AAL-1 (e.g., via incorporation of some degree of continuous monitoring, and looking beyond models to consider a wider fraction of relevant system-level and platform-level mitigations), though still not all such errors. More rigorous technical analysis will be brought to bear at this level, making use of a richer range of information. 
This includes access to safety cases provided by the company.
For AAL-2, auditors verify most or all aspects of such safety cases.

\medskip 
\begin{infobox}
\textbf{What this can detect:} Negligence and sloppiness in safety practices; gaps between stated policies and actual practice in examined areas; cherry-picking of favorable evaluation results; basic security control failures.

\medskip
\textit{Example conclusion:} ``We independently reproduced the company's internal capability evaluations and confirmed the release gate was followed for this deployment.''

\medskip
\textbf{What this cannot detect:} Problems in unsampled systems or processes; long-running concealment efforts; undisclosed training runs or ``shadow'' systems; sophisticated fraud. Auditors assume sampled areas are representative and the company is not actively deceiving them.

\medskip
\textbf{Who audits:} Accredited evaluation organization(s), potentially with subcontractors for specialized domains (security, biosafety). Lead auditor coordinates and takes responsibility. Stronger independence requirements: cooling-off periods, revenue diversification, disclosed terms of engagement.
\end{infobox}
\medskip 

\clearpage
\textbf{AAL-3 (high assurance).} All company activities are presumptively in-scope and information about these activities is fully available (i.e., white-box access); the range of possible abstraction errors is narrow and well-defined. Remaining uncertainties primarily relate to the possibility of sophisticated deception or industry-wide vulnerabilities (e.g., unknown problems in open-source code or unknown weaknesses in state-of-the-art alignment techniques). An audit at AAL-3 indicates to stakeholders that the conclusion is highly likely to be true. At this level, all material aspects of safety cases and information security protocols will have been directly verified, and continuous monitoring (over the course of multiple years) will be used extensively to ensure that audit conclusions continue to hold and material changes are unlikely to be missed by auditors. Nevertheless, the gaps that do remain are significant enough that a very skeptical party --- one who believes that the audited company would exploit any opportunity to cut corners or deceive auditors --- would still not be willing to rely on an AAL-3 in very high-stakes situations.
\medskip 
\begin{infobox}
\textbf{What this can detect:} Most concealment and corner-cutting; significant safety gaps; undisclosed major training runs (given adequate compute accounting); material misrepresentations; changes that invalidate prior findings.

\medskip
\textit{Example conclusion:} ``We have continuous visibility into all major training runs and deployments; compute accounting confirms no undisclosed runs exceeding the agreed threshold.''

\medskip
\textbf{What this cannot detect:} Extremely sophisticated deception involving compromised monitoring infrastructure or multiparty collusion; compromised hardware supply chains. Auditors assume monitoring works as intended and at least one auditor is honest and competent.

\medskip
\textbf{Who audits:} Multiple accredited auditors with long-term (multiyear) engagement. Lead auditor can subcontract specialized teams. An oversight body receives unredacted reports and can inspect the audit itself. Stringent independence requirements (e.g., more restrictive financial disclosure requirements than earlier levels; payment must come from a source other than auditees).
\end{infobox}
\medskip 

\textbf{AAL-4 (very high assurance).} At this level, any violations of load-bearing assumptions behind key safety and security claims would be quickly detected by auditors before significant harm could occur. Furthermore, few if any potential sources of abstraction error are considered to have material likelihoods (and any that do are closely monitored and well-quantified). AAL-4 audits can provide confidence in conclusions even assuming highly resourced and motivated actors aggressively exploiting opportunities to cheat (see \autocite{us_gao_iaea_2005, rockwood_verification_2016}). ``Very high assurance'' is not well-defined in existing auditing literature and represents our effort to bridge the literature on auditing with the literature on verification of international arms control agreements, where very high assurance is required \autocite{vaynman_dual_2023}. We make this connection because we believe that AAL-4 audits may ultimately be needed for very high-stakes purposes such as verifying US--China cooperation on baseline AI safety and security standards \autocite{mitre_artificial_2025}. They could also plausibly be needed for domestic regulation purposes alone, simply due to significant advances in AI capabilities and their associated risks, making significant uncertainty in frontier AI risk mitigations no longer tolerable.

\medskip 
\begin{infobox}
\textbf{What this can detect:} Deliberate deception including even relatively small hidden training runs and inference jobs, shadow systems with safeguards removed, and selective disclosure of information about systems and practices.

\medskip
\textit{Example conclusion:} ``Hardware attestation and cryptographic logs confirm compliance with the agreed restrictions on fine-tuning for dangerous capabilities, even accounting for potential evasion attempts.''

\medskip
\textbf{What this cannot detect:} Completely novel evasion techniques unknown to highly-resourced verification designers; unknown vulnerabilities in long-used, well-studied cryptographic algorithms. No verification regime provides absolute guarantees, but the remaining sources of error are very tightly circumscribed.

\medskip
\textbf{Who audits:} One or more accredited lead auditors and a range of subcontractors performing various functions. Government involvement is likely necessary for legitimacy, enforcement, and access to national security information. May require multi-jurisdictional representation and security clearances.
\end{infobox}
\medskip 

\subsubsection{Choice of AI Assurance Levels}

There is a trade-off between gaining more confidence that risks have been mitigated, and the financial costs associated with reaching that higher confidence (see \autocite{kang_client_2022, badertscher_assurance_2023}). Depending on the scope of the engagement, and with very high uncertainty, we loosely estimate that AAL-1 engagements could cost around \$300,000--\$600,000 for multi-week to few-month engagements, AAL-2 engagements might cost around \$1,000,000 or more for multi-month engagements, while AAL-3 or AAL-4 engagements requiring continuous access and specialized technical verification could cost several million dollars annually. 

These costs could potentially be reduced over time through automation and amortization of initial infrastructure investments, though as the stakes of missing key issues increase, it may be appropriate to increase investment in computing power applied to auditing, which may cancel out that effect. AAL-3 appears (at least) very difficult today and AAL-4 appears infeasible today, making research and pilots on each a priority.

In this paper, we do not aim to settle the question of which of the four AI Assurance Levels (AALs) should be applied to which subsets of frontier AI, beyond recommending AAL-1 as the floor for frontier AI as a general category, and AAL-2 as a near-term goal for the most advanced subset of frontier AI.\footnote{An illustrative operationalization of ``the most advanced subset'' might be, e.g., companies that have produced, within the prior year, any AI systems that were within three months of the state-of-the-art at the time. Making such a definition more precise would require wider discussion among stakeholders and analysis, which we hope that this paper helps encourage.} We do not think it's desirable or realistic to be much more prescriptive than that at this stage given the many factors bearing on the decision (discussed in \cref{ssec:comprehensive_organizational}). We make recommendations in \cref{sec:challenges} for how these threshold questions can be continuously updated over time based on the latest evidence.

\subsection{Deep, secure, and timely access to information and resources} \label{ssec:deep_access}

A mature frontier AI auditing ecosystem depends on auditors having deep, secure, and timely access to the information, systems, and organizational processes under examination. Ultimately access should be deep enough to assure auditors and other stakeholders, but secure enough to reassure auditees. The access provisions should depend on the specific audit and be proportional to the risks believed (pre-audit) to be posed by the company's systems and the assurance level sought (see \cref{ssec:levels}). Auditors should also be provided with sufficient tooling and compute resources, as well as channels for viewing non-public information. Access should be provided through secure evaluation environments if required for addressing privacy or security concerns. 

\begin{figure}[t]
    \centering
    \includegraphics[width=0.7\linewidth]{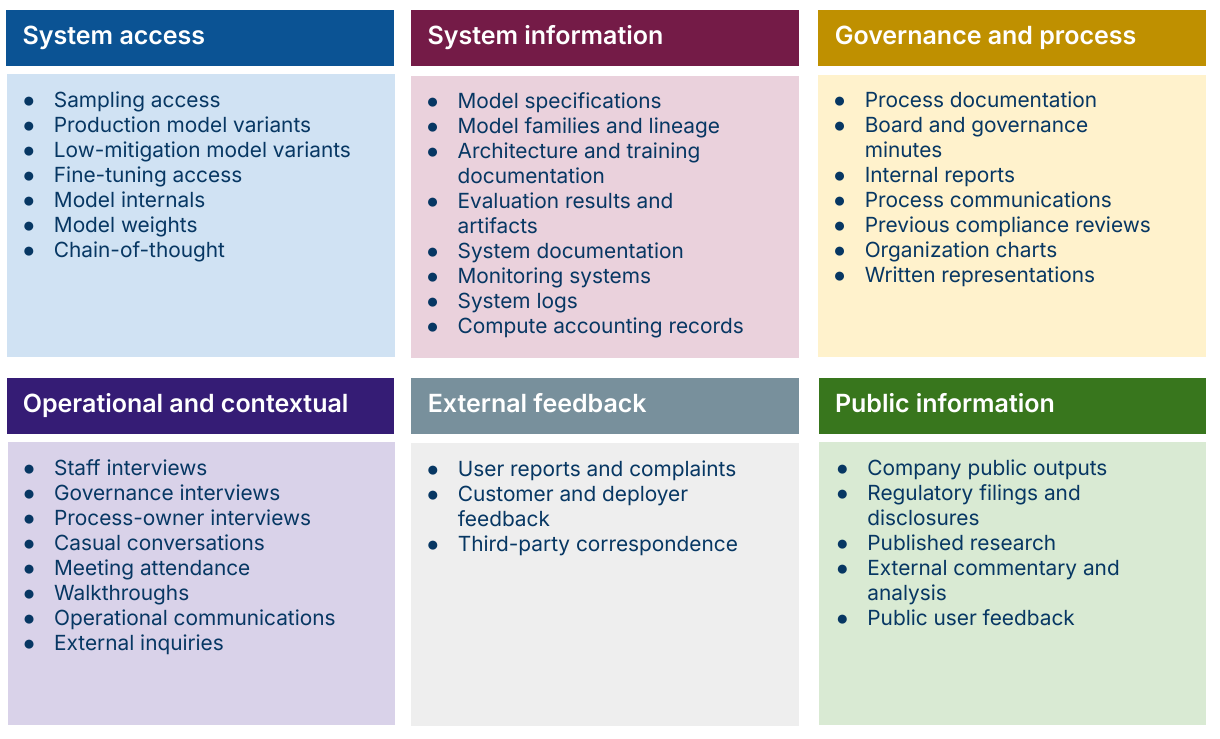}
    \caption{A non-exhaustive taxonomy of information sources that companies may provide access to, across model, system, governance, and operational domains. Public information is also included, as auditors should consider it alongside company-provided sources. The depth of access required will depend on the specific audit engagement and the assurance level sought. See \cref{apx:access} for descriptions of the different items.}
    \label{fig:list_information_resources}
\end{figure}

\textbf{Deep access.} To verify safety and security claims relating to AI systems, auditors need deep technical and organizational access (see Figure \ref{fig:list_information_resources} and \cref{apx:access}). This is essential in order to avoid the four types of abstraction error discussed in \cref{ssec:comprehensive_organizational}.

For AI models and systems, at a minimum, this should include \hyperlink{gls:blackbox}{black-box} sampling access to the system(s) through an API with permissive rate limits. However, greater levels of assurance will likely require auditors being provided with deeper access, including access to output logits, weights, activations, or the ability to modify the model (for example, through fine-tuning) \autocite{bucknall_structured_2023, casper_black-box_2024, kembery_position_2024, che_model_2025, bucknall_position_2025}. Furthermore, assessing some claims will require auditor access to systems other than those to be deployed, as well as information about the system's functioning and how it was developed. For example, establishing upper bounds on a system's potential risks will depend on assessing system versions without safety guardrails in place (e.g., helpful-only models can be used to understand worst-case scenarios if those guardrails fail). In addition to access to AI systems, auditors should be provided access to other relevant non-public information, such as compute accounting,\footnote{By compute accounting, we mean systematically tracking the use of computing power in order to verify that it was used in the manner described, and to reduce uncertainty about the possibility of undisclosed, significant training runs or inference runs \autocite{brundage_toward_2020, baker_verifying_2025, harack_verification_2025}. Notably, this is particularly relevant to higher assurance levels. It is technically challenging to do compute accounting effectively, and at lower assurance levels, claims about compute usage will likely need to be taken on trust.} incident reports, internal risk assessments, meeting notes, and decision logs. Auditors may also need to conduct interviews with relevant staff members in order to verify information or dig deeper on topics that are not well-documented.

\textbf{Secure access.} Providing deeper levels of access and sensitive information to auditors could expose commercially valuable intellectual property or national security-critical assets to leaks, theft, or sabotage, as well as legal risks.\footnote{Model weights, proprietary training data, internal tools, detailed security documentation, and even model outputs in sensitive domains like nuclear technology are sensitive assets, and are often protected by law or norms. For example, privacy law limits access to some datasets; 
trade‑secret protections constrain what can be widely shared; 
export controls may restrict whether foreign nationals can see certain information \autocite{22_cfr_2025}. Further, several AI companies sharing auditors can also raise concerns about information spillovers \autocite{kang_client_2022}.} When access to valuable or sensitive assets is required, audits should be conducted through secure evaluation environments that allow auditors to run tests, probe model behavior, examine system responses, and simulate realistic traffic, while preventing unauthorized disclosure. These environments should also prevent the auditee from observing or influencing tests, reducing the possibility of ``teaching to the test'' and ensuring auditors' findings remain meaningful over time and across companies. Comparable practices exist in other high-stakes assurance settings, such as regulator-operated test environments in food safety \autocite{us_food_and_drug_administration_fsma_2021}, medical device safety \autocite{eu_regulation_2017}, and automotive emissions \autocite{epa_about_2025}. OpenMined's PySyft framework has been tested with Anthropic and the UK AI Security Institute using NVIDIA H100 secure enclaves for mutually protected evaluations, though so far only on small, non-production systems \autocite{trask_secure_2025}.\looseness=-1

Access to sensitive information should be provided through secure channels and legal safeguards. Secure channels could involve issuing monitored and secured corporate devices to auditors, or having auditors work on-premise at company facilities under controlled conditions. Other sectors with high confidentiality requirements use similar structures to balance access and protection. For example, due diligence in mergers and acquisitions often relies on the use of ``\hyperlink{gls:cleanroom}{clean rooms}'' (physically or logically isolated digital environments for analyzing sensitive information) and ``\hyperlink{gls:cleanteam}{clean teams}'' (personnel who have access to confidential information but are insulated from decisions where that information could create conflicts of interests) \autocite{hewitt_how_2010, whitaker_establishing_2012}. Other existing mechanisms include \hyperlink{gls:compart}{audit compartmentalization}, where different auditors assess different aspects of operations, and secure enclave models in cloud security assurance \autocite{bartock_hardware-enabled_2022}. Legal safeguards may include nondisclosure agreements, contractual liability for breaches, and professional sanctions for auditors who mishandle sensitive information.\looseness=-1

\textbf{Adequate resources and tooling.} Auditors should have the resources required to conduct high-quality, multi-domain assessments of frontier AI systems. This includes access to sufficient compute (e.g., for evaluations, parallelization, verification, and re-runs); robust tooling for inspecting logs, lineage, and governance records; and technical support from the auditee regarding the operation of auditing tools.

\textbf{Timely and responsive access.} Auditors should be given adequate time to design, execute, and analyze all audit activities, whether technical evaluations, document reviews, or staff interviews. Communication channels should be provided through which auditors can receive prompt answers to follow-up questions and rapid access to updated materials after significant changes, including logs and lineage information for technical assessments, and governance artifacts, decision records, and relevant personnel.

\textbf{Governed and accountable access arrangements.} Auditor access should be structured through clear agreements that specify rights, obligations, permitted uses \autocite{stosz_aef-1_2025}, and consequences for misuse. Access arrangements should define what auditors may examine, how information may be used and stored, and the conditions under which access may be expanded, restricted, or revoked. Auditors should be subject to confidentiality obligations backed by contractual, professional, and, where appropriate, legal sanctions, with clear liability for breaches. Disputes about the scope or burden of access requests should be resolved rapidly through clear escalation pathways and independent mediation so that audits are not delayed or obstructed.\looseness=-1

\textbf{Safeguards against omission and selective disclosure.} For the highest levels of assurance, mechanisms should be in place that allow for detection of whether relevant information has been withheld, manipulated, or selectively disclosed by the auditee. This requires tools and processes such as compute accounting to assess whether there are likely any materially significant AI systems created by the company that auditors haven't assessed, random sampling of logs or lineage artifacts, structured comparisons between public and private documentation, protected whistleblowing channels for employees, and whistleblowing bounties \autocite{grunewald_whistleblower_2025}. Where companies refuse access without adequate justification, auditors should make adverse inferences consistent with established practice in other assurance systems (e.g., IAASB ISA 705 \autocite{iaasb_international_2015} and PCAOB AS 3105 \autocite{noauthor_as_2017}), with fair processes for arbitration of disagreements between auditors and auditees (e.g., by the organizations discussed in \textbf{\cref{ssec:ensuring_standards}} and \textbf{\cref{ssec:reaching_full}}).

\subsection{Continuous, risk-responsive assurance} \label{ssec:continuous_assurance}

Our vision for a mature frontier AI auditing regime is one where audit conclusions remain accurate as systems and developer practices change, risk-relevant changes can be detected and responded to in a timely fashion, and assurance processes are aligned with the times at which risks are created, amplified, mitigated, or revealed. The goal is to produce living assessments that evolve with the systems and companies they evaluate, not static documents that become stale within days or weeks of publication.

\begin{figure}[t]
    \centering
    \includegraphics[width=0.8\linewidth]{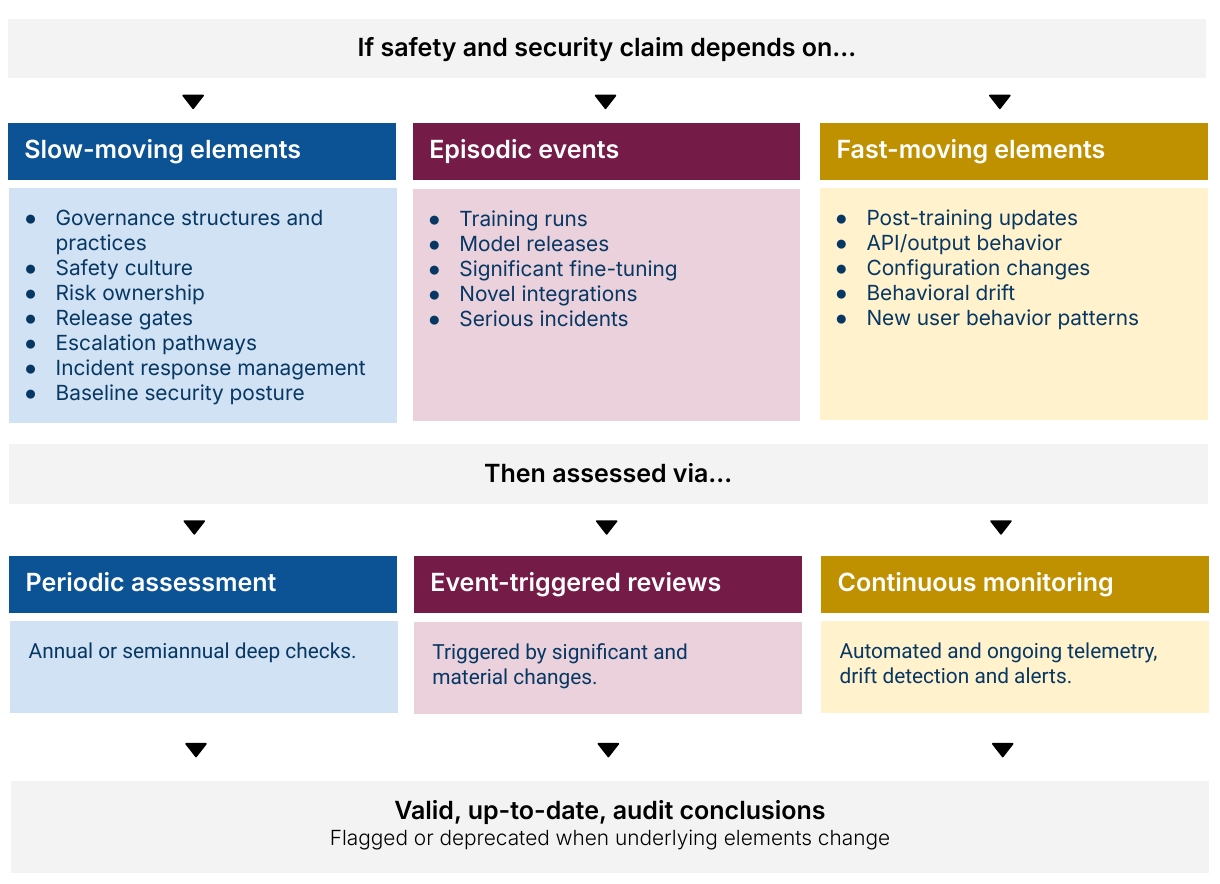}
    \caption{Matching assessment cadence to rate of change. Safety and security claims depend on elements that change at different speeds. To keep audit conclusions valid over time, the auditing ecosystem should assess each element at a cadence matching its rate of change: periodic deep assessment for slow-moving organizational elements, event-triggered reviews for episodic technical and deployment decisions, and continuous automated monitoring for fast-changing behavioral surfaces.}
    \label{fig:auditing_cadences}
\end{figure}

\textbf{Audit cadences.} The frontier AI auditing ecosystem should assess different safety and security claims at cadences that match how quickly the underlying elements change (see Figure~\ref{fig:auditing_cadences}). This ensures audit conclusions remain valid over time. Slower-moving elements of the organization (e.g., governance, safety culture, release gates, incident response, security posture) warrant less frequent, periodic deep assessment (e.g. annual or semiannual). Faster-moving and time-bound elements and decisions (e.g., training runs, releases, post-training, incidents) can trigger event-based reviews that test whether controls were implemented and risk decisions were appropriate. Claims that depend on the fastest-changing surfaces (e.g., API behavior, configuration, drift, user behavior patterns) require continuous automated monitoring that logs changes and triggers alerts when systems deviate from certified parameters. Of course, other considerations beyond the velocity of change, such as the impact on risk, uncertainties, and assumptions, as well as the assurance level should likely also determine the required audit cadence for a given claim. Within ongoing audit engagements, the lead auditor should be responsible for synthesizing evidence across domains and cadences relevant to the evaluated safety and security claim(s), and coordinating specialist subcontractors, where used, to ensure sufficient coverage for the targeted assurance level and audit's scope (\cref{ssec:comprehensive_organizational}).\looseness=-1

\textbf{Live certification and deprecation.} Assurance should remain valid only while underlying assumptions hold \autocite{nist_risk_2018, noauthor_isoiec_2019}, automatically downgrading on material changes via \hyperlink{gls:livecert}{live certification} or time-limited \hyperlink{gls:validityperiod}{validity periods}. Live certification requires maintaining current records of supporting assumptions; when material changes occur, certification is flagged or downgraded pending review. Each certification should specify the conditions on which it rests (e.g., the model version, safety configuration, and deployment pathway) along with clear criteria for what constitutes a material change requiring re-evaluation. Companies should proactively flag upcoming changes to auditors, though auditors should avoid over-reactivity. Active change monitoring aligns with companies' interest in avoiding inadvertent performance degradation.\looseness=-1 

Several assurance regimes already operate on variants of this principle: in information security, certified organizations are required to maintain controls continuously and correct weaknesses when conditions change \autocite{standardization_isoiec_2022}, and in aviation, changes in the organization or its activities that cause it to no longer meet requirements require the organization to seek an amendment to its approval or certificate \autocite{noauthor_easy_2024}. However, frontier AI auditing faces distinctive challenges in implementing continuous assurance. AI systems may be updated far more frequently than products in other regulated industries, and changes in behavior may be subtle or emerge gradually. Unlike financial materiality thresholds, there is not yet consensus on what should trigger re-certification (e.g., the magnitude of capability change \autocite{pacchiardi_framework_2025}). Addressing these challenges will require investment in automated monitoring infrastructure and the development of consensus on what constitutes materiality in changes.

\subsection{Independent, expert, and well-governed auditors} \label{ssec:independent_auditors}

Frontier AI auditing should be delivered through a mature, professionalized ecosystem of independent, technically proficient auditors subject to credible oversight. Auditors should be free from commercial or political influence; combine deep expertise in AI evaluation, safety, security, and governance with strong confidentiality practices; and provide reliable assurance to AI companies, policymakers, investors, insurers, and the wider public. Independence creates credibility, since auditors' reputations rest almost entirely on the quality and integrity of their assessments.

\textbf{Frontier AI auditing should be conducted by independent third parties.\footnote{Sometimes companies maintain ``internal audit'' functions, which have some operational independence from product teams, and these may be referred to as ``independent'' by the company in question \autocite{schuett_frontier_2024}. However, on their own, we consider these to be insufficiently independent to provide credible assurance, and below the threshold for independence in the way we use the term.\looseness=-1}} These may be non-profit or for-profit organizations. Potential audit providers (or subcontractors thereof) include AI assessment-focused companies and non-profits, law firms, accounting firms, security penetration-testing firms, government entities, and hybrid arrangements that combine these capabilities.\footnote{For related discussion regarding the benefits and limitations of different providers, see \autocite{homewood_third-party_2025}.} Individual auditors should have no direct financial stake in the companies they assess.\footnote{Many frontier AI experts hold equity in frontier AI companies. Excluding these experts from the auditing process entirely could leave the pool of qualified auditors too small, while including them without safeguards creates obvious conflicts. To mitigate these conflicts while accommodating the realities of the current AI field, we recommend that lead auditors in particular should not hold any direct equity in the frontier AI company being audited (and should also disclose any indirect conflicts), and that anyone involved in the audit process (not as the lead) should also disclose any equity they hold. See footnote \ref{fn:directindirect} on the distinction between direct and indirect equity holdings.} Audit firms should avoid material revenue dependence on any single client\footnote{Revenue from a client is ``material'' when it could reasonably be seen as threatening the auditor's independence. In other contexts, this has been defined using a quantitative benchmark, typically 10\% or 15\% of a firm's annual fees \autocite{iesba_revisions_2021, financial_reporting_council_revised_2024}. While setting an analogous benchmark may be appropriate for AI, it is not obvious what that threshold should be. Frontier AI is by definition a limited subset of the AI industry, which may make it difficult to avoid crossing 10\% and 15\% thresholds in particular, perhaps suggesting the need for a higher threshold. It could also be easier to stay below such a threshold if an audit firm also services non-frontier AI companies or carries out other non-auditing activities.} and maintain strict boundaries between assurance work and non-assurance work (e.g., consulting) \autocite{church_auditor_2015, yakubu_theoretical_2020}. Such independence requirements are standard in financial auditing under \hyperlink{gls:pcaob}{PCAOB} rules and in European aviation certification \autocite{easa_making_nodate}, and also feature prominently in the first set of standards for frontier AI evaluation, AEF-1 \autocite{stosz_aef-1_2025}. 

Importantly, payment should not depend on audit results \autocite{stosz_aef-1_2025}. But this alone is not sufficient, since an auditor might still perceive that future work, or continued access, depends on favorable findings. Allowing companies to choose their own auditors has created recurring problems in other industries \autocite{moore_conflicts_2006}. As such, it is preferable to address this risk before the sector scales much further by, for example, transitioning toward payment models that avoid direct financial dependence on audited companies. We believe research on possible alternatives should be urgently pursued, and progress toward alternatives should be made by the end of 2026. (See \cref{apx:risks} for further discussion.)

\textbf{Transparent conflict-of-interest management.} Additional safeguards to manage conflicts of interest are also needed. Mechanisms from established assurance regimes, such as public conflict‑of‑interest registers, cooling-off periods for personnel moving between auditing organizations and audited companies, restrictions on non-audit services, and disclosure of any financial or advisory ties should also be implemented in frontier AI auditing \autocite{costanza-chock_who_2022}. These measures are standard in finance and information security auditing and help reduce familiarity and self‑interest biasing audit results \autocite{pcaob_ethics_nodate}. When auditors are paid directly by clients, public disclosure of audit terms and adherence to industry standards may improve audit quality (see \autocite{ronen_corporate_2010}). The AEF‑1 standard for AI evaluations requires disclosure of potential conflicts of interest, including financial dependence \autocite{stosz_aef-1_2025}.

Frontier AI auditing may require additional or alternative safeguards given the small pool of technical experts in the space, though any deviations from these best practices should be justified and prominently disclosed. For example, if a long cooling‑off period would be prohibitively restrictive, recent frontier AI company employees could instead serve as subcontractors with a very specific technical remit on a larger audit engagement, rather than as lead auditors themselves.\footnote{One way in which such a rule might be evaded is if a lead auditor merely served as a ``front'' for a subcontractor, who does all of the real work. However, this is a possibility regardless of conflict of interest rules, so we do not view it as a decisive objection against such rules. A requirement for lead auditors to sign their findings has been shown to improve results \autocite{carcello_costs_2013} in a financial auditing context, and the ``front'' scenario is an additional reason why such signatures may also be appropriate in an AI context.\looseness=-1} Lead auditors should not have direct financial stakes in companies they audit, but more work is needed to specify granular standards for edge cases.\footnotemark \footnotetext{\label{fn:directindirect}Many frontier AI companies are included in publicly-traded stock indices. Even when an auditor holds no ``direct'' equity in the frontier AI company, they may hold ``indirect'' equity through owning units of an exchange-traded fund, or another company that in turn holds equity in the frontier AI company. Existing frameworks providing guidance on independence requirements regarding investments (see \autocite{public_company_accounting_oversight_board_et_nodate}) could be adapted for the context of frontier AI auditing. Many frontier AI companies remaining private presents another challenge, making it difficult for ex-employees to divest at short notice.} Public registers of auditor affiliations and financial stakes could help avoid either the appearance or the reality of conflict.\looseness=-1

\textbf{A mature frontier AI assurance system requires a sufficiently large and diverse pool of qualified auditors.} In contrast to relying on a small set of repeat auditors, a broad pool of qualified auditors ensures access to the technical and organizational expertise needed for credible assessment, provides an incentive for innovation that improves scalability and security, supports pluralism in methods, and reduces the likelihood of a single point of failure.\footnote{Whether the number of auditors is ``sufficient'' will depend on factors such as the number of clients, the number of audits, the breadth and depth of each audit, and the number of qualified personnel. When introducing auditing requirements, policymakers should monitor the capacity of the ecosystem, and consider interventions (e.g., training programs or encouraging new entrants) if concentration risks emerge.}

\textbf{Auditors need to have deep expertise.} At a minimum, auditors must have deep expertise in AI evaluation, safety, security, and governance. Specific audit functions (e.g., concerning highly specialized knowledge, like CBRN risks) require additional, deep domain expertise. If a single organization lacks the diversity and depth of expertise required, subcontracting should be used --- a common scenario in the near to medium term given the breadth of skills required. In such cases, a lead auditor should coordinate and assume responsibility for the whole audit, including the audit's scope, assessments, conclusions, and reporting. The ecosystem should enable collaboration across specialist firms, research institutions, and civil society experts so that engagements can ``mix and match'' complementary skill sets. For example, large professional services firms could partner with specialized AI assessment organizations through subcontracting and \hyperlink{gls:flowdown}{``flow-down'' agreements}, or alternatively by forming a \hyperlink{gls:consortium}{consortium}. For information security auditing specifically, there is a more well-established ecosystem of penetration testers and other assessment providers to draw on. It will be critical to ensure that, even if such contracting is done separately from a lead frontier AI auditor, the lead auditor has access to the findings of information security assessments and can make sure that no gaps exist between assessment of ``traditional'' information security risks (e.g., theft of intellectual property) and AI-specific security risks (e.g., data poisoning for language models).\looseness=-1

\textbf{Transparent and standardized terms for auditing contracts.} Currently, third‑party AI assessments are generally performed under bespoke, negotiated contracts between developers and evaluators that are subject to strict confidentiality. There is significant variation in transparency, with third-party evaluators' identities sometimes not even disclosed. Standardized terms of engagement can help prevent companies from shopping for favorable auditors and ensure a consistent baseline of auditor access, independence, and reporting obligations. OpenAI recently published excerpts from agreements they use for pre-deployment testing \autocite{openai_strengthening_2025}, and the AI Evaluator Forum is recommending a baseline set of standards to be used in drafting such agreements \autocite{ai_evaluator_forum_transparency_nodate, stosz_aef-1_2025}.\footnote{Details about, for example, the scope of an audit would often not be appropriate to disclose publicly, but procedural details about who, when, why, and how assessments are performed are needed in order to properly and contextually interpret their results and ensure that individual engagements are not performed in an ad hoc way.} Future regulation may also standardize the terms of audits that cover regulation requirements, as is the case in other industries (e.g., \autocite{noauthor_regulation_2014, eu_regulation_2017}).

\textbf{Independent oversight and quality assurance mechanisms.} Having an independent oversight board charged with raising standards in the sector can be valuable in tracking developments and punishing egregious practices (e.g., \autocite{public_company_accounting_oversight_board_spotlight_2024-1}). For financial auditors of publicly traded companies in the US, this takes the form of the Public Company Accounting Oversight Board (PCAOB), which was created under the Sarbanes–Oxley Act in the wake of accounting-related scandals at Enron and elsewhere. Such a body could serve several functions: developing auditing standards, certifying auditors, examining the quality of audits themselves, and revoking credentials where circumstances warrant. We discuss considerations bearing on the design of a ``PCAOB-for-AI'' in \cref{sec:challenges}. Looking ahead, a mature ecosystem could include a live \hyperlink{gls:tracker}{AI assurance tracker}: a public platform maintained by an oversight body showing each company's stated policies, applicable regulations, lead auditor, and recent audit conclusions, updated as material changes occur.\looseness=-1

\subsection{Rigorous, traceable, and adaptive processes} \label{ssec:rigorous_processes}

A mature frontier AI auditing ecosystem depends on audits following a rigorous, reproducible process that produces reliable, traceable, and defensible evidence appropriate for the given assurance level. This requires adaptivity to balance competing demands: consistency and comparability across engagements, scalability as audit volume and system complexity increase, methodological flexibility to keep pace with rapidly evolving technology, and procedural fairness to audited companies without compromising independence.\looseness=-1

\textbf{Standardized criteria.} Each frontier AI audit should apply predefined criteria describing what counts as sufficient evidence of safety, security, and sound risk management. These criteria may draw on regulatory frameworks, international standards, industry best practices, and companies' own commitments, supplemented by expert-developed criteria for emerging risk areas where existing frameworks provide insufficient guidance. Currently, there is no universally accepted or sufficiently granular set of auditing criteria for frontier AI systems. Developing, testing, and refining these criteria will be an important task for companies, auditors, regulators, and standards bodies. Effective criteria should be comparable across engagements, flexible enough to accommodate evolving technology, grounded in technical validity through rigorous measurement design, supported by stakeholder processes that confer legitimacy across jurisdictions, and designed to minimize gaming while aspiring to genuine safety rather than bare minimum compliance.

\begin{figure}[t]
    \centering
    \includegraphics[width=1\linewidth]{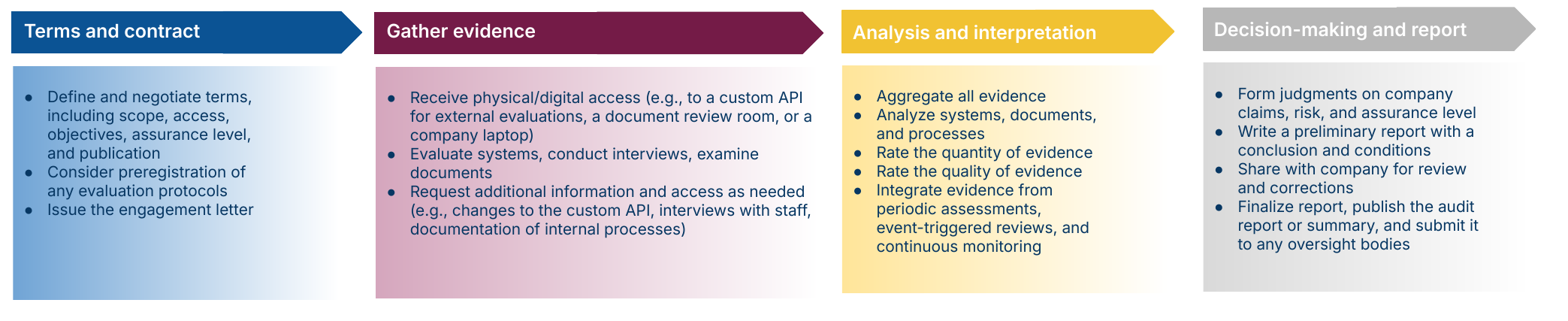}
    \caption{Standard frontier AI auditing workflow.}
    \label{fig:standard_workflow}
\end{figure}

\textbf{Standardized auditing process.} Audits should follow a clear and consistent process model that supports rigor and reproducibility. Frontier AI audits involve many moving parts, and inconsistent methods risk producing variable or unreliable results. A standardized process sets out expected steps for scoping, access, evidence gathering, analysis, verification, continuous monitoring, and reporting. (See Figure \ref{fig:standard_workflow} for a potential workflow for a frontier AI audit that builds on a standard auditing workflow.) By using these steps, auditors can consistently provide high-quality audits. If auditing methods are consistent, auditors can be overseen more effectively, and different companies' results can be compared. In addition, wherever feasible and appropriate, audit procedures should make use of auditable automated processes to enable consistent application of methods across engagements, continuous monitoring as systems evolve, and oversight of automated methods. Importantly, a commitment to systematic auditing processes does not diminish the value of flexible, minimally structured red-teaming efforts, which remain essential for discovering novel vulnerabilities and failure modes that more structured approaches might miss.

\textbf{Auditor flexibility and autonomy.} At the same time, standardization should not lock in immature practices or constrain the methodological autonomy that makes independent evaluation meaningful. Auditors must have freedom in deciding their methods --- including defining metrics, determining how to elicit target properties, and establishing criteria for success or failure --- rather than being constrained to validate the company's pre-selected approaches \autocite{stosz_aef-1_2025}.\footnote{Of course, this autonomy should accompany adherence to standardized best practices, transparency to oversight bodies about methods, and independent oversight of auditors themselves (see \cref{sec:challenges}).} Auditors should also retain flexibility to adjust the scope of their inquiry as the evaluation proceeds, since important issues may only become apparent during the course of an assessment. Narrow scopes can allow organizations to meet commitments in form but not in substance, leaving critical risks unexamined.

\textbf{Reproducibility of results.} Audit processes and methods should be documented sufficiently that another auditor or oversight body could reproduce the approach and verify the results \autocite{paskov_toward_2025}. This includes recording the specific procedures used, the evidence gathered, the criteria applied, and the reasoning behind key judgments. Reproducibility supports quality assurance, enables oversight bodies to inspect audit work, and builds confidence that findings reflect genuine properties of the system rather than idiosyncrasies of a particular engagement. For each audit, a contract should define the scope, objectives, and responsibilities of an auditor, and pre-registration of evaluation protocols can further strengthen reproducibility and protect against bias. Pre-registration is an accepted best practice in the sciences \autocite{nosek_preregistration_2018, angelis_clinical_2004}. By specifying the questions to be answered, the metrics to be used, and the criteria for success before results are known, pre-registration ensures that audit conclusions reflect genuinely confirmatory tests rather than post hoc rationalizations shaped by preliminary findings \autocite{carro_prep-eval_2026}.

\textbf{Evaluation validity.} Audits can only provide meaningful assurance if the underlying evaluations are methodologically sound and protected from compromise. Auditors should ensure that evaluations measure what they claim to measure (construct validity) and reflect real-world deployment conditions rather than artificial test settings (ecological validity) \autocite{bean_measuring_2025, eriksson_can_2025, salaudeen_measurement_2025, schellaert_evaluation_2025, wallach_position_2025, zhou_general_2025, weidinger_sociotechnical_2023, bowman_what_2021, raji_ai_2021, paskov_toward_2025, khan_randomness_2025, cooper_machine_2025}. Elicitation methods should be sufficient to surface a model's true capabilities \autocite{shevlane_model_2023, patricia_paskov_preliminary_2025}, test coverage should span a representative range of inputs derived from the real-world \autocite{tamkin_clio_2024} and failure modes (e.g., \autocite{jeune_realharm_2025}), and mitigations should be tested under strong adversarial pressure rather than only benign conditions or under weak attack \autocite{carlini_evaluating_2019, uesato_adversarial_2018}. Auditors should also verify that the evaluated system matches what is actually deployed, including its configuration and context of use.

Auditors and companies should cooperate to identify and address system behaviors that may undermine evaluation validity, including overfitting to known evaluation sets, reward hacking, sandbagging (i.e., deliberately underperforming to avoid triggering safety thresholds), and other forms of gaming \autocite{weij_ai_2025}. Companies should commit not to view, retain, or train on evaluation inputs and outputs without explicit auditor consent, ensuring that private evaluation sets remain effective over time and cannot be gamed. Auditors should consider adding canary data to evaluation sets to enable later detection of whether models have been trained on evaluation materials. Furthermore, in sensitive domains like biosecurity, even the evaluation sets and answer keys themselves may contain information that should not be widely published, adding another layer of confidentiality requirements beyond protecting evaluations from the auditee.\looseness=-1 

\textbf{Procedural fairness and dispute resolution.} Audits should incorporate procedural safeguards that ensure fairness to the audited company while preserving the independence and integrity of audit conclusions. Companies should have structured opportunities to provide input at defined points in the audit process, for example, to correct factual errors in draft findings, provide additional context or evidence that auditors may have missed, and respond to preliminary conclusions before they are finalized. However, these opportunities must be carefully controlled to prevent companies from unduly influencing audit outcomes. Company responses should be documented and, where appropriate, included in the final audit record so that readers can understand what objections were raised and how auditors addressed them. When disagreements cannot be resolved through the standard review process, they should be addressed through structured escalation pathways that do not significantly delay or obstruct the overall audit, such as arbitration.\looseness=-1

\subsection{Clear communication of audit results} \label{ssec:clear_communication}

A mature frontier AI auditing ecosystem requires that stakeholders are able to understand the results of the audit. In other industries, the auditor provides the results in the form of an \hyperlink{gls:report}{audit report} (see \autocite{pcaob_as_2024-1}). The report should contain the scope, level of assurance, conclusions, and reasoning \autocite{staufer_audit_2025}. A redacted or summarized version of the report should be shared with external stakeholders.

\textbf{Content of the audit report.} The audit report should contain the following:
\begin{itemize}[leftmargin=0.25in,topsep=0.01in]
    \item \textbf{Scope}. The audit report should contain a ``Scope'' section describing the risk categories (or subsets thereof) that were assessed. The section should also describe any exclusions (i.e., whether certain risks or types of information were excluded) and the rationale (e.g., resource constraints, inapplicability, or denial of access). Where appropriate, the report should describe any limitations of the engagement (see \autocite{iaasb_international_2013}).\looseness=-1

    \item \textbf{Assurance level statement}: The audit report should explicitly state the level of assurance at which the auditor provides their conclusion. This allows stakeholders to understand how much confidence they should have about the audit results.

    \item \textbf{Conclusions}. The audit report should clearly state the overall conclusions of the auditor (e.g. ``we agree with the company that the risk posed by their models is currently low''). The report should state any reservations the auditor had.

    \item \textbf{Reasoning}. The auditor should describe their reasoning for each conclusion in the audit report. Separately, the auditor should document (where it is possible to do so without compromising the integrity of the evaluation) the evidence and analysis they used to arrive at their conclusion. With the exception of evaluation techniques that are unique to the auditor, it should be possible for other auditors with access to the unredacted report to reproduce key steps of the audit and to arrive at similar conclusions given the same access.

    \item \textbf{Recommendations}. The auditor should describe their recommendations for remediation of any issues that arise.

    \item \textbf{Documentation}. Where it's possible to do so without compromising the integrity of the audit and confidential information, detailed documentation of the auditor's analysis and findings should be provided.\looseness=-1

\end{itemize}

\textbf{Sharing results appropriately.} Different stakeholders likely need different levels of detail about the results of the audit. The company's board and relevant executives, and in some cases relevant regulatory bodies, should receive the full report, as is standard in other industries \autocite{kovynev_2014,fdic_2025}. Other employees and relevant government bodies could receive a lightly redacted version of the report. The company could publish a summary of the audit report, along with an attestation from the auditor that the summary is fair. This approach can protect sensitive details about the company or their systems. External stakeholders should have access to a public standardized summary of scope, assurance level, and conclusions. Disagreements about appropriate redaction should be addressed through arbitration or, when established (see \cref{ssec:ensuring_standards}), a relevant standard-setting body.

\newpage 
\section{Challenges and Next Steps}\label{sec:challenges}

This section explores four challenges that must be addressed to achieve effective and universal frontier AI auditing:\looseness=-1

\begin{itemize}[leftmargin=0.25in]
    \item \textbf{Ensuring high quality standards} for frontier AI auditing, so it does not devolve into a checkbox exercise or lag behind changes in the AI industry.
    \item \textbf{Growing the ecosystem} of audit providers at a rapid pace without compromising quality.
    \item \textbf{Accelerating adoption} of frontier AI auditing by clarifying and strengthening incentives.
    \item \textbf{Achieving technical readiness} for high AI Assurance Levels so they can be applied when needed.
\end{itemize}

\subsection{Ensuring high quality standards} \label{ssec:ensuring_standards}

One central challenge for frontier AI auditing is straightforward to state but difficult to solve: audits must be rigorous enough to provide meaningful assurance to skeptical stakeholders, yet adaptive enough to keep pace with one of the fastest-evolving industries in history. If standards become too rigid, they risk devolving into procedural box-ticking that misses actual risks. If they become too flexible, they invite opinion shopping by auditees and corner-cutting by auditors.

Two specific risks illustrate these tensions. First, \hyperlink{gls:goodhart}{Goodhart's Law} holds that when metrics become compliance targets, actors optimize for targets rather than underlying outcomes. Second, temporal mismatch occurs when traditional standards update over years while AI capabilities transform within months. Auditing practice must be designed with durable goals but evolving practices. Goals should be general: principles that outline the outcomes audits are supposed to achieve, much as the PCAOB outlines in AS 1000 \autocite{pcaob_as_2024} what financial auditors are supposed to do and general guidelines on how they should do it. Practices should be flexible and carefully tailored to AI capabilities \autocite{paskov_toward_2025}. This mirrors principles-based financial regulation, where private standard setters operate under public oversight with industry input.

As in financial auditing, quality control can benefit from an independent oversight body that sets standards, inspects auditors, investigates failures, and enforces norms. This oversight body would be responsible for ``auditing the auditors''; specifically, for ensuring that auditing services are provided and outcomes are achieved as expected, and for creating accountability mechanisms if an auditor fails to do so, including potentially through the loss of its auditing credential. Oversight is a critical element of building a robust, accountable auditing ecosystem.

An effective auditing regime and independent oversight body with enforcement power may take time to build given the challenge of passing relevant laws in key jurisdictions. In the near-term, soft pressure to comply with standards, imposed by AI companies, their customers, or other third parties like investors and insurers, can serve a stopgap function before formal requirements are in place. Companies seeking trust should respond to this pressure by publishing verifiable, scoped claims assessed by credible auditors.

\begin{infobox}
\textbf{Recommendation 1:} AI companies, philanthropists, investors, and insurers should fund analysis of the quantity and quality of audits and auditors, and make these assessments available to the public.
\end{infobox}

These stakeholders should particularly support and invest in independent watchdog efforts evaluating companies' public claims, with rubrics for:
\begin{itemize}[leftmargin=0.25in]

    \item The total amount of audit capacity in the ecosystem according to various metrics, and its rate of growth over time;

    \item Whether claims made in audit reports and self-assessments by companies include sufficient methodological detail to be reproduced by those with similar access (e.g., \autocite{reuel_betterbench_2024, bordes_eval_2025}) and to be mapped onto relevant frameworks such as the AI Assurance Level (AAL) framework;

    \item Whether assessments meet the AEF-1 standard (evaluator identity, scope, access limits, publication constraints) \autocite{stosz_aef-1_2025}; and

    \item Audit quality metrics, such as \hyperlink{gls:interrater}{inter-rater reliability} (i.e., do auditors come to similar conclusions given the same evidence).

\end{itemize}

Investments in research and development, decisions by insurers and procurers, and regulatory mandates can then be informed by realistic estimates of future auditing supply at a given quality. A balance is needed between driving growth in the market over time through demand pressure, on the one hand, and, on the other, asking the impossible, which encourages corner-cutting.

Regular public reports on ecosystem health (at least quarterly) can help to create soft pressure for improvement by auditors and the companies working with them, but ultimately this pressure will reach its limits and more formal incentives will be needed.

\begin{infobox}
    \textbf{Recommendation 2:} Policymakers should implement a PCAOB-style non-profit ``auditor of auditors'' that has legitimacy through final government approval of its standards, the authority to hold auditors accountable through revoking accreditation or other means, and the ability to innovate at the pace of the private sector.
\end{infobox}

An audit-quality oversight institution would set baseline standards for frontier AI auditors, evaluate compliance, track market evolution over time and publish relevant analysis, accredit/revoke credentials, and fine auditors or refer cases for enforcement in the event of serious violations of auditing standards. 

Lessons can be drawn from a range of hybrid organizations that combine some of the benefits of the private sector (such as greater-than-government salaries, which is critical for competing for talent while overseeing a lucrative industry) and the public sector (namely enforcement power and democratic oversight). We emphasize the PCAOB --- the Public Company Accounting Oversight Board --- since it specifically governs financial audit quality, but precedents for government-authorized but private standard-setting bodies exist across critical industries. FINRA (in securities), NERC (for the US electrical grid), and other institutions provide a wealth of lessons on how government and private entities can collaborate on standard-setting while maintaining appropriate accountability \autocite{rudder_public_2016}.

Policy analysis on the appropriate design for a PCAOB-for-AI should identify hard-to-game audit quality indicators and effective inspection methods (e.g., re-performance of an audit by a separate auditor given equivalent access), and more rigorously specify the AALs that we provided preliminary sketches of in this paper. 

To reduce capture risk by a particular government, the AI auditing ecosystem should be designed to be legitimate in multiple jurisdictions. The oversight body should have diverse funding sources as well as transparent rules for conflicts of interest. The body should aim for its standards to be globally credible and endorsed by multiple governments. This could be accomplished by varying standards depending on the jurisdiction (see \cref{ssec:reaching_full}), and giving multiple governments the power to appoint and remove board members of the body.

To reduce capture risk from the AI industry or the auditing industry itself, a PCAOB-for-AI should have significant representation from outside both sectors on its board.

\subsection{Growing the ecosystem} \label{ssec:growing}

Growing the ecosystem requires pulling together expertise from a wide range of disciplines as well as ensuring that auditors do not face undue legal risks when doing their work.

Effective auditing requires auditors with a broad range of expertise across disciplines. Providing that expertise may require utilizing large firms alongside specialized AI assessment organizations and domain experts (e.g., in alignment, bioweapons, or cybersecurity). These multi-organization teams face coordination challenges: preventing accountability dilution \autocite{nissenbaum_accountability_1996, kroll_accountable_2015, cooper_accountability_2022, lee_talkin_2024}, ensuring consistent quality and clear communication across different disciplinary and organizational boundaries, and managing conflicts of interest.

Addressing talent bottlenecks also requires investment in human capital through on-the-job growth in multi-organizational teams and formal training programs. Accreditation standards --- formal, standardized credentials indicating that auditors possess certain knowledge and skills and are familiar with certain professional standards --- can establish and then progressively raise the bar for competency and incentivize skill investment across the ecosystem.

\begin{infobox}
\textbf{Recommendation 3:} The AI evaluation ecosystem should establish a Frontier AI Auditor Accreditation Program with tiered certifications and specialty endorsements, as well as meaningful accountability mechanisms.    
\end{infobox}

The foundational tier of this accreditation would establish baseline competencies in AI systems, audit methodology, and ethics. Specialty endorsements might certify expertise in capability evaluation, alignment, control, information security, cybersecurity, or biosecurity. Organizations could be accredited based on credentialed staff and quality management systems. Nonaccredited individuals should still participate in audits with circumscribed roles (e.g., technical analysis); this allows the ecosystem to develop capabilities and train talent while creating additional accountability for auditors. Organization-level accreditation could be required for certain contexts, such as government procurement.

A particularly promising application of this accreditation program would be providing supplemental training to academic researchers (including graduate students, professors, and post-docs) who have many of the technical skills needed to conduct audits, and who could benefit from additional income as well as relevant work experience during their studies. While this raises a range of implementation questions (e.g., the revolving-door question discussed previously), the large talent pool in academia is impossible to ignore when thinking about quickly growing a skilled auditing ecosystem. Without active measures to tap this pool of talent, it might be difficult to build a sufficiently large pool of disinterested auditors with sufficient speed.\looseness=-1

In addition to actively growing the supply of qualified auditors, there also needs to be attention to preventing roadblocks on a scalable ecosystem. One potential roadblock is liability faced by auditors. Auditors --- who are typically far less capitalized than either developers or insurers --- may face disproportionate liability relative to their fees and their actual contribution, given the massive scale of the AI market. This may cause them to rationally decline engagements involving the highest-risk systems, precisely where independent assessment is most valuable. For example, an auditor might fear a lawsuit from a company contesting good-faith audit results that caused a backlash to that company's products or a fall in their share price.\looseness=-1

There could therefore be cases where regulators should provide liability protections to auditors, such as through a legal ``\hyperlink{gls:safeharbor}{safe harbor}'' \autocite{longpre_safe_2024, kumar_ignore_2024, kumar_legal_2020, xiao_when_2025}, though getting these right requires care.

Terms of service and enforcement strategies used by AI companies to deter model misuse can inadvertently disincentivize good-faith safety evaluations, causing researchers to fear that conducting such research or releasing their findings will result in account suspensions or legal reprisal. \autocite{kumar_legal_2020, kumar_ignore_2024, xiao_when_2025, longpre_safe_2024}. Safe harbor provisions can address this chilling effect while maintaining accountability for genuinely harmful conduct. In the case of cybersecurity, explicit corporate policies to promote responsible third-party disclosure of vulnerabilities is a common practice, and similar norms are needed in AI.

Again, we urge caution regarding such changes, but see value in a carefully scoped safe harbor that avoids creating a liability gap in which neither the developer nor the auditor is liable for harms within the scope of the audit.

\begin{infobox}
\textbf{Recommendation 4:} Policymakers and developers should implement targeted safe harbors that protect good-faith safety research and auditing while avoiding a liability gap, and that are conditional on auditor compliance with established best practices.
\end{infobox}

Such safe harbors should be designed with several principles in mind:
\begin{itemize}[leftmargin=0.25in]
    \item They should protect good-faith testing conducted under disclosed rules of engagement, drawing on established norms from cybersecurity vulnerability disclosure. Best practices from a body such as the PCAOB-for-AI discussed above could be referenced in the design of a safe harbor. 
    
    \item They should be conditional on researchers and auditors following responsible disclosure practices, refraining from exfiltration or misuse of sensitive information, and avoiding conduct that would independently violate applicable laws. 
    
    \item They should extend to both legal liability (providing protection against causes of action that might otherwise apply to authorized safety testing) and technical enforcement (protecting researchers from account suspensions or access revocation for legitimate assessment activities consistent with best practices). 
    
    \item They should not provide blanket immunity but rather require that protected parties demonstrate adherence to specified conditions --- determinations that can be made ex post in the event of a dispute.
\end{itemize}

Developers can unilaterally approximate these safe harbors through contractual commitments before any legislation is passed: as with coordinated vulnerability discovery policies in information security, frontier AI companies can provide explicit testing permissions on their website and designated disclosure channels that span a wide spectrum of harms, and can provide non-retaliation commitments for researchers who identify vulnerabilities. Such commitments would help ensure that auditors and independent researchers can conduct rigorous assessments without fearing that their findings --- particularly critical findings --- will expose them to retaliation.

\subsection{Accelerating adoption} \label{ssec:reaching_full}

Achieving the full promise of frontier AI auditing requires industry-wide adoption (\cref{sec:motivations}). This requires both domestic adoption within countries such as the US --- ensuring all frontier AI developers, not merely the most safety-conscious, submit to rigorous assessment --- and international adoption, including in countries where independent assessment is currently more sparse (the most notable example of a country producing a significant amount of frontier AI systems while lagging on third-party assessment is China). Weak links in the industry can cause incidents that reflect negatively on the industry as a whole.

Markets have a key role to play in driving auditing adoption, as discussed in \cref{ssec:enabling}, but markets alone face limitations. Some frontier AI developers do not prioritize deployment to enterprise customers or even external deployment at all (e.g., Safe Superintelligence Inc.), and some may see a short-term market advantage in cutting corners. In a nascent industry where profitability may be years away, it is unclear how much market discipline alone can address safety and security absent binding regulatory requirements.

We recommend two steps, one focused on ensuring that insurance can play the same constructive role it historically has played in new technology adoption \autocite{kvist_underwriting_2025} and the other on creating a regulatory requirement for auditing.

First, there is growing concern that the scale of risks from current and future frontier AI systems go well beyond those that were originally envisioned when writing most companies' insurance policies. This has caused some insurers to begin moving toward excluding AI-related harms from their policies and others to provide AI-focused insurance \autocite{yang_insurance_2025, noauthor_insurers_2025}. However, exclusions are often bottlenecked by government approval, and insurers with particularly strong market power may be tempted to postpone difficult decisions longer than is ideal for the market as a whole. A similar development occurred with cybersecurity, and ultimately the US government helped drive clarity. 

Ambiguity on AI risk coverage could forestall the constructive role that insurance could play in driving AI risk mitigation, as it has in many other industries \autocite{trout_when_2025}. This role, however, is most plausibly concentrated on insurable operational risks associated with the deployment and use of AI systems rather than on fundamentally uncertain risks arising from frontier model development itself. Greater clarity on what is and isn't covered under a given policy would accelerate the development of specialized insurance options for AI-related risks where they are appropriate, incentivize general insurers to pay closer attention to AI-related risks, or both, depending on the nature of the determinations made (see \cref{apx:additional_motivations} for further discussion).

\begin{infobox}
\textbf{Recommendation 5:} National governments should quickly resolve outstanding and near-term requests from insurers regarding exclusions one way or the other, and in government procurement contexts, they should specify that frontier AI companies need explicit coverage of AI-related risks (whether through a specialized or general policy).    
\end{infobox}

If audits do in fact provide valuable information on AI safety and security and there are existing uninsured risks, greater clarity on ownership of AI-related risk will help create market pressure throughout the supply chain: downstream businesses gain financial reason to prefer audited models, which in turn incentivizes frontier AI developers to pursue auditing. 

Second, governments are unlikely to sit idly while private governance solutions form; the political salience of AI is rapidly increasing \autocite{kennedy_how_2025}, and for the reasons discussed above and in both \cref{ssec:improving_outcomes} and \cref{ssec:enabling}, it would be undesirable for them to do so. The question is what exactly this regulatory involvement should entail.

Regulation can codify and universalize practices adopted by market leaders, while market dynamics can reveal more efficient ways to achieve regulatory goals \autocite{hadfield_regulatory_2023}. Direct mandates for frontier AI auditing hedge against a scenario in which market forces are not sufficiently strong to drive safety and security alone (e.g., due to unpriced negative externalities), though they also could overstep and burden innovators without a commensurate safety and security benefit.

Procurement policies represent a potential middle-ground between industry-wide mandates, on the one hand, and government inaction, on the other. For example, governments could require frontier AI auditing --- at a specified level of assurance, for a given scope of risks, and by auditors with certain accreditations --- before purchasing AI services in high-stakes sectors such as health and defense. This could accelerate the growth of the auditing market as a whole and accelerate the pace of safety and security improvement in the frontier AI sector.

\begin{infobox}
\textbf{Recommendation 6:} Policymakers should incorporate frontier AI auditing requirements into procurement processes, with particularly strong requirements for systems that will be deployed in high-stakes domains such as health and defense.
\end{infobox}

However, government procurement policies, no matter how strong, are not likely to fully address risks from frontier AI. Not all frontier AI developers sell services to government agencies, and it is uncertain how much other factors will fill the gap in demand (e.g., insurance, investor due diligence, demand from enterprise customers). Still, procurement policies are a starting point for improving safety and security outcomes involving government use of AI, and could produce key findings that inform further steps in other contexts.

Beyond procurement, a more comprehensive approach would be to require auditing for frontier AI systems or companies meeting a certain threshold (or, require auditing at different assurance levels corresponding to different thresholds). This could be layered onto existing threshold-based transparency requirements and whistleblower protections, using the same threshold or thresholds. A key question in formulating such a statutory requirement is: what safety and security standards should companies be audited against (besides their own policies and existing regulations)? Notably, there are few substantive standards in existing regulations --- companies are generally required simply to have some kind of safety and security policies, which may or may not need to be detailed or meet certain substantive criteria, depending on the jurisdiction. While we don't recommend a specific path forward here, we note that some have proposed that governments articulate a risk level to stay below in certain domains, and have private sector institutions identify efficient means of achieving those ends~\autocite{hadfield_regulatory_2023, fathom_independent_nodate}. Auditing requirements could then either be directly required for certain classes of AI systems and companies, or strongly incentivized as part of a larger regulatory strategy for AI.\footnote{One policy mechanism that has been proposed to improve legal predictability and risk mitigation is statutory liability shields for frontier AI developers who take certain actions (including, e.g., submitting to auditing). There are various arguments for and against such proposals \autocite{weil_tort_2024, weil_instrument_2025, ball_framework_2025}, but here we simply note that any such shields should be very carefully scoped to avoid undercutting positive incentives for mitigating avoidable risks. Insofar as there is a case for such shields, it is strongest for knowing misuse by users in cases where such misuse is not plausibly preventable even after the application of best practices. Giving companies ``something for nothing'' (e.g., a broad liability shield in exchange for submitting to shallow audits by unqualified auditors) would make things worse from almost all stakeholders’ perspectives.}

\subsection{Achieving technical readiness for high AALs} \label{ssec:achieving}

Three interconnected challenges must be addressed to make AAL-3 and AAL-4 technically feasible, cost-effective, and sufficiently protective of companies' sensitive information: completeness, continuous monitoring, and the transparency-security trade-off.

At high assurance levels, auditors need not only analyze information presented to them, but gain confidence that they are receiving complete information --- i.e., that there aren't material omissions that would change the audit conclusions. Gaining complete-enough information in order to rule out most possible errors and fraud is very difficult. Small, simple changes (e.g., disabling a safety feature) and difficult-to-detect actions (e.g., taking model weights out of a datacenter on a USB stick) can have big consequences. This \hyperlink{gls:completeness}{completeness problem} means that it is inherently difficult --- if not impossible --- to have high or very high confidence in a company's risk posture based on analysis of a single AI system in isolation.\looseness=-1

These small-footprint, big-consequence changes and actions could be mistakes on the part of the frontier AI company or intentional efforts to deceive; in either case, though particularly for deception, proving their absence will be challenging. While it seems that we have at least a few months before AI systems are capable of very sophisticated scheming and planning to undermine audit results \autocite{metr_details_2025}, eventually auditors will need to plan for such possibilities, and will need proportionally stronger auditing techniques that can rule out sophisticated deception by human or machine.

Achieving completeness likely involves a combination of ``\hyperlink{gls:lowtech}{low-tech}'' mechanisms such as whistleblower bounties, so those with knowledge of material omissions or deceptions have an incentive to come forward \autocite{grunewald_whistleblower_2025}, and ``\hyperlink{gls:hightech}{high-tech}'' mechanisms such as compute accounting and ``proof-of-training'' techniques, so that the amount of unaccounted-for compute and models of unknown provenance can be carefully circumscribed.\footnote{This dichotomy is intended to convey the basic idea of multiple options for achieving the same goals, though we gloss over many details. For more granular frameworks, and a more detailed discussion of several themes in this subsection, see \autocite{harack_verification_2025, baker_verifying_2025}.\looseness=-1}

A second challenge is the \hyperlink{gls:transparencysecurity}{transparency-security trade-off}, a concept developed in the literature on arms control \autocite{coe_why_2020}. The trade-off is that the very same information that third parties desire in order to have confidence that the audited organization is abiding by their commitments is also often the same type of information that is very sensitive itself or mixed up with sensitive information in complex ways that are difficult to share externally. In cases where countries have needed to cooperate via arms control treaties, they have often had to develop sophisticated technologies to navigate this trade-off \autocite{toivanen_regulating_2017}. 

Again, there are low-tech and high-tech ways of addressing this challenge. A low-tech path is to rely on human institutions such as rigorous background checks and personnel screening in order to vet the auditors before giving them access to the most sensitive information, and dividing up the sensitive information among different auditors. A high-tech path is to develop or apply technologies that are specifically intended to address trade-offs like this, such as cryptography, or technologies that are general-purpose in nature but can be applied in a way that alleviates the trade-off (e.g., applying AI itself to summarize or paraphrase information in a way that removes sensitive details \autocite{anthropic_clio_2024, trask_beyond_2024}). Another approach is to use \hyperlink{gls:flexhegs}{FlexHEGs} or other hardware-enabled governance mechanisms that can keep track of some properties but not others. In each case, use of formal verification to mathematically prove certain properties of the software used, as well as open-source hardware design, could improve confidence on the part of both auditors and auditees in the technology used. Lastly, as mentioned in \cref{sec:vision}, early pilots have shown the ability to conduct evaluations of (non-production) language models while assuring all parties that excess information will not be revealed.\looseness=-1

A final challenge is \hyperlink{gls:continuousmonitoring}{continuous monitoring}. Even if completeness is established at a single point in time, high AALs require ongoing assurance. Again, small changes can have big consequences, and these can happen nearly instantaneously for some aspects of AI safety and security. AAL-3 and AAL-4 require ongoing monitoring that produces change-detection signals and triggers re-examination when prior validity conditions no longer hold.

Elaborating on one promising theme, cryptographic certificates provide a promising way forward. When technically feasible, developing and deploying AI systems within cryptographic protocols could enable AAL-3 and AAL-4 while guaranteeing privacy protections for all stakeholders. Specifically, \hyperlink{gls:zkp}{zero-knowledge proofs (ZKPs)} allow an auditee to prove statements about their private data or private system without revealing any further information:
ZKPs provide a certificate that could be made public, avoiding the need for an independent auditor or trusted third party. 
ZKPs provide secure ``white-box'' access to a set of pre-specified data or operations while protecting the intellectual property or sensitive data of an auditee; an appropriate cryptographic commitment (roughly, a tamper-evident digital fingerprint of the system that is fixed during development and deployment) by an auditee to their system enables continuous monitoring or change-detection signals that cannot be surreptitiously altered.
Current work is ongoing for ZKP-based certificates for properties such as privacy, fairness, and uncertainty calibration. More research is needed to achieve sufficient computational efficiency to enable practical use at scale.

Low-tech pathways for addressing continuous monitoring include embedded auditors who have a similar level of access to information as normal company employees do, or even enriched access relative to normal employees (this approach is used for regulation of systemically important financial institutions as well as nuclear power plant safety in the US). High-tech pathways include data diodes \autocite{wikipedia_unidirectional_nodate} that could continuously emit a small amount of information, negotiated in advance and perhaps formally verified, \textit{without} also being capable of sending malicious commands back into the server. Through such means, auditors could have confidence that --- for example --- a certain datacenter is still being used for running an audited model rather than training of unaudited models, without the company putting any additional confidential information at risk.\looseness=-1

\begin{infobox}
\textbf{Recommendation 7:} Philanthropists, governments, and frontier AI companies should invest in an ambitious ``Auditability R\&D and Pilots'' portfolio aimed at making AAL-3 and AAL-4 technically feasible and cost-effective.    
\end{infobox}

Priority areas include:
\begin{itemize}[leftmargin=0.25in]
    \item \textbf{Confidential evaluations at scale:} As mentioned in Section 5, the PySyft framework has been used --- in combination with secure enclaves --- to conduct mutually confidential evaluations of a small language model. But we need significant efficiency improvements for techniques like this to be viable for frontier AI models and systems. 

    \item \textbf{Proof-of-training and proof-of-learning:} Confidence in how large amounts of computing power were used can help rule out (or put limits on the possible size of) unaudited systems produced by a given set of computing hardware. For evaluations, proof-of-training and proof-of-learning, ZKPs are a promising approach.
    
    \item \textbf{Change-detection infrastructure:} Monitoring technologies (including open-source data diodes, FlexHEGs, and other hardware-enabled governance mechanisms) can produce auditable signals when systems change in ways that could invalidate prior assessments, while ensuring that additional information will not be transmitted.
    
    \item \textbf{Adversarial testing of verification mechanisms:} Independent teams attempting to spoof attestations, bypass monitoring, or create shadow systems can ensure that auditability infrastructure actually works.
    
    \item \textbf{Model and system fingerprinting techniques:} Emerging techniques may make it possible to detect significant changes to model weights through black-box interfaces, but their limits need to be more rigorously understood \autocite{finlayson_every_2025}. Additionally, there has been little research on whether other types of changes could be reliably detected (e.g., changes to system prompts, new inference optimizations or classifiers for inputs and outputs).
    
    \item \textbf{Formal verification fundamentals and applications:} Research advancing AI systems' ability to assist with engineering system design, formal specifications, and theorem proving can accelerate development of high assurance audit infrastructure. In addition to improving general techniques, specific, immediately usable applications of formal methods should also be pursued (e.g., to give confidence in an application of privacy-preserving AI for analysis of sensitive documents).
\end{itemize}

In parallel with building strong technical foundations, there is a need to learn from experience. Given substantial uncertainty about how quickly AI and its risks will evolve, pilots for AAL-3 and later AAL-4 should begin with urgency.

\begin{infobox}
\textbf{Recommendation 8:} Companies closest to the state-of-the-art should work with auditors, researchers, governments, and other stakeholders to conduct early pilots of AAL-3 and later AAL-4 auditing in order to accelerate the maturity of relevant technologies and processes.
\end{infobox}

These pilots should:
\begin{itemize}[leftmargin=0.25in]
    \item Test a range of ``low-tech'' procedural approaches that can support AAL-3 using methods available today, although these will likely not be scalable in all respects (e.g., requiring intensive vetting of the personnel involved).
    
    \item Trial emerging ``high-tech'' mechanisms (such as proof-of-training, on-server use of privacy-preserving AI methods, and change-detection for internal deployment) in realistic settings to identify gaps between current capabilities and frontier-scale requirements.
    
    \item Document what works and what doesn't, contributing to public knowledge about minimum access requirements, trust assumptions, and practical obstacles.

\end{itemize}

Since the companies closest to the state-of-the-art will create AI systems that pose the greatest risks before others, it is appropriate that they should be first movers in this area, and show how their technology can be trusted with confidence by skeptical parties. Early experimentation will ensure that AAL-3 and AAL-4 are ready when they are needed most, including in the most difficult cases, such as US--China cooperation on baseline safety and security norms \autocite{brundage_unbridled_2025}.

\newpage 
\section{Conclusion}\label{sec:conclusion}

Today, no mechanism exists to confidently confirm that AI companies' safety and security claims are accurate or that their practices meet relevant standards, forcing a difficult choice between taking all companies at their word and relying on a combination of unreliable signals (third-party assessments based on very limited information, the apparent trustworthiness of senior leadership in the company, etc.). Frontier AI auditing provides an alternative: rigorous, independent scrutiny of technical systems and organizational practices by qualified third parties.

We presented a vision for what effective frontier AI auditing requires: comprehensive scope covering intentional misuse, unintended system behavior, information security, and emergent social phenomena; an organizational perspective that assesses companies holistically rather than focusing narrowly on specific models; calibrated assurance levels that clearly communicate warranted confidence; deep access to non-public information combined with rigorous security measures; continuous, rather than one-off, verification that updates automatically as systems change; auditor independence enforced by disclosure, industry standards, and oversight of auditors; rigorous, traceable, and adaptive audit processes; and clear communication of audit results. These principles draw on more established domains where societies have repeatedly built independent assurance mechanisms for high-stakes activities.

Substantial challenges remain: ensuring high quality standards, growing the ecosystem, accelerating adoption, and achieving technical readiness for the higher assurance levels. But these challenges are neither unprecedented nor insurmountable.

This paper has several important limitations:
\begin{itemize}[leftmargin=0.25in]
    \item \textbf{We focus primarily on frontier AI developers and closed-weight models}. The auditing challenges for open-weight models, fine-tuning providers, and downstream deployers differ in ways we do not directly address. 
    
    \item \textbf{We only make interim, near-term recommendations on the appropriate levels of assurance} (AAL-1 for frontier AI as a whole, and AAL-2 for the leading subset thereof). Judgments about how to proceed after that will involve various complex considerations that will be informed by the further research and pilots we propose, and we propose institutions for helping to make and implement such decisions. 
    
    \item \textbf{Many of our recommendations implicitly depend on or at least strongly benefit from institutions other than auditing} --- e.g., robust whistleblower protections at AI companies, detailed transparency requirements for frontier AI companies (so that they are making significant claims that merit verification in the first place), and clear allocation of liability for AI-related harms. These are important but can be pursued in parallel to frontier AI auditing.

    \item \textbf{Much of our analysis implicitly assumes a certain context (namely, developed, largely Western countries)}, and scaling frontier AI auditing to countries such as China raises various challenges. We discussed technical challenges in achieving high degrees of assurance, but additional cultural and political challenges were not addressed.
\end{itemize}

As frontier AI systems grow more capable --- possibly at an accelerating rate --- the cost of getting safety and security wrong rises sharply. The time to invest in frontier AI auditing is today.

{
\small
\printbibliography
}

\newpage
\appendix \label{sec:appendices}
\crefalias{section}{appsec}
\addappheadtotoc
\addtocontents{toc}{\protect\setcounter{tocdepth}{1}}

\appsection{Glossary}\label{apx:glossary}

\hypertarget{gls:abserror}{\textbf{Abstraction error}}

Forming the wrong conclusion by treating a partial or simplified unit of analysis (e.g., evaluating a specific component in isolation) as if it were sufficient to assess overall system and organizational risk.

\hypertarget{gls:accidents}{\textbf{Accidents}}

Harms arising from AI systems behaving in unintended ways \autocite{zwetsloot_thinking_2023}. 

\hypertarget{gls:ai}{\textbf{AI (artificial intelligence)}}

Digital systems that are capable of performing tasks commonly thought to require intelligence, with these tasks typically learned via data and/or experience \autocite{brundage_toward_2020}.

\hypertarget{gls:levels}{\textbf{AI Assurance Levels (AALs)}}

A proposed standardized vocabulary for expressing how much weight to place on audit findings --- i.e., how confident a skeptical third party can be in an auditor’s conclusions given the engagement’s access, evidence, and methods.

\hypertarget{gls:tracker}{\textbf{AI assurance tracker}}

A proposed public platform (maintained by the independent oversight body proposed within \cref{ssec:achieving}) that would show, in a standardized format, items such as each frontier AI company’s stated policies, applicable regulations, incident reports, lead auditor, and recent safety/security publications --- updated as relevant changes occur.

\hypertarget{gls:alignment}{\textbf{Alignment}}

Effort to ensure that “AI systems behave in line with human intentions and values” \autocite{ji_ai_2024}.

\hypertarget{gls:assessment}{\textbf{Assessment}}

Any effort to determine the properties of an AI system or an AI company/developer, whether based on public or non‑public information and whether rigorous or informal. 

\hypertarget{gls:level}{\textbf{Assurance level}}

The degree of confidence that can reasonably be placed in an audit’s conclusions, determined in part by the audit’s scope, access, and methods.

\hypertarget{gls:audit}{\textbf{Audit}}

A systematic, evidence-based process in which a qualified party examines an organization’s activities, records, technologies, and claims to provide assurance that stated information is accurate and/or that applicable standards are being met.

\hypertarget{gls:report}{\textbf{Audit report}}

A document produced by an independent auditor that communicates the results of a frontier AI audit in a way that external stakeholders can rely on. It should include the audit’s scope, assurance level (AAL), auditor’s conclusions, the reasoning behind those conclusions, conditions under which the conclusion is valid, and recommendations for remediation. A full, unredacted version may be shared with a company’s board/executives, with a redacted or summarized version released for external stakeholders.

\hypertarget{gls:blackbox}{\textbf{Black-box access}}

A level of model access in which evaluators can query the model through an external interface (e.g., API) and observe its outputs, but cannot inspect internal components such as weights, activations, or intermediate computations. Black-box access is typical of most current third-party evaluations \autocite{casper_black-box_2024}.

\hypertarget{gls:cot}{\textbf{Chain-of-thought}}

Intermediate reasoning steps generated by a model when solving problems, either explicitly through prompting techniques or captured through access to internal model traces. Chain-of-thought can reveal capabilities and risks not apparent from final outputs alone.

\hypertarget{gls:cleanroom}{\textbf{Clean rooms}}

Physically or logically isolated digital environments where sensitive information can be analyzed with reduced risk of external exposure.

\hypertarget{gls:cleanteam}{\textbf{Clean teams}}

Personnel who have access to confidential information but are insulated from competitive decision-making within their organization \autocite{hewitt_how_2010}.

\hypertarget{gls:closedweight}{\textbf{Closed-weight model}}

AI models that can only be accessed through API or similar means, to ensure their code and weights are not directly accessible.

\hypertarget{gls:compart}{\textbf{Compartmentalization (of an audit)}}

A structuring approach where different auditors learn about and assess different aspects of a company’s operations \autocite{bartock_hardware-enabled_2022}.

\hypertarget{gls:completeness}{\textbf{Completeness problem}}

At high assurance levels, the core verification challenge of confirming that all relevant systems, training runs, and governance processes have been surfaced --- i.e., nothing material has been omitted.

\hypertarget{gls:consortium}{\textbf{Consortium (of auditors)}}

Two or more organizations jointly conducting audits, with no organization serving as the lead auditor. 

\hypertarget{gls:continuousmonitoring}{\textbf{Continuous monitoring}}

Ongoing monitoring of a frontier AI system (or other aspects of a frontier AI company) in order to detect changes that might invalidate previous audit conclusions.

\hypertarget{gls:defense}{\textbf{Defense in depth}}

The idea that assessment is needed at multiple different lifecycle stages.

\hypertarget{gls:emergent}{\textbf{Emergent social phenomena}}

Risks that arise from interaction between humans and AI systems and do not fit neatly into ``misuse'' or ``unintended behavior,'' but can nevertheless cause significant harm if left unaddressed. Examples include addiction to or emotional dependence on AI systems, AI-induced or AI-enabled psychosis, and facilitation of self-harm.

\hypertarget{gls:evaluation}{\textbf{Evaluation}}

Any activity that measures, characterizes, or analyzes properties of AI models or systems and the organizations operating them. 

\hypertarget{gls:event}{\textbf{Event-triggered reviews}}

Audits conducted following key points such as major training runs, model releases, significant incidents, or notable novel integrations.

\hypertarget{gls:expectations}{\textbf{Expectations gap}}

The gap between what the public believes audits guarantee (e.g., the absence of fraud) and the more limited mandate auditors actually operate under.

\hypertarget{gls:flexhegs}{\textbf{FlexHEGs (Flexible Hardware-Enabled Guarantees)}}

A technical direction for on-chip, privacy-preserving verification of claims (for example, verifying that training runs have not exceeded certain compute thresholds without revealing proprietary details).

\hypertarget{gls:flowdown}{\textbf{Flow-down agreements}}

A contract provision under which certain obligations (such as those related to confidentiality) “flow down” from a lead contractor to subcontractors. In this context, we are referring to domain specialists hired by a lead auditor. 

\hypertarget{gls:frontierai}{\textbf{Frontier AI}}

General-purpose AI models and systems whose performance is no more than a year behind the state-of-the-art on a broad suite of general capability benchmarks.

\hypertarget{gls:frontieraiauditing}{\textbf{Frontier AI auditing}}

Rigorous third-party verification of frontier AI developers’ safety and security claims and evaluation of their systems and practices against relevant standards, drawing on deep, secure access to non-public information.

\hypertarget{gls:frontieraidevelopers}{\textbf{Frontier AI developers}}

Companies that train models from scratch themselves or significantly extend their capabilities (e.g., via further training or creation of an agentic scaffold), producing AI systems that qualify as frontier-level per the definition above.

\hypertarget{gls:goodhart}{\textbf{Goodhart’s Law}}

The phenomenon that when specific metrics become targets for compliance, actors optimize for those targets rather than the underlying outcomes the metrics were meant to capture.

\hypertarget{gls:graybox}{\textbf{Gray-box access}}

A level of model access that provides evaluators with some visibility into internal model components beyond what is available through standard interfaces, such as chain-of-thought outputs, logits, or sampling of internal logs, but without full access to model weights or complete system internals. Gray-box access is intermediate between black-box and white-box access \autocite{casper_black-box_2024}.

\hypertarget{gls:hardwareattest}{\textbf{Hardware attestation}}

A verification mechanism in which hardware components provide cryptographically signed evidence about their identity, configuration, and operating state. 

\hypertarget{gls:hightech}{\textbf{High-tech pathway}}

One pathway toward higher assurance that focuses on developing new technical infrastructure for secure information sharing that reduces how much any single party must be trusted.

\hypertarget{gls:independence}{\textbf{Independence}}

A core principle for audits: results should be trustworthy because auditors are genuinely independent third parties and conflicts of interest are carefully managed.

\hypertarget{gls:infosec}{\textbf{Information security}}

A risk category covering failures of confidentiality or integrity affecting critical AI assets, including theft of model weights, sensitive research, or customer data; risks to user privacy; sabotage of highly capable AI systems; and unauthorized use of compute resources.

\hypertarget{gls:intentionalmisuse}{\textbf{Intentional misuse}}

The use of frontier AI systems by malicious actors to enable or scale harmful activities (examples include cyberattacks; chemical, biological, radiological, or nuclear weapons development; large-scale disinformation; violent and criminal activity; fraud; and generation of CSAM or NCII).

\hypertarget{gls:interrater}{\textbf{Inter-rater reliability}}

The extent to which different auditors converge on the same conclusions given the same evidence.

\hypertarget{gls:livecert}{\textbf{Live certification and deprecation}}

The idea that audit certifications should remain valid only while their underlying assumptions hold, and should automatically downgrade (or be flagged for review) when material changes occur.

\clearpage
\hypertarget{gls:textbf}{\textbf{Logits}}

The raw, unnormalized output values produced by a neural network before they are converted to probabilities (e.g., through a softmax function). Access to logits provides evaluators with more fine-grained information about model behavior and confidence than observing only the final outputs.

\hypertarget{gls:lowtech}{\textbf{Low-tech pathway}}

One pathway toward higher assurance that brings auditors into the organization’s trust boundary using existing legal and physical infrastructure (e.g., corporate devices, clean room arrangements, and confidentiality obligations).

\hypertarget{gls:misuse}{\textbf{Misuse}}

A broad category of risks related to AI systems, specifically those that stem from the use of an AI system in a way that is different from its intended purpose. Misuse may or may not be malicious.

\hypertarget{gls:checkpoints}{\textbf{Model checkpoints}}

Saved snapshots of a model's parameters at specific points during training or fine-tuning. Access to checkpoints enables evaluators to examine how model capabilities and behaviors evolve over the training process.

\hypertarget{gls:open}{\textbf{Open-weight model}}

AI models whose weights are publicly released and can be freely copied or redistributed.

\hypertarget{gls:orgperspective}{\textbf{Organizational perspective}}

The principle that culture, governance, and security matter --- not just specific AI systems.

\hypertarget{gls:pcaob}{\textbf{PCAOB}}

The Public Company Accounting Oversight Board, a US non-profit corporation established by the Sarbanes–Oxley Act of 2002. The PCAOB oversees audits of public companies by setting auditing standards, registering and inspecting audit firms, and disciplining auditors for misconduct. It coordinates with audit regulators in over 50 jurisdictions. In this paper, the PCAOB model is referenced as a potential template for independent oversight of frontier AI auditors.

\hypertarget{gls:reasonable}{\textbf{Reasonable assurance}}

A term of art (from financial auditing) used to indicate a higher degree of confidence in an audit’s conclusions compared to one involving only “limited” assurance. In our framework, this corresponds to AAL-2, and we also consider still higher degrees of assurance beyond this at AAL-3 and AAL-4.

\hypertarget{gls:safeharbor}{\textbf{Safe harbor provisions}}

Conditional protections designed to encourage auditing and disclosure --- modeled in the paper on regimes where entities that discover violations through systematic auditing, disclose promptly, and correct issues can receive reduced penalties or immunity under specified conditions (with exclusions for certain serious or bad-faith cases). 

\hypertarget{gls:safeguard}{\textbf{Safeguard}}

A technical measure or process designed to prevent AI systems from causing harm.

\hypertarget{gls:safety}{\textbf{Safety}}

The functioning of AI systems in a way that avoids causing significant harm, ranging from accident risks (unintended harmful behavior due to factors such as misspecified goals, operator error, or system bias) to misuse risks (harms caused intentionally by the deployer or user of an AI system).

\hypertarget{gls:safetycase}{\textbf{Safety case}}

Structured arguments supported by evidence that justify the safety of a system \autocite{uk_ministry_of_defence_defence_2007}.

\hypertarget{gls:security}{\textbf{Security}}

The protection of the AI system itself as well as surrounding infrastructure, intellectual property, and user data against unauthorized access, exfiltration, manipulation, or disruption. In the frontier AI context this includes, for example, protecting model weights and other sensitive artifacts from theft, as well as preventing adversaries from hijacking an AI system to cause harm.

\hypertarget{gls:structural}{\textbf{Structural risks}}

Risks emerging from how AI systems reshape systems, incentives, and environments in which they are deployed \autocite{zwetsloot_thinking_2023}. We intend for “emergent social phenomena” to be distinct from this category in the sense that emergent social phenomena, while distributed across society, are nevertheless directly connected to the deployment of specific, identifiable AI systems.

\hypertarget{gls:systemcards}{\textbf{System cards}}

Descriptions of the properties and risk profile of AI systems. Introduced here and commonly used as a label for reports produced by frontier AI companies about their latest model or system releases. ``Model card,'' an earlier term, is often used for such documents as well, and while technically this denotes a model rather than system level of analysis, in practice, model cards often discuss system-level components and vice versa.

\hypertarget{gls:transparencysecurity}{\textbf{Transparency-security trade-off}}

The tension between making information open (which enables accountability, trust, and collaboration) and keeping it hidden (which protects against those who would exploit that knowledge to cause harm).

\hypertarget{gls:vericationassessments}{\textbf{Treaty-grade verification}}

Very high assurance in which one can have confidence in audit conclusions even assuming the audited party will take every available opportunity to cut corners and deceive.

\hypertarget{unintended}{\textbf{Unintended system behavior}}

AI systems behaving in ways unintended or unsafe from the perspective of developers and users that are serious enough to risk large-scale harm, including accidents in high-stakes deployment contexts caused by misaligned behavior (e.g., reward hacking), capability failures, biased outputs, or behaviors that circumvent human intent and effective human oversight.

\hypertarget{gls:validityperiod}{\textbf{Validity period}}

An explicit time period during which an audit finding or conclusion is treated as valid, with the expectation that conclusions should be deprecated or revisited when assumptions no longer hold due to system change.

\hypertarget{gls:verification}{\textbf{Verification}}

The activity of confirming whether a specific claim, commitment, or property (e.g., an evaluation result, a training compute figure, or a claim about mitigation effectiveness) is true. 

\hypertarget{gls:whitebox}{\textbf{White-box access}}

The most comprehensive level of model access, granting evaluators full visibility into model weights, architecture, training data, and all internal components \autocite{casper_black-box_2024}. White-box access enables the deepest forms of technical analysis but requires the strongest intellectual property protections and is typically reserved for higher assurance levels.

\hypertarget{gls:zkp}{\textbf{Zero-knowledge proofs (ZKPs)}}

A cryptographic protocol that enables one party (the prover) to demonstrate to another party (the verifier) that a statement is true without revealing any information beyond the statement's validity itself.

\newpage 
\appsection{Additional motivations for frontier AI auditing} \label{apx:additional_motivations}

\textbf{Enabling risk price discovery through insurance}

Quantifying the actual risks of frontier AI systems remains a fundamental challenge. Expert judgment is valuable but struggles to aggregate dispersed information into actionable signals. Insurance markets offer a complementary mechanism: insurers have strong financial incentives to price risk accurately. They can do this by translating private assessments of safety practices, loss histories, and exposure pathways into premiums, coverage terms, and exclusions that function as observable signals of risk.

These markets cannot function without reliable information. Third-party audits give insurers the verified, standardized data they need to differentiate risk profiles across companies and systems. Without this, adverse selection prevents meaningful coverage or pricing.\footnote{This mirrors early cyber insurance markets, which struggled until standardized security assessments provided sufficient data.}

The societal value extends well beyond risk transfer. Insurance pricing is one of the few mechanisms that can translate diffuse uncertainty about AI risk into a single, continuously updating number. This gives policymakers an independent measure of risk to inform regulation. It gives the public a legible signal of whether safety is improving or deteriorating over time. And it creates a common reference point around which developers, regulators, insurers, and civil society can coordinate. A credible auditing ecosystem is the foundation on which such a market can be built.

\textbf{Maintaining international stability}

The development of frontier AI has profound implications for international security and stability. Intense competition between nations can create a “race to the bottom” dynamic, where actors may feel pressured to cut corners on safety to accelerate development. A lack of verified information about competitors’ AI capabilities can fuel destabilizing arms-race dynamics.

Third-party auditing offers a crucial mechanism for de-escalation. By providing a trusted, neutral means to verify that all parties are adhering to shared safety commitments, audits can build confidence and reduce the incentive for competitive risk-taking. Competing governments are unlikely to grant each other deep access for meaningful verification. Independent auditors can serve as trusted intermediaries to confirm compliance with agreed-upon ground rules.\footnote{It is not inconceivable that direct bidirectional access could be granted to some AI systems or components thereof --- indeed, there is precedent for this in arms control contexts (e.g., mutual inspection of missile facilities). We are making the more conservative assumption that this is off the table, in order to be prepared for a wider range of possible scenarios.}

Looking ahead, a well-developed international auditing regime can serve as a foundational platform for future treaties governing AI. Just as arms control agreements rely on inspection mechanisms to verify compliance, international agreements on AI safety will require a credible verification system. A shared audit framework creates a common vocabulary and benchmarks for risk, facilitating international cooperation and providing the tools needed to ensure that commitments are being met. This extends to security concerns like counter-proliferation, where audits of a developer’s operational security and cybersecurity provide assurance that powerful models will not be stolen by hostile actors.

\textbf{Ensuring accountability for risk creation}

As AI's influence over society expands, the public requires strong evidence that frontier models do not place them at undue risk. Public skepticism toward corporate “safety‑washing” is rising, creating demand for credible, external validation of safety claims --- demand that auditing can directly address. Recent polling suggests that the American public supports \autocite{vigers_americans_2025} independent expert evaluation of AI systems over self-assessment, and even prefers it to direct government testing. Research across multiple industries consistently shows that auditing improves the perceived credibility of organizational claims compared with self-reporting alone \autocite{power_auditbook_1997, paracini_fear_2014,Cheng2015-mg,Holt2019-or}. This is intuitive: organizations naturally seek to present themselves positively, whereas third parties will --- assuming conflicts of interest are well‑managed --- have better incentives to provide accurate risk assessments, as this capacity represents the entirety of their institutions’ identities and reputations.

As policymakers develop and enact AI governance frameworks, they require reliable mechanisms to verify compliance. Currently, compliance with voluntary commitments is uneven \autocite{wang_ai_2025}. As risks increase, companies will have an increasing need to comply with safety and security standards, alongside government regulation. Governments typically do not conduct audits directly, although they have a key role to play in standard-setting and enforcement after a violation has emerged. Frontier AI audits could demonstrate that companies are complying with laws, allowing regulators to hold companies accountable. Auditing can facilitate documentation and subsequent accurate and proportional assignment of liability in cases of safety and security incidents post-deployment.\footnote{While we generally emphasize ways incentives for responsible behavior can be increased where they might be lacking by default, the converse motivation is also relevant, namely avoiding excessive blame directed to a company that in fact behaved responsibly. Favorable evidence from an audit could help exculpate a company in a lawsuit or regulatory context by establishing compliance with relevant best practices, and suggest that blame can be found elsewhere, such as on the user of a product.}

There are also structural reasons to prefer that accountability be mediated through third-party auditors, rather than through direct government audits. Distributing authority reduces the concentration of power in any single entity and makes politically-motivated investigations less likely. Private sector auditors can offer specialized technical expertise and higher salaries, and scale more readily than government agencies.\looseness=-1

\newpage
\appsection{Access types (non-exhaustive)}\label{apx:access}
{\footnotesize
\begin{longtable}{|>{\RaggedRight}p{2.5cm}|>{\RaggedRight}p{4cm}|>{\RaggedRight\arraybackslash}p{9cm}|}
\caption{A non-exhaustive taxonomy of information sources that companies may provide access to, across model, system, governance, and operational domains, adapted from \autocite{homewood_third-party_2025}. Public information is also included, as auditors should consider it alongside company-provided sources. The depth of access required will depend on the specific audit engagement and the assurance level sought.}\\ 

\hline
\rowcolor{headergray}
\textbf{Category} & \textbf{Access type} & \textbf{Description} \\ \hline
\endfirsthead

\multicolumn{3}{c}%
{{\bfseries \tablename\ \thetable{} -- continued from previous page}} \\
\hline
\rowcolor{headergray}
\textbf{Category} & \textbf{Access type} & \textbf{Description} \\ \hline
\endhead

\hline
\multicolumn{3}{|r|}{\emph{Continued on next page}}\\
\hline
\endfoot

\hline
\endlastfoot

% \hline
% \endlastfoot

\multirow{12}{=}{System access} 
& Sampling interfaces & Ability to query the model via API, specify sampling parameters, and access output probabilities and logits \\ \cline{2-3}
& Production model variants & Access to deployed model versions with all safety mitigations in place, to assess real-world behavior \\ \cline{2-3}
& Low-mitigation model variants & Access to versions of the model with minimal safety mitigations (e.g., ``helpful-only'' variants) to avoid refusal contamination during capability evaluation \\ \cline{2-3}
& Fine-tuning & Ability to fine-tune models through supervised learning, reinforcement learning, or custom loss functions \\ \cline{2-3}
& Model internals & Access to activations, attention patterns, gradients, embeddings, chain-of-thought traces (when available), and raw outputs \\ \cline{2-3}
& Routing algorithms & For mixtures-of-experts (MoEs) and other multi-model systems, access to routing policies and related algorithms \\ \cline{2-3}
& Model weights & Privacy-preserving access to model weights for interpretability and verification \\ \hline

\multirow{15}{=}{System information} 
& Model specifications & System prompts, architectural details, hyperparameters, and training data summaries \\ \cline{2-3}
& Model families and lineage & Collections of models of varying sizes and fine-tuning levels, including checkpoint histories \\ \cline{2-3}
& Architecture and training documentation & Detailed model architecture, training procedures, and design decisions \\ \cline{2-3}
& Evaluation results and artifacts & Results from internal and third-party evaluations of model capabilities, limitations, and potential risks, including methodology documentation, test datasets, prompts, scoring rubrics, and logs of model outputs \\ \cline{2-3}
& System documentation & Documents explaining how production systems work and are used \\ \cline{2-3}
& Monitoring systems & Tools and dashboards for observing model behavior in production \\ \cline{2-3}
& System logs & Records of events, operations, and state changes within deployed systems \\ \cline{2-3}
& Compute accounting records & Compute allocation logs, training run records, access controls, hardware configuration, and declared vs. logged compute reconciliation \\ \hline

\multirow{3.5}{=}{Governance and process} 
& Process documentation & Internal documents describing procedures and responsibilities \\ \cline{2-3}
& Board and governance minutes & Records of discussions, decisions, and actions from board and governance meetings \\ \cline{2-3}
& Internal reports & Documents describing experiment results, incidents, or process outcomes \\ \cline{2-3}
& Process communications & Approval emails, request tickets, escalation threads, and decision logs \\ \cline{2-3}
& Previous compliance reviews & Reports and notes from prior audits or compliance reviews \\ \cline{2-3}
& Organization charts & Diagrams showing roles, reporting lines, and responsibilities \\ \cline{2-3}
& Written representations & Signed management statements confirming responsibilities, activities, or factual matters \\ \hline

\multirow{12}{=}{Operational and contextual} 
& Staff interviews & Structured interviews with personnel on how processes function in practice \\ \cline{2-3}
& Governance interviews & Interviews with senior executives and board members \\ \cline{2-3}
& Process-owner interviews & Interviews with employees leading specific processes \\ \cline{2-3}
& Casual conversations & Informal conversations with employees \\ \cline{2-3}
& Meeting attendance & Observation of meetings between employees, leadership, or external parties \\ \cline{2-3}
& Walkthroughs & Physical or virtual observation tracing a workflow from start to finish \\ \cline{2-3}
& Operational communications & Emails, message threads, and call summaries on safety-relevant events \\ \cline{2-3}
& External inquiries & Inquiries with external parties to clarify the existence and extent of engagement \\ \hline

\multirow{4}{=}{External feedback} 
& User reports and complaints & Bug reports, safety complaints, vulnerability disclosures, and feedback from users \\ \cline{2-3}
& Third-party correspondence & Communications with external researchers, civil society, regulators, or other stakeholders raising concerns or sharing findings \\ \hline

\multirow{8}{=}{Public information} 
& Company public outputs & System cards, model cards, blog posts, press releases, social media posts, and interviews \\ \cline{2-3}
& Regulatory filings and disclosures & Mandatory regulatory disclosures, incident reports filed with authorities, and compliance certifications \\ \cline{2-3}
& Published research & Academic papers, technical reports, and preprints by company employees \\ \cline{2-3}
& External commentary and analysis & Third-party analyses, media coverage, civil society reports, and independent research on the company's systems \\ \cline{2-3}
& Public user feedback & Public reviews, complaints, and feedback
\end{longtable}
}

\newpage 
\appsection{Frontier AI auditing in context}
\label{apx:frontier}

Frontier AI safety and security requires three elements: standards for sufficient safety and security, companies having incentives to follow such standards, and the public and other stakeholders having evidence that these standards are being followed. Auditing is independent of standard-setting, meaning companies can be audited against their own policies, applicable laws, or industry best practices. 

When discussing auditing, we often emphasize verifying companies’ safety and security claims. However, if a company makes only trivial claims (e.g., “we thought about safety before deploying this system”) or none at all, verification alone has limited value. We therefore advocate that frontier AI auditing should assess companies against relevant standards --- including informal best practices --- not just their own claims. Additionally, government-mandated or private sector-based transparency requirements can enable more effective auditing by ensuring that there are meaningful claims to audit in certain key areas.

Audits provide evidence that companies follow relevant standards. Companies can also voluntarily or mandatorily disclose information to the public --- what we call transparency --- which serves as a baseline that auditing builds upon.\footnote{The term transparency is sometimes used to refer to companies privately disclosing information to a regulator or downstream providers, as in the case of the EU General-Purpose AI Code of Practice \autocite{european_commission_general-purpose_nodate}. In this paper, we reserve the term transparency for cases where information is disclosed fully publicly.} Auditing can verify information and evaluate properties that are too sensitive to be fully disclosed publicly, and ensure that publicly disclosed information is accurate and only redacted where appropriate. This can create an additional incentive to follow these standards, from the public sector (e.g., fines and litigation risks if laws are violated) as well as from the private sector (e.g., lower insurance premiums if certain practices are followed).

Additionally, company whistleblowing policies and government-imposed protections for whistleblowers are complementary to auditing and transparency requirements. If company staff believe that they will be protected if they inform government agencies that a statement by the company is misleading or that a company practice is dangerous, such misleading statements and dangerous behavior are less likely to occur in the first place.

\newpage 
\appsection{Lessons from assessment in diverse domains}\label{apx:lessons}

In domains from finance to food safety, third-party auditors remain independent of the companies they assess while analyzing non-public information and enabling public trust in the industrial sectors they audit. In this appendix, we outline a sample of such domains. The sample spans various assessed scales, from whole organizations (e.g., financial auditing) to specific, tangible outputs (e.g., consumer products). 

Although risk assurance practices in these domains generally exceed those in frontier AI (in rigor, maturity, and scale), we do not claim that they are gold standards to emulate in every respect. The lessons are not uniformly positive and do not solve the distinctive challenges of frontier AI auditing. We aim to extend best practices from diverse domains while avoiding their failures.

\subsection{Food safety testing} \label{ssec:food_safety}

Food safety focuses on preventing harm from products consumed by the public through food safety standards, systematic testing, and ongoing monitoring from production to consumption \autocite{fda_standardization_2025}. When functioning properly, this defense-in-depth approach operates through multiple independent checkpoints. Farmers and dairy cooperatives conduct initial quality tests on raw milk; processing facilities perform intake testing before production; and government regulators conduct random sampling throughout the supply chain. Because each layer tests independently using different methods and incentives, contamination is likely to be caught at one stage even if it evades another.

The importance of food safety is illustrated by its failures, such as the 2008 Chinese milk scandal \autocite{gossner_melamine_2009}. A baby formula producer sold products contaminated with melamine, leading to the hospitalization of thousands of babies and at least a decade of distrust for Chinese milk products \autocite{li_consumer_2021}. Subsequent government milk recalls --- a sign of a functioning food safety system --- have yet to restore consumer trust in milk products in China. The crisis is so remembered among Chinese consumers that foreign milk products are still highly sought after \autocite{pak_foreign_2018}. Safety system failures are thus also a crisis for the implicated industry.

Frontier AI systems can also have safety regressions, causing backlash against the whole AI enterprise \autocite{openai_sycophancy_2025, difeliciantonio_california_2025}. Regressions may go undetected without regular testing analogous to food safety testing. Unlike food, which benefits from a well-understood body of scientific work over the last century \autocite{blum_poison_2019}, the science of frontier AI safety and security is still rapidly developing.

\textbf{Key lessons for frontier AI auditing are:}
\begin{itemize}[leftmargin=0.25in]
    \item Effective safety culture involves “defense in depth” with product testing entering at several different stages and looking for multiple types of failure, at different levels of granularity.
    
    \item Safety system failures produce widespread distrust and product avoidance that propagate across companies and can last for many years.
    
    \item Modern food safety benefits from more than a century of research and development. The breadth, scale, and impact of frontier AI systems require a similar or greater level of investment in assurance methods in a far shorter time period.

\end{itemize}

\subsection{Consumer product safety} \label{ssec:consumer_safety}

Independent testing organizations like Underwriters Laboratories (UL) \autocite{underwriters_laboratories_inc_engineering_2016} and independent government agencies like the Consumer Product Safety Commission \autocite{cpsc_celebrates_2022} have developed rigorous standards and protocols over many decades. These protocols cover tens of thousands of product types, enabling firms around the globe to bring the latest technologies to market safely. The standards instituted by these organizations and others like them address categories ranging from children’s toys to electronics, identifying potential hazards before products reach consumers. The UL mark, applied after products have been certified to meet safety standards, appears on 22 billion products annually \autocite{ul_ulmarks_2026}. 

In the EU, regimes such as CE marking and the Radio Equipment Directive (RED) show that self-declaration can be effectively combined with third-party assessment \autocite{noauthor_ce_2025, european_commission_radio_nodate}. Manufacturers are allowed to declare that their product complies with the relevant safety directives, but they must base this declaration on testing against harmonized standards, keep detailed technical documentation, and accept legal responsibility if the product is later deemed unsafe or causes harm. For high-risk products, a “notified body” (an accredited third-party conformity assessor) must review the design, perform tests, and issue reports that verify the manufacturer’s declaration \autocite{european_commission_notified_nodate}.

This creates aligned incentives across participants. Manufacturers need CE/RED compliance to access the EU market and satisfy the requirements of retailers and insurers \autocite{european_commission_ce_nodate, clarke_williams_insurance_brokers_ce_nodate}. Retailers require proper documentation, and insurers price coverage based on conformity assessment evidence. Regulators and market-surveillance authorities spot-check products and can order recalls or fines, giving teeth to self-declaration. End users benefit from clear safety marks and multiple actors incentivized to keep noncompliant products off the market.

While external testing labs are sought after for the trust and marketing credibility they provide, the proliferation of independent test organizations has also been driven by regulatory requirements. The Testing, Inspection, and Certification (TIC) landscape demonstrates that various incentives can compel voluntary participation, but regulatory mandates have often been necessary to ensure full participation.

New specialized testing labs also contribute to participation. Each year, startups are launched to address niche safety risks in specific and often emerging product categories. Despite this long-tail diversity, the TIC space remains largely dominated by a handful of large organizations that cover many categories. Beyond classic factors like brand recognition and shared support costs, these firms benefit from economies of scale that only broad-ranging testing organizations can achieve: joint processes to gain and maintain many accreditations, larger networks of field engineers spanning wider geographic footprints, and shared labs and equipment deployable across many product types. Because test labs must constantly demonstrate competencies that are expensive for product manufacturers to replicate, technical expertise tends to concentrate in organizations that can apply it across many categories simultaneously.
Finally, while product safety testing is proactive, other mechanisms exist that are naturally reactive to emerging or unforeseen safety hazards. Consumers can report when they are harmed by a product \autocite{us_consumer_product_safety_commission_saferproductsgov_nodate}, and product recalls may be issued as a result \autocite{us_consumer_product_safety_commission_recalls_nodate}. Such practices prevent repeated harms by removing hazardous products from the market and, importantly, by informing the design of safer next-generation products. These practices generate data detailing real‑world impacts and circumstances. Similar approaches are already being applied to AI products, but their scope and integration with auditing practices are limited \autocite{raji_closing_2020}.\looseness=-1

\clearpage
\textbf{Key lessons for frontier AI auditing are:}
\begin{itemize}[leftmargin=0.25in]
    \item While there are immediate concerns regarding the dearth of qualified frontier AI auditing organizations, the growth of third-party product test labs shows that a large enough market demand for assessment can eventually produce qualified market actors.
    
    \item The competency of third-party test labs can be certified by external actors (“accreditation bodies”) against recognized standards and may be mandated by regulators.
    
    \item Product safety’s “trust marks” and government pre-clearance provide demand drivers elsewhere that could become relevant to frontier model audits, while post-market surveillance serves as a reactive complement, revealing the successes and failures of the broader audit community.

\end{itemize}

\subsection{Safety-critical systems engineering} \label{ssec:safety-critical}

Safety-critical systems engineering is used in domains like aviation, nuclear power, and civil infrastructure where failures can have catastrophic consequences \autocite{leveson_introduction_2023}. Modern safety-critical systems engineering treats the safety of a high-stakes system as an emergent property arising from interactions and control relationships within complex sociotechnical systems \autocite{perrow_normal_1999, vaughan_challenger_2016, cooper_accuracy-efficiency_2021, rilinger_failure_2024}. The discipline employs structured methodologies --- including hazard analysis techniques --- to proactively identify hazards, quantify risks’ severity and probability, and maintain continuous risk management tracking systems.

A key belief in safety-critical systems engineering is that catastrophic accidents are often anticipatable and avoidable with sufficient attention to process and sufficient incorporation of lessons from near-misses along the way. For instance, the International Atomic Energy Agency found the 1986 Chernobyl disaster to be caused by flaws with the reactor design, but also by “a remarkable range of human errors and violations of operating rules” \autocite{noauthor_chernobyl_2025}. To avoid such organizational causes of disasters, industries with safety-critical systems have thus developed rigorous safety cultures, which emphasize independent verification, continuous monitoring, and strong incentives for identifying potential failures early and learning from failures and near-misses \autocite{office_of_safety_and_mission_assurance_safety_nodate}.\footnote{Safety engineering research identifies a progression of organizational safety cultures, from “pathological” (characterized by blame and denial) through “reactive” and “bureaucratic” stages, to “proactive” and ultimately “generative” or high-reliability cultures. Auditors assessing frontier AI organizations should evaluate not just whether safety policies exist, but whether the organization’s culture actively seeks out “latent pathogens” and “error traps” before they manifest as incidents \autocite{cairn_risk_consulting_decoding_2024}.}

\textbf{Key lessons for frontier AI auditing are:}
\begin{itemize}[leftmargin=0.25in]
    \item Analyzing specific technical systems is important but needs to be paired with auditing of company practices.
    
    \item Continuous lifecycle risk management with formal acceptance and verification helps manage systems as they change over time better than one-off certifications. 
    
    \item Near-misses and incidents are often early warning signs of failures that eventually result in significant harm. 

    \item Structured artifacts such as hazard analyses and safety cases show the value of documented, evidence-based risk arguments --- going beyond measurements of a system’s properties, to proper arguments about the implications of these properties. Analogously, frontier AI auditors should review informal analyses and/or formal safety cases produced internally by companies, in addition to the raw evaluation results.

\end{itemize}

\subsection{Aviation safety} \label{ssec:aviation_safety}

Aviation safety involves systematic oversight of aircraft design, manufacturing, operation, and maintenance. Safe aviation has long been viewed as necessary for commercial air travel’s viability --- people would be far less likely to fly if it were not both faster and safer than other forms of transit \autocite{noauthor_perception_2025}. The well-aligned interests of the aviation industry and public safety have helped produce a track record of safety that is exceptional when compared to other means of transportation \autocite{national_safety_council_deaths_nodate}.\footnote{Note that here we focus on a specific subset of aviation, though we use the term aviation for shorthand. The subset we focus on is commercial scheduled aviation in developed nations. Other forms of aviation tend to have much weaker safety track records.} The strong safety record includes many interlocking elements providing defense in depth, including pre-approval of new design elements, simulation and flight testing requirements with real human operators, mandatory reporting of accidents and major incidents, extensive government-funded and government-housed safety research, and criminal charges in some instances \autocite{federal_aviation_administration_design_nodate, faa_how_nodate, faa_safety_nodate, federal_aviation_administration_key_nodate, us_department_of_transportation_aviation_2025, noauthor_united_2025}. 

Despite this extensive safety apparatus, the Boeing 737 MAX disasters (2018–2019) exposed critical weaknesses in the certification ecosystem, resulting in 346 fatalities across two crashes \autocite{house_transportation_final_2020}. These events highlighted the dangers of excessive reliance on manufacturer self-certification. Over time, the FAA’s delegation of authority had expanded to the point where Boeing self-certified 96\% of the parts for the 737 MAX \autocite{house_transportation_final_2020}. While Boeing employees serving as “Authorized Representatives” were intended to represent the FAA’s interests, internal surveys revealed that 39\% perceived “undue pressure” from management and 29\% feared consequences for reporting safety concerns \autocite{house_transportation_final_2020}. When technical experts did raise concerns, they were often overruled by management prioritizing the manufacturer's timeline, creating what an internal investigation later termed “an environment of mistrust” \autocite{house_transportation_final_2020}.

Following these events, the Aircraft Certification, Safety, and Accountability Act of 2020 introduced reforms limiting self-certification of safety-critical systems and strengthening protections for whistleblowers \autocite{aircraft_certification_act_2020}. But this was too late to prevent not only substantial loss of life, but also a decline in trust in both Boeing and the FAA. 

Overall, despite decades of development and many strengths compared to current AI practices, periodic crashes underscore that even strong auditing and safety regimes leave residual risk and that expectations for AI auditing in \cref{sec:challenges} should be ambitious but also realistic \autocite{noauthor_out_2018, icao_latest_2025}.

\textbf{Key lessons for frontier AI auditing are:}
\begin{itemize}[leftmargin=0.25in]
    \item Delegating certification to the entities being certified creates dangerous conflicts of interest, particularly when commercial pressures are high. When employees face internal pressure not to raise concerns, self-assessment regimes can become self-dealing. For frontier AI, this underscores why auditors must be genuinely independent third parties.
    
    \item Auditors and regulators cannot effectively oversee what they do not understand; if public or third-party auditors lack the resources to maintain independent technical expertise, they risk becoming “rubber stamps” for decisions made by the companies they oversee.
    
    \item Government agencies often face structural limitations in technical oversight, including an inability to match private sector salaries and retain specialized expertise. This supports the case for a private-sector auditing ecosystem with public oversight as discussed in (\cref{ssec:government_vs_private}).

\end{itemize}

\subsection{Penetration testing} \label{ssec:penetration_testing}

Penetration testing consists of hiring skilled, adversarial-minded experts to actively probe a complex digital system --- like a company’s network, applications, or infrastructure --- as if they were real attackers, but in a controlled and permissioned way. Instead of checking only whether documented requirements are met, these testers creatively search for unexpected failure modes, chain together subtle weaknesses, and try to achieve concrete, high-impact goals such as stealing sensitive data or taking control of critical operations. Their findings don’t just result in a pass/fail grade; they produce detailed reports that prioritize vulnerabilities by severity, demonstrate realistic attack paths, and recommend targeted fixes, often iterated on over multiple rounds. Over time, this kind of structured, adversarial evaluation becomes a recurring discipline: independent teams are brought in, rules of engagement are defined, safeguards are put in place, and the results feed into broader risk management and governance, especially for systems whose failures could have serious consequences beyond the organization itself.

Penetration testing demonstrates that security attributes are often best assessed through active adversarial testing. Mitigations that ostensibly provide defense in depth may prove inadequate when subjected to realistic attack strategies, since adversaries can adaptively route through the holes in each layer of defense. Penetration testing can reveal critical vulnerabilities missed by even highly qualified in-house security teams, though notably, it complements and builds on rather than substitutes for in-house security capacity. It also shows that creativity and realism matter more than static checklists, and that an adversarial analytical posture need not imply an adversarial relationship with the organization.

The field also illustrates challenges in measuring defensive strength, such as simulating the considerable effort and resources of a motivated nation-state attacker. While easily finding and exploiting a vulnerability shows defenses are unlikely to withstand real attacks involving similar expertise and effort, the converse is not necessarily true (e.g., even if a realistic simulation of state-level attacking skills may fail, a real state-level attacker might succeed). Furthermore, simulating higher-level attackers is correspondingly more difficult and expensive. Penetration testing best practices are codified in standards such as the Penetration Testing Execution Standard, CBEST, TIBER-EU, PCI‑DSS, and NIST SP 800‑115 \autocite{ptes_penetration_nodate, bank_of_england_cbest_nodate, european_central_bank_tiber-eu_2025, pci_security_standards_council_penetration_nodate, nist_nist_2020}. 

\textbf{Key lessons for frontier AI auditing are:}
\begin{itemize}[leftmargin=0.25in]
    \item Active, adversarial testing should be core to audits for security- and misuse-related risks, rather than relying only on checklist-style reviews.
    
    \item Legal safe harbors for good-faith researchers are essential in order to unlock constructive engagement with companies on high-risk aspects of products.
    
    \item Penetration-style engagements should complement, not substitute for, in-house security work, with audits focused on realistic attack paths and prioritized remediation guidance.
    
    \item An adversarial analytical posture can coexist with a collaborative relationship, where auditors and companies iteratively fix issues rather than treating audits as one-off pass/fail exercises.
    
    \item Bug bounty-style programs can complement formal penetration tests by providing continuous, incentive-aligned scrutiny from a broad pool of researchers, with payment tied to impact and clear expectations for rapid remediation.
    
    \item While private engagement is valuable in providing space to mitigate risks, it is eventually important for findings to be surfaced publicly in order to drive ecosystem-wide improvements.

\end{itemize}

\subsection{Financial auditing} \label{ssec:financial_auditing}

While “audit” has many meanings, for many people it is synonymous with financial auditing. Financial audits may seem far removed from safety and security audits, but the key parallel with frontier AI auditing is that independent reviewers examine highly sensitive, non-public information to judge both whether an organization’s public claims are credible and whether its internal safeguards are effective.

Modern financial auditing emerged in the 19th century alongside unprecedented cross-border capital flows (notably to finance US railroads) and evolved again after scandals like Enron’s collapse prompted the Sarbanes--Oxley Act, which raised expectations around auditor independence and executive accountability. Today, financial auditing is a highly structured and professionalized ecosystem spanning firms of many sizes, private standard-setting alongside public regulation, and a large body of case studies --- both successes and failures --- from which to learn.

Financial auditing is therefore both a source of inspiration and a source of cautionary tales for frontier AI. On the positive side, it demonstrates that societies can build professional processes and standards that enable independent parties to review extremely sensitive information; develop relatively standardized ways to compare risks and controls across very different organizations; and combine private and public “demand signals” into an ecosystem that supports high-stakes decisions and large investments.

Financial auditing also offers a mature set of conceptual tools that frontier AI auditing can borrow rather than reinvent. These include norms for managing conflicts of interest and evaluating evidence; sharp distinctions between error and fraud (with fraud demanding more rigorous detection approaches); recognition that professional judgment is indispensable; a view of auditing as a profession with duties to the public and investors that supersede obligations to the client; attention to organizational culture as a driver of wrongdoing; close collaboration with domain experts; and the separation between prepared statements and independent verification. These analogies translate naturally to AI: system cards can be treated as the analogue of financial statements, and safety, security, and compute-provenance measures as internal controls whose effectiveness auditors should assess.

Financial auditing’s failures also illustrate what can go wrong when independence is compromised. The Enron and Wirecard scandals show how heavy reliance on a small number of large clients --- and pressure to retain them --- can distort auditor incentives. The Enron scandal also illustrates two distinct mechanisms of capture: Arthur Andersen (the auditing firm) faced impaired objectivity after performing extensive consulting that positioned it to audit work it helped design, and internal incentives favored client satisfaction, with Enron paying roughly comparable sums for consulting and audit services (\$27M vs. \$25M) \autocite{mary_locatelli_good_2002}. Even after the Sarbanes–Oxley Act reduced some advisory–audit conflicts, dependence on clients persisted, and PCAOB inspections (“auditing the auditors”) have repeatedly raised concerns about lax attention to detail and overly procedural, box-checking approaches that risk missing systemic problems \autocite{pcaob_pcaob_2023}. The sector has also struggled with an “expectations gap” between public belief that audits guarantee the absence of fraud and auditors’ actual mandate to provide only reasonable assurance.

\textbf{Taken together, these case studies suggest several lessons for frontier AI auditing:}
\begin{itemize}[leftmargin=0.25in]
    \item Professional standards for how audits are conducted and communicated can establish a common language for understanding findings related to many highly diverse companies.
        
    \item Given the right incentives, a very large ecosystem can be built to provide well-defined assurance services.
    
    \item Since audits have a weak track record at detecting deliberate fraud, regimes must either explicitly scope some deception risks out --- or invest in unusually deep access, high-effort testing, and strong disincentives for misrepresentation. 
    
    \item Audit criteria and documentation must avoid devolving into gameable box-ticking (Goodhart’s Law) by pairing standardized evidence requirements with professional judgment about systemic risk.
    
    \item Standards often lag innovation, so AI auditing frameworks should be more adaptive than traditional financial-audit rules while still drawing on institutional precedents (e.g., oversight bodies). 
    
    \item As the frontier AI ecosystem scales on a far shorter timeline than finance did, it will likely need to lean heavily on automation to achieve adequate coverage, while keeping the core professional ethic intact: frontier AI auditors’ duties should be framed not only to clients, but centrally to AI users and the broader public.
    
    \item Scandals such as Enron and Wirecard showed that auditor independence and disclosure of conflicts of interests are essential.

\end{itemize}

\newpage 
\appsection{Contemporary third-party frontier AI assessment}\label{apx:contemporary}

Third-party frontier AI assessment has grown significantly in recent years, providing a foundation on which to build (\cref{sec:lessons}).

We assess the current state of third‑party assessment along nine dimensions: reporting, access, rigor, standardization, continuous monitoring, scope, scale, independence, and ecosystem maturity. \Cref{tab:gap_appF} summarizes the gap between current practices and the vision introduced in \cref{sec:vision}. We specifically focus on current vs. future practices in assessment of AI companies’ technical systems, rather than assessment of organizations’ risk culture or internal processes, for which we are not aware of established precedents. Where possible, we cite published literature; where sources are unavailable given the nascent state of AI assessment, we rely on direct experience and author expertise.

Contemporary assessment efforts provide an important foundation, but realizing the proposed vision will require closing substantial gaps along those nine dimensions, as discussed below.

\renewcommand{\arraystretch}{1.6}

\begin{longtable}{|>{\bfseries\RaggedRight}p{2.8cm}|>{\RaggedRight}p{6cm}|>{\RaggedRight\arraybackslash}p{6cm}|}
\caption{The gap between contemporary third-party frontier AI assessment and our vision for future third-party frontier AI auditing.} \\

\hline
\rowcolor{headergray}
\textbf{Dimension} & \textbf{Today (January 2026)} & \textbf{Future Vision} \\ \hline
\endfirsthead

\multicolumn{3}{c}%
{{\bfseries \tablename\ \thetable{} -- continued from previous page}} \\
\hline
\rowcolor{headergray}
\textbf{Dimension} & \textbf{Today (January 2026)} & \textbf{Future Vision} \\ \hline
\endhead

\hline
\multicolumn{3}{|r|}{\emph{Continued on next page}}\\
\hline
\endfoot

\hline
\endlastfoot

Reporting & 
Sparse, inconsistent public reporting \par\medskip Details depend on auditor-auditee agreements & 
Standardized, rigorous public reporting frameworks with justified redactions \\ \hline

Access & 
Mostly public-level access \par\medskip Limited pilots with deeper access & 
Deep access comparable to trusted internal engineers \par\medskip Structured secure environments \\ \hline

Rigor & 
Fraction of effort applied by the most sophisticated internal teams \par\medskip Significantly less than other safety-critical domains & 
Rigor matching or exceeding other safety-critical contexts \\ \hline

Standardization & 
Emerging norms and proposals \par\medskip Bespoke contracts & 
Clear professional norms backed by consensus and incentives \\ \hline

Continuous monitoring & 
One-off snapshots with unknown shelf-life & 
Continuous monitoring with automatic downgrading based on drift \\ \hline

Scope & 
Predominantly model-centric capability evaluation & 
Whole-organization assessment including security, platform controls, and governance \\ \hline

Scale & 
Voluntary participation by few developers & 
Universal adoption across frontier developers \\ \hline

Independence & 
Evaluators depend on company goodwill & 
Access and financial standing secure regardless of findings \\ \hline

Ecosystem maturity & 
The third-party evaluation ecosystem is growing but currently consists of a small number of specialized private evaluators (e.g., METR, Apollo Research, SecureBio, Irregular), often with focuses on particular risks, and a small number of government agencies (e.g., US CAISI and UK AISI). & 
A mature regime of private and public evaluators conduct audits of frontier AI systems. Some specialize in evaluations for niche risks while others perform holistic evaluations. Auditors coordinate to collaboratively articulate best practices. %\\ %\hline
\label{tab:gap_appF}
\end{longtable}

\subsection{Reporting} \label{ssec:reporting}

Public reporting on third-party audits remains inconsistent both across and within frontier AI developers. Reporting templates, substance, and style vary substantially by audit, auditor, and developer \autocite{staufer_audit_2025, reuel_who_2025}. By some analyses, reporting quality has declined over time \autocite{reuel_who_2025, wang_2025_foundation_2025}. To date, audit results are most commonly communicated through system cards and related publications, both of which we draw from to inform this section. 
Frequently, system cards only mention third-party evaluators in the abstract and provide little detail about methodological details of third-party audits \autocite{google_deepmind_2024}. In some cases, system cards mention third-party evaluators by name \autocite{openai_gpt-4o_2024, noauthor_system_2025}. Sometimes, third-party assessors themselves (e.g., Irregular, METR \autocite{irregular_irregular_2025, metr_details_2025}) or assessed companies (e.g., OpenAI, Anthropic, Amazon \autocite{openai_gpt-4_2023, noauthor_system_2025, amazon_evaluating_2025}) share additional public details about specific evaluations that have been conducted, as a complement to briefer discussions in system cards.\footnote{Some companies do not typically share information about which third parties they work with --- for example, Google DeepMind frequently mentions third-party assessment but does not name the individuals or organizations in question. It is typically unclear from the outside whether, in these cases, assessors are allowed to discuss their work with these companies, and how rigorous the practices are relative to cases that are documented in more public detail.} 
This lack of transparency hinders understanding and advancement of the third-party auditing landscape.\footnote{Legitimate reasons, including information hazards, may exist to exclude select information from public documentation. However, we posit that current transparency and reporting gaps are far from fully accounted for by these reasons. As discussed in Section 5.4, auditors’ ability to review unredacted safety information --- and other non-public information more generally --- and to attest to the reasonableness of redactions in public versions is one key component of avoiding pure self-assessment while protecting sensitive information.\looseness=-1}\looseness=-1

\subsection{Access} \label{ssec:access}

To assess safety-relevant properties of frontier AI deployments with reasonable or high assurance, third parties need access to various types of information \autocite{bucknall_structured_2023, casper_black-box_2024, kembery_position_2024, che_model_2025, bucknall_position_2025}. At a minimum, this includes timely “black-box” access to model outputs via an API, preferably with configurable settings (e.g., around reasoning effort, temperature, etc. as applicable). More comprehensive assessments require access to richer interfaces --- such as variants with reduced safety mitigations, “helpful‑only” models, the ability to fine-tune models in a custom manner, or bespoke testing endpoints --- as well as non‑public information about how systems were trained and which mitigations and monitoring are in place. For high-stakes questions, third‑party evaluators may need “gray-box” or “white-box” access to model internals or “outside-the-box” access to additional resources (see Table \ref{tab:access_level}).

\begin{table}[h]
\caption{The first three entries are cumulative in that gray-box includes black-box access and white-box includes gray-box access, whereas outside-the-box is separate from these.}
\label{tab:access_level}
\centering
\renewcommand{\arraystretch}{1.8}
\setlength{\tabcolsep}{10pt}  

\begin{tabularx}{\textwidth}{|p{4cm}|X|}
\hline
\rowcolor{headergray}
\textbf{Access level} & \textbf{Information accessed} \\ 
\hline

Black-box & 
The ability to query a system with inputs and analyze the resulting outputs. \\ 
\hline

Gray-box & 
Partial visibility into a system's operations such as chain-of-thought, sampling probabilities, or some activation patterns. \\ 
\hline

White-box & 
Access to full activations, model weights, and architecture. \\ 
\hline

Outside-the-box & 
Access to relevant training data, training details, source code, documentation, logs, ``helpful-only'' models, ability to fine-tune models in a custom manner, and organizational artifacts. \\ 
\hline

\end{tabularx}
\end{table}

In practice, third-party access to frontier AI systems (often limited to black-box API access) remains dependent on developer discretion \autocite{casper_black-box_2024}. Many independent organizations conduct evaluations using public APIs or short access windows to pre-deployment APIs (only rarely has this access occurred more than a few weeks before launch). Pre‑deployment testing exercises for Anthropic's and OpenAI’s models by the US Center for AI Standards and Innovation (CAISI) and the UK AI Security Institute (AISI) provide examples of deeper access in practice \autocite{anthropic_strengthening_nodate, openai_working_2025}, as do pilots conducted between industry and the non-profit organizations METR and Apollo Research \autocite{metr_review_2025, openai_detecting_2025}.\footnote{To date and to our knowledge, this remains limited to API-level testing, limited non-public documentation, access to chain-of-thought, one staff interview, access to helpful-only model variants, and private company attestations to specific claims.}

To date, evaluators rarely --- if ever --- receive access to model training data, chain-of-thought \autocite{openai_o1_2024}, model internals, or even basic (let alone detailed) training and deployment documentation, even while developers themselves acknowledge that such information is important for their own confidence in their safety and security mitigations \autocite{anthropic_tracing_2023, anthropic_tracing_nodate, openai_detecting_2025, korbak_chain_2025}. Existing pilots that expose chain-of-thought tend to be limited in duration and scope, e.g., sharing static examples rather than continuous access for each query, and companies often decline to guarantee not training on evaluation data, thereby risking contamination of future analyses \autocite{bucknall_position_2025}. Gray- and white-box access to model internals, and outside-the-box access to source code, training data, and internal evaluation results are currently highly limited but will be increasingly important for rigorous external assessment as the limits of black-box analysis are reached \autocite{casper_black-box_2024}.\looseness=-1

Evaluators also face practical obstacles, including model providers breaking previously-safe assumptions with no warning (e.g., new reasoning settings); API bugs that only manifest after hours of benchmark execution; bugs specific to extremely large and slow models (e.g., backends timing out); providers swapping in quantized or weaker models when overloaded without notification; crushingly low rate limits; model updates during evaluation periods that invalidate prior results; limited ability for third parties to verify when model changes are occurring due to quantization or other factors; and time constraints that prevent thorough assessment.\looseness=-1

\subsection{Rigor} \label{ssec:rigor}

Methodology and rigor in third-party assessments vary substantially:
\begin{itemize}[leftmargin=0.25in]
    \item Benchmark-based assessments struggle with issues related to quality, design, elicitation methods, simplifying assumptions, data contamination, and differences between evaluation conditions and the real world \autocite{reuel_betterbench_2024, eriksson_can_2025, salaudeen_measurement_2025, bean_measuring_2025, mcgregor_risk_2025, wallach_position_2025, chouldechova_shared_2024, raji_aiwideworldbenchmark_2021, McIntosh_2026, ren_safety_2024, weidinger_sociotechnical_2023}.

    \item Red-teaming-based methods are skill-dependent and frequently fail to be rigorous in practice \autocite{chouldechova_comparison_2025, feffer_redteaming_2024}.

    \item Empirically, evaluations often struggle to identify failures, and the worst things identified in an evaluation can only offer a lower bound on the system’s worst possible case harms \autocite{gal_towards_2024, openai_o1_2024, che_model_2025}.

    \item It’s hard to do audits with full construct validity that accurately capture real-world, often subtle types of risks \autocite{weidinger_sociotechnical_2023, wallach_position_2025, bean_measuring_2025, salaudeen_measurement_2025}.

    \item Audits usually focus on a limited number of risk domains (\cref{ssec:risk_scope}) and typically stick to evaluating harms that manifest in single uses of a system rather than extended uses in real-world applications.
\end{itemize}

Time and resource constraints may be the most significant barrier to rigorous assessment. Evaluators typically operate under severe time pressure, with assessment windows rarely exceeding a few weeks and often compressed to days. This contrasts with months-long certification processes in aviation, nuclear safety, and even lower-risk consumer products. Resource asymmetries compound these temporal constraints: third-party assessors generally operate with a fraction of the computational budget, personnel, and specialized tooling available to frontier AI developers, making it difficult to match the depth and sophistication of internal safety teams.

Methodological standardization remains a persistent challenge. Without consensus frameworks, different assessors employ divergent threat models, scoring rubrics, and evaluation protocols, rendering cross-assessment comparisons difficult or meaningless. What constitutes a “dangerous capability” or an acceptable risk threshold varies substantially across organizations, and the criteria for determining whether a model “passes” or “fails” an evaluation are often implicit rather than codified. This heterogeneity undermines the field's ability to establish baselines, track progress over time, or provide stakeholders with consistent signals about relative risk levels.

Reproducibility issues further undermine confidence in assessment outcomes. Many evaluations rely on proprietary prompts, specialized human expertise, or particular API configurations that are not fully documented or shared. When evaluators publish their methodologies, subtle differences in implementation, model versions accessed, or even API call timing can produce substantially different results. The lack of standardized reporting requirements means critical methodological details are often omitted from public reports, making independent verification nearly impossible.

\subsection{Standardization} \label{ssec:standardization}

Standards for frontier AI assessment remain nascent but are evolving rapidly. Recent developments include proposed frameworks for the design, implementation, and reporting of evaluations \autocite{patricia_paskov_preliminary_2025, reed_what_2025, reuel_betterbench_2024}; an initial statement of best practices from the industry-led Frontier Model Forum; the recent announcement of a forum for third-party evaluators, AEF \autocite{noauthor_ai_nodate}; and an initial statement from that forum (AEF-1) \autocite{stosz_aef-1_2025}. While promising, these developments lag behind evaluation regimes in more established industries. AI evaluations are almost always conducted under bilaterally negotiated, confidential, ad hoc contracts. The terms of these contracts, including scope, access provisions, and publication rights, are rarely visible to regulators or the public. The absence of standardized contractual and reporting requirements means that critical methodological details are often omitted from public reports, making independent verification nearly impossible.

\subsection{Continuous monitoring} \label{ssec:continuous_monitoring}

AI developers frequently make both incremental and substantial changes to their systems without providing early access to third parties to conduct updated risk assessments, or else publish information and grant access only after changes have already been deployed \autocite{zeff_openai_2025,hashim_google_2024}. Such changes can occur for several reasons. Some stem from system-level modifications, such as altering how multiple instances of a model are coordinated within an agentic product. Others result from inference-time optimizations aimed at improving efficiency, or from new post-training updates to the model itself.

A balance is needed between excessive third-party review of all changes and insufficient checks and balances to prevent severe risks. The EU General-Purpose AI Code of Practice articulates criteria for ``similarly safe'' models \autocite{european_commission_general-purpose_nodate}, and analogous criteria will eventually be needed for other purposes and contexts for the size of changes to a product that uses substantial amounts of test-time compute with a fixed underlying model. At minimum, greater third-party use of automated measurement for system changes is needed (at least those that can be measured in a relatively resource-efficient manner), thereby lessening reliance on company self-reporting. A nascent effort in this direction is \href{https://www.stampr-ai.com/}{stampr-AI}, which checks APIs for changes in “model fingerprints.” By having continuous public and auditor insight into (some subset of) significant changes, it will be easier to determine whether a prior third-party assessment’s “shelf life” has been exceeded.

\subsection{Scope} \label{ssec:scope}

Evaluations are conducted on a wide range of topics without access to non-public information (e.g., as happens in academic research or customer testing of products they are considering adopting). Such evaluations are not our focus here, though they are important, and critically, they are inherently easier to scale than evaluations that involve non-public information.

When frontier AI developers provide non-public information to third-party evaluators, their assessments generally focus on capability evaluation and, increasingly, propensity evaluation (e.g., whether models tend to act deceptively under certain circumstances) with a predominant focus on biological, chemical, cyber, nuclear, and deception-related risks. 

Some assessments evaluate safety and security mitigations (e.g., jailbreak robustness), including by specialized organizations such as FAR.AI, Gray Swan, and Haize Labs, though these efforts often focus on system-level robustness rather than platform-level assessment of controls (e.g., considering efforts to break harmful activity down into benign-looking components across multiple API accounts, as has been shown in academic research and later discovered in the wild \autocite{glukhov_censorship_2023, openai_disrupting_2025}). One step toward assessing the organization as a whole rather than just individual systems is METR’s analysis of GPT-5.1 Codex-Max, which incorporated a forward-looking extrapolation of OpenAI model capabilities into its risk analysis, given private statements from OpenAI regarding their future expectations and plans \autocite{metr_details_2025}. We are not aware of explicit third-party assessments of frontier AI companies’ safety and security cultures. 

Assessments of different aspects of safety and security require different operating conditions. For example, system-level assessments may require unfettered access and rate limits, whereas the efficacy of platform-level assessments would be undermined by being given “special treatment,” and instead the more important bottleneck may be establishing safe harbor protections for researchers who must violate terms of use to conduct security research. There has also been little third-party investment in assessing whether mitigations are sufficient for a clearly defined threat model --- a gap that is becoming increasingly important as models are approaching or crossing dangerous capability thresholds \autocite{openai_chatgpt_2025}, and as companies routinely report misuse of their product by state and non-state actors \autocite{openai_detecting_2025, anthropic_detecting_2025, google_threat_intelligence_group_adversarial_2025}.

Current assessments focus predominantly on technical consumer-facing systems, particularly the model itself, rather than the full organizational stack. Assessors typically lack visibility into internal processes, safety and security culture, governance structures, platform-level abuse mitigations, and AI systems that are deployed internally within an AI company \autocite{stix_ai_2025}. Some third parties grade public safety and security frameworks from industry labs (\href{https://ailabwatch.org/}{AI Lab Watch}, \href{https://ratings.safer-ai.org/}{SaferAI Ratings}, \href{https://futureoflife.org/ai-safety-index-summer-2025/}{AI Safety Index}), though to date there has been only relatively limited (disclosed) efforts to augment such grading with non-public information. Current third-party work is primarily assessment (measuring claims) rather than verification (confirming specific claims), in part because companies have not yet made sufficiently specific claims that would warrant verification (e.g., companies' safety and security frameworks often set very high levels for what constitutes unacceptable risks, and claims regarding the effectiveness of mitigations are often vague).\footnote{The AI Safety Index’s methodology includes subjective evaluation of companies’ performance by relevant experts. These subjective evaluations will tend to draw on, among other things, non-public information known to these experts, and likely has a role to play in improving public understanding of how companies compare, but this is different in nature from the structured, explicit use of such information that we focus on here.}

One recent example of a step beyond viewing models and systems themselves as the only unit of analysis is OpenAI soliciting third-party assessment of their gpt-oss model \autocite{openai_gpt-oss-120b_2025} --- third parties submitted critiques of and recommendations for the risk assessment and mitigation process, drawing in part on non-public information (namely earlier drafts of risk assessments, particularly focused on fine-tuning the model to increase certain dangerous capabilities) in order to provide such input. Another recent example is Anthropic soliciting input from METR on their pilot sabotage risk report on Claude Opus 4 and 4.1 \autocite{samuel_r_bowman_anthropics_2025}, where METR reviewed company methodology and model-centric evidence including evaluation results, deployment information, and safeguard descriptions. METR also produced an analysis that was based in part on review of an unredacted version of a pertinent safety artifact, allowing METR to speak publicly to the reasonableness of the redactions in the public version \autocite{metr_review_2025}.

\subsection{Scale} \label{ssec:scale}

While competitive, reputational, and legal pressures motivate most leading developers to conduct some safety testing, participation is neither universal nor consistent. OpenAI and Anthropic have established ongoing relationships with third-party evaluators and government-backed institutes. Google DeepMind has also engaged external parties, though with less public detail on these arrangements. Other developers, particularly fast followers and open-weight model developers like Meta, Mistral, and xAI, have been more variable in their engagement with external assessment \autocite{saferai_methodology_nodate, fli_ai_nodate, chang_death_2025}.\footnote{An emerging driver of greater consistency is the EU General-Purpose AI Code of Practice, which requires signatories to undergo independent external evaluations, at least to the extent they are able to find qualified assessors, including for monitoring the model after it has been placed on the market. While signing and complying with the Code of Practice is voluntary, doing so grants developers a presumption of conformity with the EU AI Act. At the time of writing, many --- but not all --- developers have signed the Code of Practice. Notable non-signatories include Meta and Chinese AI companies. The requirement for independent evaluation also depends on signatories being able to find qualified assessors, which may be a constraint given the limited scale of the current ecosystem.}

Another major gap is limited third-party assessment of frontier Chinese AI systems, despite a growing number being built and deployed there. Some efforts have emerged --- for example, Shanghai AI Lab’s Frontier Risk Framework mentions third-party auditing \autocite{shanghai_ai_lab_frontier_2025} --- but these remain exceptions rather than the norm, and it is unclear whether such assessments involve significant use of non-public information analogous to the current (admittedly still limited) practices at American companies. For example, we are not aware of pre-deployment third-party testing of Chinese systems. This highlights challenges in scaling frontier AI auditing globally: legal barriers may restrict foreign auditors' access to domestic AI systems; language and cultural differences can impede understanding of organizational practices and safety culture; and geopolitical sensitivities may limit willingness to grant access to external parties, particularly across rival jurisdictions.\footnote{For example, data localization requirements and cross-border data transfer restrictions can create legal barriers to foreign auditors remotely accessing domestic AI systems. China’s Cybersecurity Law requires that personal information and “important data” be stored locally, with cross-border transfers requiring security assessments and government approval \autocite{sacks_data_2020}. The US has imposed analogous restrictions in the other direction through CFIUS reviews and export controls that can block transactions giving “countries of concern” access to AI-related data and technology \autocite{wang_regulatory_2024}.}

If all frontier AI developers demanded frontier AI auditing at the highest assurance levels, third-party organizations could not meet this demand immediately, though broad coverage of the lower assurance levels is likely achievable by tapping into the talent and networks discussed in \cref{ssec:reaching_full}. Currently, since universal coverage is not required and there is a trade-off between quality and quantity of coverage, assessors make prioritization decisions based on factors like the expected risk of a system and the learning value to assessors of conducting a given assessment (e.g., whether a new type of access or analysis can be pioneered during an engagement).

\subsection{Independence} \label{ssec:independence}

Developers participate in evaluations voluntarily and retain substantial contractual leverage in setting evaluation terms. Because access to future models depends on maintaining cooperative relationships with developers, third-party assessors may face implicit pressure to avoid findings or disclosures that could jeopardize continued access. Public reporting of safety incidents by companies themselves or their assessors can lead to media backlash \autocite{liv_mcmahon_ai_2025}, which could incentivize developers to obscure important safety information or assessors to soften critical conclusions.
While similar dynamics exist in other auditing domains, these regimes have developed greater institutional safeguards such as mandatory audits, standardized terms, and regulatory oversight that partially mitigate these pressures.

\subsection{Ecosystem Maturity} \label{ssec:ecosystem_maturity}

Ad hoc algorithmic audits, often conducted by academics and interest groups, have a significant history of precedent and impact \autocite{raji_actionable_2019}. However, the current ecosystem of specialized third party auditing organizations is much more nascent. Dedicated safety evaluation capacity is concentrated in a handful of specialized non-profits and government-backed institutes, likely comprising only a few hundred full-time employees specifically focused on frontier AI.\footnote{Our estimate draws on publicly available information about the main organizations conducting third-party frontier AI safety assessment. Among non-profits, METR employs approximately 21--50 staff and Apollo Research approximately 15--20. Government-backed institutes are larger but vary considerably: the UK AI Security Institute reports over 100 technical staff and £66 million in annual funding, while the US Center for AI Standards and Innovation operates with substantially fewer resources. These figures are approximate and subject to change.} In contrast, for security evaluation, traditional enterprise security auditing practices (penetration testing, SOC 2 compliance, ISO 27001 certification) are mature and widely adopted, and some leading developers do engage conventional security auditors (e.g., SOC 2) \autocite{openai_security_2025, anthropic_anthropic_2025, google_cloud_google_nodate}. However, AI-specific risk assessment --- evaluating protections for model weights, adversarial robustness, and novel attack surfaces unique to machine learning systems --- remains far less developed \autocite{tabassi_artificial_2023}. This asymmetry means that while many developers may have robust conventional security postures, the specialized security challenges posed by frontier AI systems receive comparatively less external scrutiny.\looseness=-1

Private evaluation organizations are typically non-profits that specialize in a specific type or domain of evaluation. While they perform evaluations separately, private auditing organizations are beginning to publicly coordinate. The \href{https://aievaluatorforum.org/about/members}{AI Evaluator Forum} was established in December 2025 with the goal of allowing evaluation organizations to coordinate on shared standards \autocite{ai_evaluator_forum_our_nodate}. Its founding members are \href{https://transluce.org/}{Transluce}, \href{https://metr.org/}{METR}, \href{https://www.rand.org/}{RAND}, \href{https://www.averi.org/}{AVERI}, \href{https://securebio.org/}{SecureBio}, Princeton \href{https://hal.cs.princeton.edu/}{HAL}, \href{https://www.cip.org/}{The Collective Intelligence Project}, and \href{https://meridianlabs.ai/}{Meridian Labs}. All are located in the United States and the United Kingdom, although Shanghai AI Lab’s Frontier Risk Framework mentions third-party auditing \autocite{shanghai_ai_lab_frontier_2025}, indicating the geographic distribution may result more from an absence of third-party evaluation than an absence of political support for the practice.

Evaluation organizations often differentiate themselves by domain focus or methodological approach. For example, \href{https://www.apolloresearch.ai/}{Apollo Research} focuses primarily on evaluations for deception, \href{https://transluce.org/}{Transluce} emphasizes white-box evaluations, and \href{https://securebio.org/}{SecureBio} assesses biorisks. While private evaluation organizations frequently publish research outputs, relatively little information is publicly available about specific audits. This opacity largely reflects the fact that evaluations are typically conducted under bespoke contractual terms negotiated with auditees, which often constrain disclosure.

Some national governments also conduct third-party evaluations of frontier AI systems. In 2024, an informal International Network of AI Safety Institutes (later renamed the Network for Advanced AI Measurement, Evaluation, and Science) was established, in part to coordinate on evaluations and to conduct joint testing exercises \autocite{noauthor_international_2024}. Network members included Australia, Canada, the European Commission, France, Japan, Kenya, the Republic of Korea, Singapore, the United Kingdom, and the United States \autocite{noauthor_international_2024}.\footnote{In December 2025, the network was renamed the Network for Advanced AI Measurement, Evaluation and Science \autocite{department_for_science_innovation_and_technology_efforts_2025}.} National governments increasingly conduct their own evaluations independently, often specializing in risks that align with national priorities and security expertise. However, there is some precedent for multinational coordination. In July 2025, a joint agent evaluation effort was announced involving Singapore, Japan, Australia, Canada, the European Commission, France, Kenya, South Korea, and the United Kingdom \autocite{aisi_international_nodate}. Public reporting on government-led assessments remains uncommon and limited. The UK’s AI Security Institute stands out in how it publicly shares a relatively large (but still limited) amount of information about evaluation methods and findings \autocite{aisi_aisi_nodate}.\footnote{Notably, not all information about assessments should be published (at least immediately after they are conducted), since such information could hurt the integrity of future assessments. See discussion in \cref{sec:vision} regarding the need for auditors to keep (some of) their methods private.}

\newpage 
\appsection{Risks of and alternatives to ``auditee pays'' models}\label{apx:risks}

Today, AI assessment organizations are generally funded by a combination of philanthropy and frontier AI company payments. Several alternative funding models are foreseeable, including auditor payment by insurers, regulator-administered funding pools, payments from downstream enterprise users in high-stakes or regulated sectors, industry-wide levies administered by an industry body or independent entity, or hybrid approaches combining these mechanisms. These alternatives could provide stronger independence guarantees while ensuring adequate funding. In some cases, even a small levy could support a substantial expansion of the AI auditing ecosystem.

When auditors compete for contracts awarded by the companies they evaluate, there is a significant risk that auditing devolves into ``rubber stamping,'' as occurred in financial auditing prior to the 2008 crisis. In such arrangements, conscious or unconscious bias toward outcomes favorable to the client can emerge. Funding models in which auditors are paid by insurers or regulators therefore merit particular attention, as they better align auditor incentives with accurate risk assessment rather than client satisfaction. Insurers have strong financial incentives for accurate risk quantification, making them natural principals for auditing services.

Empirical evidence on the effectiveness of these alternative models remains limited (with few exceptions, such as \autocite{duflo_truth-telling_2013}), and the ideal model for the frontier AI industry remains unclear. Accordingly, we do not recommend any specific payment model at this stage but recommend research toward answering this question.\looseness=-1

\newpage
\appsection{Frontier AI definitions and different thresholds for triggering audits}\label{apx:defs}

Our definition of frontier AI --- ``general-purpose AI models and systems whose
performance is no more than a year behind the state-of-the-art on a broad suite of general capability benchmarks'' --- is similar to the definition used by the Frontier Model Forum \autocite{fmf_frontier_nodate}, which is ``a general-purpose model that outperforms, based on a range of conventional performance benchmarks or high-risk capability assessments, all other models that have been widely deployed for at least 12 months.'' The main difference is that we include systems, not just models, as a central part of the definition. 

We use the 12 month threshold, but we acknowledge there are problems with relying on a single threshold, and with temporal thresholds generally. The reason one might want to use this kind of approach, at least for exploratory research and policy discussions, is to convey that actual system capabilities are the focus, rather than inputs, which are only a proxy for those capabilities. However, imperfect proxies are often easier to administer and communicate, and can be more predictable for (potentially) regulated companies. This helps explain why computing power-based thresholds, revenue-based thresholds, and expenditure-based thresholds are more common in regulatory contexts.

We don’t think the choice of threshold significantly changes our basic proposal for auditing processes and incentive design, but it’s important to ensure that any codification of a frontier AI threshold anchored to auditing requirements has the capacity to be changed over time \autocite{casper_practical_2025}, given the fast-moving nature of AI. A metric that works today may not work as well a year from now. This is important to consider for both public sector actors such as legislators writing AI legislation, as well as private sector actors such as insurers writing standard policies.

A good example of how a proxy can go wrong is the training required in order to ``pre-train'' a language model, which was an early metric used to determine which AI models or systems were subject to frontier AI regulations. Computing power and design decisions at each stage of the supply chain provide concrete, quantitatively precise opportunities for regulatory intervention compared to some alternatives \autocite{sastry_computing_2024, lee_talkin_2024}, but anchoring on a specific way that computing power is used can be perilous. While pre-training compute remains important, two other factors --- reinforcement learning, another type of training which is different from pre-training, and ``test-time compute,'' the amount of computing power used when running a model, which can be increased in order to give better results --- have increased in relative importance. See also \autocite{heim_training_2024} and \autocite{casper_practical_2025}. Likewise, major technical developments can cause distinct but similar issues with temporal thresholds, in that a slowdown in general capability scores would lead to very few systems being in scope, and vice versa.

A different challenge with our definition is that it specifically focuses on general capabilities. Some types of AI systems (e.g., trained on certain kinds of data) might have dangerous capabilities in certain areas despite scoring poorly on general capability benchmarks. While such systems are out of scope of our definition, that does not mean that they should not be considered for auditing, and one could imagine adapting our approach to encompass some such systems (e.g., through multiple sufficient thresholds, one based on general capabilities and others based on specific ``risk verticals'').

Different stakeholders might be more willing to tolerate ``false positives'' from a given threshold (unduly burdensome audits given a system or company’s real risk profile) or ``false negatives'' (unaudited or weakly audited systems or companies that are more dangerous than the level of scrutiny applied to them would suggest). There are ways to calibrate risk judgments in an adaptive fashion in order to reduce the total amount of errors, such as building ``triaging'' ---technically knowledgeable regulators having the ability to grant exemptions rapidly --- into the audit process itself, but each has its own challenges, and inevitably there will be some imprecision. 

There are various trade-offs to consider when setting these thresholds. First, thresholds should strike a balance between stimulating demand (i.e., causing more, and more rigorous, audits to occur than would have existed otherwise) and incentivizing corner-cutting (i.e., encouraging auditors to ``churn out'' low-quality audits). These are reasons why we emphasize the need for market analysis and quality standards in our recommendations, but these will at best soften, rather than eliminate, this basic tension. Second, thresholds should strike a balance between mitigating safety and security risks from frontier AI (and, possibly, particularly dangerous narrow AI systems) and enabling beneficial AI innovations. There is complexity on both sides of this ledger. It is important to consider the direct safety and security risks from a given company, as well as the ``horizontal'' and ``downward'' learning from audits discussed in \cref{sec:motivations}. Small startups that lack the capacity to undergo even low assurance level audits might forgo launching products. Even larger companies who can afford higher assurance levels might view audits as too burdensome.  

Note also that it is difficult to reason about the burdensomeness of audits in isolation --- the safety and security standards that are audited against are also important to consider, which we treat as distinct from the assurance level at which a system and company are evaluated against a given set of standards. A range of other factors are potentially relevant, as well, such as the context in which a frontier AI system might be deployed.

\newpage
\section*{Acknowledgments}
Many gave us valuable feedback on earlier versions of the ideas discussed here and earlier versions of this paper, including but not limited to Mark Greaves, John Bailey, Tyler Cowen, Eileen Donahoe, Nathan Lambert, Geoff Ralston, Gopal Sarma, Adam Woodhall, Larissa Schiavo, Shahar Avin, Andrew Gamino-Cheong, Tom Zick, and Vijay Bolina. We’re also grateful to Stone Addington and Carly Tryens for general support and Eden Beck and Erol Can Akbaba for assistance with formatting and copyediting. None of those listed here necessarily endorse the contents of the paper.

\end{document}